\documentclass[english,prx,twocolumn,groupedaddress,reprint]{revtex4}
\usepackage[T1]{fontenc}
\usepackage[latin9]{inputenc}
\usepackage{color}
\usepackage{amsmath}
\usepackage{amssymb}
\usepackage{graphicx}
\usepackage{esint}

\makeatletter

\providecommand{\tabularnewline}{\\}

\@ifundefined{textcolor}{}
{%
 \definecolor{BLACK}{gray}{0}
 \definecolor{WHITE}{gray}{1}
 \definecolor{RED}{rgb}{1,0,0}
 \definecolor{GREEN}{rgb}{0,1,0}
 \definecolor{BLUE}{rgb}{0,0,1}
 \definecolor{CYAN}{cmyk}{1,0,0,0}
 \definecolor{MAGENTA}{cmyk}{0,1,0,0}
 \definecolor{YELLOW}{cmyk}{0,0,1,0}
}

\usepackage{bbold}
\usepackage{tikz}

\usepackage{dsfont}
\RequirePackage[colorlinks=true,linkcolor=blue]{hyperref}

\makeatother

\usepackage{babel}
\begin{document}

\title{Universal Quantum Transducers based on Surface Acoustic Waves }

\author{M. J. A. Schuetz,$^{1}$ E. M. Kessler,$^{2,3}$ G. Giedke,$^{1,4,5}$
L. M. K. Vandersypen,$^{6}$ M. D. Lukin,$^{2}$ and J. I. Cirac$^{1}$ }

\affiliation{$^{1}$Max-Planck-Institut für Quantenoptik, Hans-Kopfermann-Str.
1, 85748 Garching, Germany}

\affiliation{$^{2}$Physics Department, Harvard University, Cambridge, MA 02318,
USA}

\affiliation{$^{3}$ITAMP, Harvard-Smithsonian Center for Astrophysics, Cambridge,
MA 02318, USA}

\affiliation{$^{4}$Donostia International Physics Center, Paseo Manuel de Lardizabal
4, E-20018 San Sebasti\'an, Spain}

\affiliation{$^{5}$Ikerbasque Foundation for Science, Maria Diaz de Haro 3, E-48013
Bilbao, Spain}

\affiliation{$^{6}$Kavli Institute of NanoScience, TU Delft, P.O. Box 5046, 2600
GA Delft, The Netherlands}

\date{\today}
\begin{abstract}
We propose a universal, on-chip quantum transducer based on surface
acoustic waves in piezo-active materials. Because of the intrinsic
piezoelectric (and/or magnetostrictive) properties of the material,
our approach provides a universal platform capable of coherently linking
a broad array of qubits, including quantum dots, trapped ions, nitrogen-vacancy
centers or superconducting qubits. The quantized modes of surface
acoustic waves lie in the gigahertz range, can be strongly confined
close to the surface in phononic cavities and guided in acoustic waveguides.
We show that this type of surface acoustic excitations can be utilized
efficiently as a quantum bus, serving as an on-chip, mechanical cavity-QED
equivalent of microwave photons and enabling long-range coupling of
a wide range of qubits. 
\end{abstract}
\maketitle

\section{Introduction}

The realization of long-range interactions between remote qubits is
arguably one of the greatest challenges towards developing a scalable,
solid-state spin based quantum information processor \cite{hanson08}.
One approach to address this problem is to interface qubits with a
common \textit{quantum bus} which distributes quantum information
between distant qubits. \textcolor{black}{The transduction of quantum
information between stationary and moving qubits is central to this
approach. }A particularly efficient implementation of such a quantum
bus can be found in the field of circuit QED where spatially separated
superconducting qubits interact via microwave photons confined in
transmission line cavities \cite{schoelkopf08,walraff04,blais04}.
In this way, multiple qubits have been coupled successfully over relatively
large distances of the order of millimeters \cite{sillanp=0000E4=0000E407,majer07}.
Fueled by the dramatic advances in the fabrication and manipulation
of nanomechanical systems \cite{oconnell10}, an alternate line of
research has pursued the idea of coherent, long-range interactions
between individual qubits mediated by mechanical resonators, with
resonant \textit{phonons} playing the role of cavity photons \cite{kolkowitz12,rabl10,soykal11,habraken12,ruskov13,cleland04}. 

\begin{figure}[b]
\includegraphics[width=0.95\columnwidth]{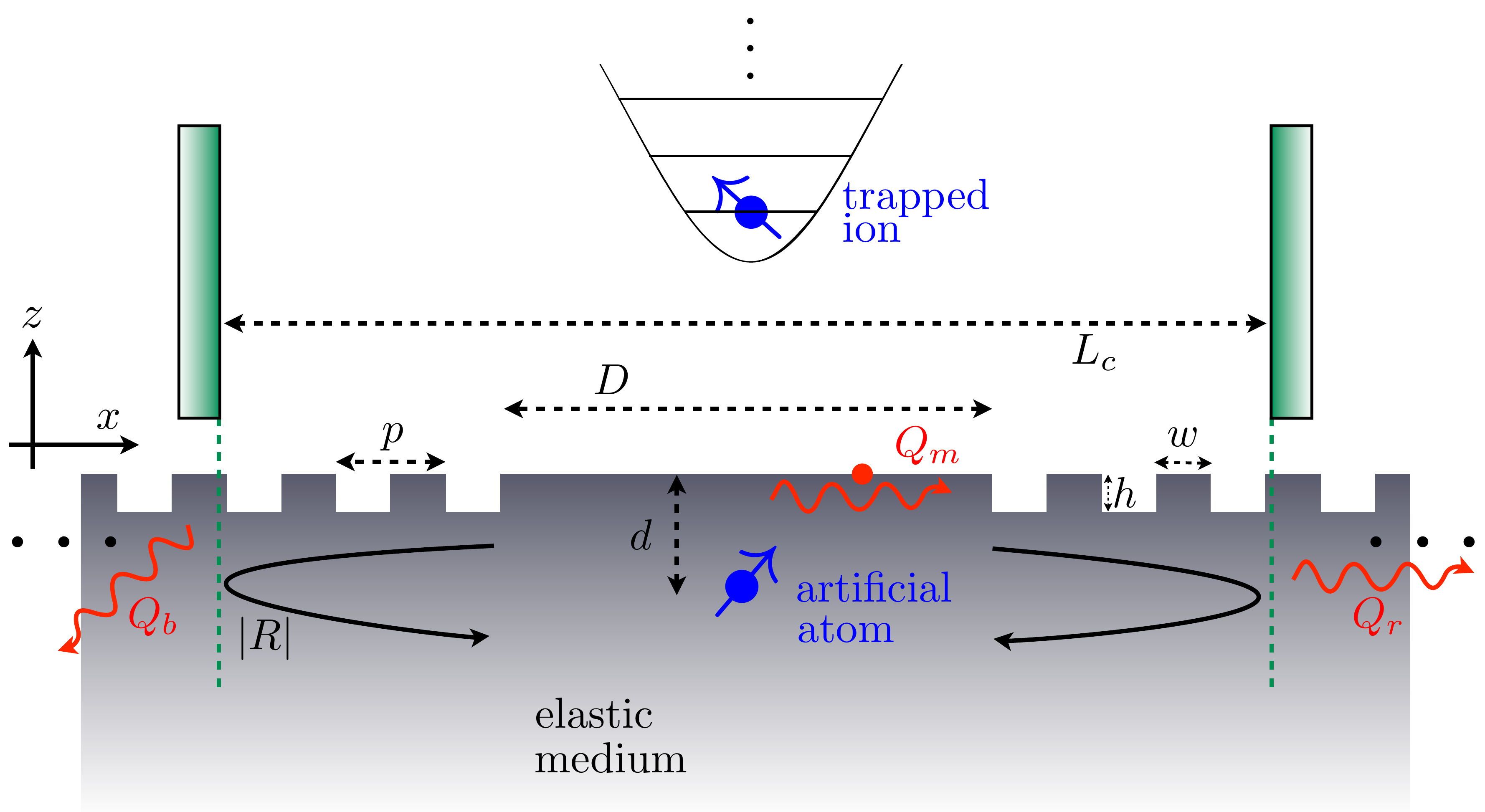}

\caption{\label{fig:SAW-system-sketch}(color online). SAW as a universal quantum
transducer. Distributed Bragg reflectors made of grooves form an acoustic
cavity for surface acoustic waves. The resonant frequency of the cavity
is determined by the pitch $p$, $f_{c}=v_{s}/2p$. Reflection occurs
effectively at some distance inside the grating;\textcolor{black}{{}
the fictitious mirrors above the surface are not part of the actual
experimental setup, but shown for illustrative purposes only.} Red
arrows indicate the relevant decay channels for the cavity mode: leakage
through the mirrors, internal losses due to for example surface imperfections,
and conversion into bulk modes. Qubits inside and outside of the solid
can be coupled to the cavity mode. In more complex structures, the
elastic medium can consist of multiple layers on top of some substrate. }
\end{figure}

\textcolor{black}{In this paper, we propose a new realization of a
quantum transducer and data bus based on surface acoustic waves (SAW).
SAWs involve phonon-like excitations bound to the surface of a solid
and are widely used in modern electronic devices e.g. as compact microwave
filters \cite{morgan07,datta86}. Inspired by two recent experiments
}\cite{gustaffson12,gustafsson14}\textcolor{black}{, where the coherent
quantum nature of surface acoustic waves (SAWs) has been explored,
here we propose and analyze SAW phonon modes in piezo-active materials
as a universal mediator for long-range couplings between remote qubits.
Our approach involves qubits interacting with a localized SAW phonon
mode, defined by a high-$Q$ resonator, which in turn can be coupled
weakly to a SAW waveguide serving as a quantum bus; as demonstrated
below, the qubits can be encoded in a great variety of spin or charge
degrees of freedom. We show that the Hamiltonian for an individual
node (for a schematic representation see Fig.\ref{fig:SAW-system-sketch})
can take on the generic Jaynes-Cummings form $\left(\hbar=1\right)$,
\begin{equation}
H_{\mathrm{node}}=\frac{\omega_{\mathrm{q}}}{2}\sigma^{z}+\omega_{c}a^{\dagger}a+g\left(\sigma^{+}a+\sigma^{-}a^{\dagger}\right),\label{eq:generic-Jaynes-Cummings-Hamiltonian}
\end{equation}
where $\vec{\sigma}$ refers to the usual Pauli matrices describing
the qubit with transition frequency $\omega_{\mathrm{q}}$ and $a$
is the bosonic operator for the localized SAW cavity mode of frequency
$\omega_{c}/2\pi\sim\mathrm{GHz}$ \cite{JC-comment}. The coupling
$g$ between the qubit and the acoustic cavity mode is mediated intrinsically
by the piezo-properties of the host material, it is proportional to
the electric or magnetic zero-point fluctuations associated with a
single SAW phonon and, close to the surface, can reach values of $g\sim400\mathrm{MHz}$,
much larger than the relevant decoherence processes and sufficiently
large to allow for quantum effects and coherent coupling in the spin-cavity
system as evidenced by cooperativities \cite{cooperativity} of $C\sim10-100$
{[}see Section \ref{sec:Universal-Coupling} and Tab.\ref{tab:g-cooperativity}
for definition and applicable values{]}. For $\omega_{\mathrm{q}}\approx\omega_{c}$,
$H_{\mathrm{node}}$ allows for a controlled mapping of the qubit
state onto a coherent phonon superposition, which can then be mapped
to an itinerant SAW phonon in a waveguide, opening up the possibility
to implement on-chip many quantum communication protocols well known
in the context of optical quantum networks \cite{cirac97,habraken12}. }

The most pertinent features of our proposal can be summarized as follows:
(1) Our scheme is not specific to any particular qubit realization,
but---thanks to the plethora of physical properties associated with
SAWs in piezo-active materials (strain, electric and magnetic fields)---provides
a common on-chip platform accessible to various different implementations
of qubits, comprising both natural (e.g., ions) and artificial candidates
such as quantum dots or superconducting qubits. In particular, this
opens up the possibility to interconnect dissimilar systems in new
electro-acoustic quantum devices. (2) Typical SAW frequencies lie
in the gigahertz range, closely matching\textcolor{magenta}{{} }transition
frequencies of artificial atoms and enabling ground state cooling
by conventional cryogenic techniques. (3) Our scheme is built upon
an established technology \cite{morgan07,datta86}: Lithographic fabrication
techniques provide almost arbitrary geometries with high precision
as evidenced by a large range of SAW devices such as delay lines,
bandpass filters or resonators etc. In particular, the essential building
blocks needed to interface qubits with SAW phonons have already been
fabricated, according to design principles familiar from electromagnetic
devices: (i) SAW resonators, the mechanical equivalents of Fabry-Perot
cavities, with low-temperature measurements reaching quality factors
of $Q\sim10^{5}$ even at gigahertz frequencies \cite{elHabti96,manenti13,magnusson14},
and (ii) acoustic waveguides as analogue to optical fibers \cite{morgan07}.\textcolor{magenta}{{}
}(4) \textcolor{black}{For a given frequency in the gigahertz range,
due to the slow speed of sound of approximately $\sim10^{3}\mathrm{m/s}$
for typical materials, device dimensions are in micrometer range,
which is convenient for fabrication and integration with semiconductor
components, and about $10^{5}$ times smaller than corresponding electromagnetic
resonators. }(5) Since SAWs propagate elastically on the surface of
a solid within a depth of approximately one wavelength, the mode volume
is intrinsically confined in the direction normal to the surface.
Further surface confinement then yields large zero-point fluctuations.
(6) Yet another inherent advantage of our system is the intrinsic
nature of the coupling. In piezoelectric materials, the SAW is accompanied
by an electrical potential $\phi$ which has the same spatial and
temporal periodicities as the mechanical displacement and provides
an intrinsic qubit-phonon coupling mechanism. For example, recently
qubit lifetimes in GaAs singlet-triplet qubits were found to be limited
by the piezoelectric electron-phonon coupling \cite{kornich14}. \textcolor{black}{Here,
our scheme provides a new paradigm, where coupling to phonons becomes
a highly valuable asset for coherent quantum control rather than a
liability. }

\textcolor{black}{In what follows, we first review the most important
features of surface acoustic waves, with a focus on the associated
zero-point fluctuations. Next, we discuss the different components
making up the SAW-based quantum transducer and the acoustic quantum
network it enables: SAW cavities, including a detailed analysis of
the achievable quality factor $Q$, SAW waveguides and a variety of
different candidate systems serving as qubits. Lastly, as exemplary
application, we show how to transfer quantum states between distant
nodes of the network under realistic conditions.}\textcolor{blue}{{}
}\textcolor{black}{Finally, we draw conclusions and give an outlook
on future directions of research.}

\section{SAW Properties}

Elastic waves in piezoelectric solids are described by 
\begin{eqnarray}
\rho\ddot{u}_{i}-c_{ijkl}\partial_{j}\partial_{l}u_{k} & = & e_{kij}\partial_{j}\partial_{k}\phi,\label{eq:Newton-third-law-PE-material}\\
\epsilon_{ij}\partial_{i}\partial_{j}\phi & = & e_{ijk}\partial_{i}\partial_{k}u_{j},\label{eq:Poisson-PE-material}
\end{eqnarray}
where the vector ${\bf u}\left({\bf x},t\right)$ denotes the displacement
field (${\bf x}$ is the cartesian coordinate vector), $\rho$ is
the mass density and repeated indices are summed over $\left(i,j=x,y,z\right)$;
$\underline{c}$, $\underline{\epsilon}$ and $\underline{e}$ refer
to the elasticity, permittivity and piezoelectric tensors, respectively
\cite{simon96}; they are largely defined by crystal symmetry \cite{royer00}.
For example, for cubic crystals such as GaAs there is only one non-zero
component for the permittivity and the piezoelectric tensor, labeled
as $\epsilon$ and $e_{14}$, respectively \cite{simon96}. Since
elastic disturbances propagate much slower than the speed of light,
it is common practice to apply the so-called quasi-static approximation
\cite{royer00} where the electric field is given by $E_{i}=-\partial_{i}\phi$.
When considering surface waves, Eq.(\ref{eq:Newton-third-law-PE-material})
and (\ref{eq:Poisson-PE-material}) must be supplemented by the mechanical
boundary condition that there should be no forces on the free surface\textcolor{black}{{}
(taken to be at $z=0$ with $\hat{z}$ being the outward normal to
the surface), that is $T_{zx}=T_{zy}=T_{zz}=0$ at $z=0$ (where $T_{ij}=c_{ijkl}\partial_{l}u_{k}+e_{kij}\partial_{k}\phi$
is the stress tensor), }and appropriate electrical boundary conditions
\cite{simon96}. 

If not stated otherwise, the term SAW refers to the prototypical (piezoelectric)
Rayleigh wave solution as theoretically and experimentally studied
for example in Refs.\cite{simon96,gustaffson12,gustafsson14,aizin98}
and used extensively in different electronic devices \cite{datta86,morgan07}.
It is non-dispersive, decays exponentially into the medium with a
characteristic penetration depth of a wavelength and has a phase velocity
$v_{s}=\omega/k$ that is lower than the bulk velocities in that medium,
because the solid behaves less rigidly in the absence of material
above the surface \cite{royer00}. As a result, it cannot phase-match
to any bulk-wave \cite{white70,morgan07}.\textcolor{black}{{} As usual,
we consider specific orientations for which the piezoelectric field
produced by the SAW is strongest \cite{morgan07,white70}, for example
a SAW with wavevector along the {[}110{]} direction of a (001) GaAs
crystal {[}cf. Refs.\cite{simon96,gustaffson12} and Appendices \ref{sec:Nonpiezoelectric-SAW}
and \ref{sec:Piezoelectric-SAW}{]}.}

\textit{SAWs in quantum regime.}---\textcolor{black}{In a semiclassical
picture, an acoustic phonon associated with a SAW creates a time-dependent
strain field, $s_{kl}=\left(\partial_{l}u_{k}+\partial_{k}u_{l}\right)/2$
and a (quasi-static) electrical potential $\phi\left({\bf x},t\right)$.
}Upon quantization, the mechanical displacement becomes an operator
that can be expressed in terms of the elementary normal modes as $\hat{{\bf u}}\left({\bf x}\right)=\sum_{n}\left[{\bf v}_{n}\left({\bf x}\right)a_{n}+\mathrm{h.c.}\right]$,
where $a_{n}\left(a_{n}^{\dagger}\right)$ are bosonic annihilation
(creation) operators for the vibrational eigenmode $n$ and the set
of normal modes ${\bf v}_{n}\left({\bf x}\right)$ derives from the
Helmholtz-like equation $\mathcal{W}{\bf v}_{n}\left({\bf x}\right)=-\rho\omega_{n}^{2}{\bf v}_{n}\left({\bf x}\right)$
associated with Eq.(\ref{eq:Newton-third-law-PE-material}) and (\ref{eq:Poisson-PE-material}).\textcolor{black}{{}
The mode normalization is given by $\int d^{3}{\bf x}\rho{\bf v}_{n}^{*}\left({\bf x}\right)\cdot{\bf v}_{n}\left({\bf x}\right)=\hbar/2\omega_{n}$
\cite{stroscio96,kornich14}.} An important figure of merit in this
context is the amplitude of the mechanical zero-point motion $U_{0}$.
Along the lines of cavity QED \cite{schoelkopf08}, a simple estimate
for $U_{0}$ can be obtained by equating the \textit{classical} energy
of a SAW $\sim\int d^{3}{\bf x}\rho\dot{{\bf u}}^{2}$ with the \textit{quantum}
energy of a single phonon, that is $\hbar\omega$. This leads directly
to 
\begin{equation}
U_{0}\approx\sqrt{\hbar/2\rho v_{s}A},\label{eq:mechanical-zero-point-motion}
\end{equation}
where we used the dispersion relation $\omega=v_{s}k$ and the intrinsic
mode confinement $V\approx A\lambda$ characteristic for SAWs. The
quantity $U_{0}$ refers to the mechanical amplitude associated with
a single SAW phonon close to the surface. It depends only on the material
parameters $\rho$ and $v_{s}$ and follows a generic $\sim A^{-1/2}$
behaviour, where $A$ is the effective mode area on the surface. The
estimate given in Eq.(\ref{eq:mechanical-zero-point-motion}) agrees
very well with more detailed calculations presented in Appendix \ref{sec:Mechanical-Zero-Point-Fluctuation}.
Several other important quantities\textcolor{black}{{} which are central
for signal transduction between qubits and SAWs} follow directly from
$U_{0}$: The (dimensionless) zero-point strain can be estimated as
$s_{0}\approx kU_{0}$. The intrinsic piezoelectric potential associated
with a single phonon derives from Eq.(\ref{eq:Poisson-PE-material})
as $\phi_{0}\approx\left(e_{14}/\epsilon\right)U_{0}$ \cite{numerical-values-phi0}.
Lastly, the electric field amplitude due to a single acoustic phonon
is $\xi_{0}\approx k\phi_{0}=\left(e_{14}/\epsilon\right)kU_{0}$,
illustrating the linear relation between electric field and strain
characteristic for piezoelectric materials \cite{cleland04}. \textcolor{black}{In
summary, we typically find $U_{0}\approx2\mathrm{fm}/\sqrt{A\left[\mu\mathrm{m}^{2}\right]}$,
yielding $U_{0}\approx2\mathrm{fm}$ for micron-scale confinement
(cf. Appendix \ref{sec:Zero-Point-Estimates}). This is comparable
to typical zero-point fluctuation amplitudes of localized mechanical
oscillators \cite{aspelmeyer14}.} Moreover, for micron-scale surface
confinement\textcolor{black}{{} and GaAs material parameters, }we obtain
$\xi_{0}\approx20\mathrm{V/m}$ which compares favorably with typical
values of $\sim10^{-3}\mathrm{V/m}$ and $\sim0.2\mathrm{V/m}$ encountered
in cavity and circuit QED, respectively \cite{schoelkopf08}. 

For the sake of clarity, we have focused on piezo-\textit{electric}
materials so far. However, there are also piezo-\textit{magnetic}
materials that exhibit a large magnetostrictive effect. In that case,
elastic distortions are coupled to a (quasi-static) magnetic instead
of electric field \cite{nowacki06,cai14}; for details see Appendix
\ref{sec:Zero-Point-Estimates}. For typical materials such as Terfenol-D
the magnetic field associated with a single phonon can be estimated
as $B_{0}\approx(2-6)\mu\mathrm{T}/\sqrt{A\left[\mu\mathrm{m}^{2}\right]}$.
Finally, we note that composite structures comprising both piezoelectric
and piezomagnetic materials can support magneto-electric surface acoustic
waves \cite{pang08,liu08}.

\section{SAW Cavities \& Waveguides}

\begin{figure}
\includegraphics[width=1\columnwidth]{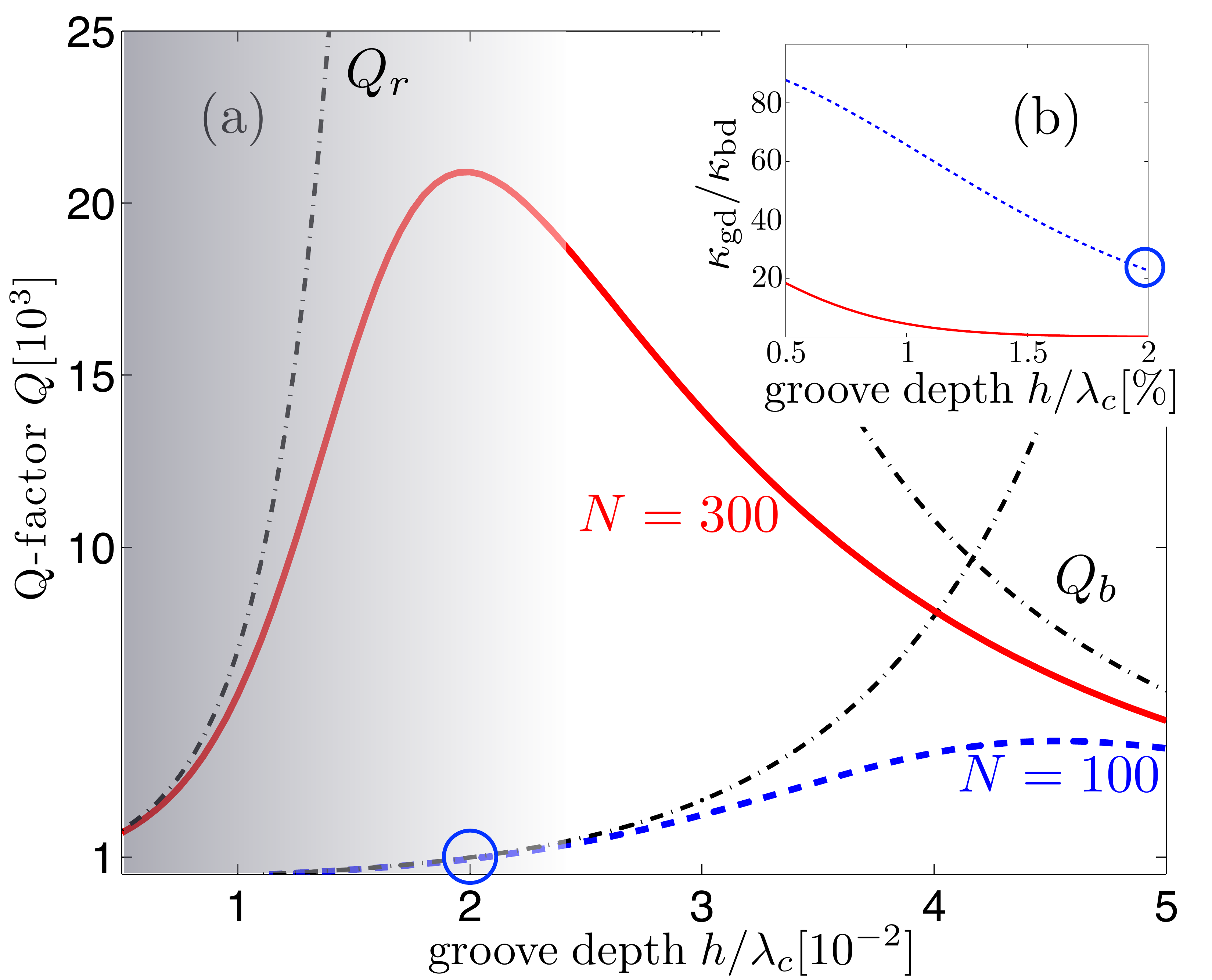}

\caption{\label{fig:SAW-cavity}(color online). Characterization of a groove-based
SAW cavity. (a) Quality factor $Q$ for $N=100$ (dashed blue) and
$N=300$ (red solid) grooves as a function of the normalized grove
depth $h/\lambda_{c}$. For shallow grooves, $Q$ is limited by leakage
losses due to imperfect acoustic mirrors ($Q_{r}$-regime, gray area),
whereas for deep grooves conversion to bulk modes dominates ($Q_{b}$-regime);
compare asymptotics (dash-dotted lines). (b) Ratio of desired to undesired
decay rates $\kappa_{\mathrm{gd}}/\kappa_{\mathrm{bd}}$. The stronger
$Q$ is dominated by $Q_{r}$, the higher $\kappa_{\mathrm{gd}}/\kappa_{\mathrm{bd}}$.
Here, $w/p=0.5$, $D=5.25\lambda_{c}$, and $f_{c}=3\mathrm{GHz}$;
typical material parameters for $\mathrm{LiNbO}_{3}$ have been used
{[}cf. Appendix \ref{sec:SAW-Cavities}{]}. }
\end{figure}

\textit{SAW cavities.}---To boost single phonon effects, it is essential
to increase $U_{0}$. In analogy to cavity QED, this can be achieved
by confining the SAW mode in an acoustic resonator. The physics of
SAW cavities has been theoretically studied and experimentally verified
since the early 1970s \cite{morgan07,bell76}. Here, we provide an
analysis of a SAW cavity based on an on-chip distributed Bragg reflector
in view of potential applications in quantum information science;
for details see Appendix \ref{sec:SAW-Cavities}. SAW resonators of
this type can usually be designed to host a single resonance $f_{c}=\omega_{c}/2\pi=v_{s}/\lambda_{c}$
($\lambda_{c}=2p$) only and can be \textcolor{black}{viewed as an
acoustic Fabry-Perot resonator with effective reflection centers,
sketched by localized mirrors in Fig.\ref{fig:SAW-system-sketch},
situated at some effective penetration distance into the grating }\cite{morgan07}\textcolor{black}{.
Therefore, the total effective cavity size along the mirror axis is
$L_{c}>D$, where $D$ is the physical gap between the gratings}.
The total cavity line-width $\kappa=\omega_{c}/Q=\kappa_{\mathrm{gd}}+\kappa_{\mathrm{bd}}$
can be decomposed into desired (leakage through the mirrors) and undesired
(conversion into bulk modes and internal losses due to surface imperfections
etc.) losses, labeled as $\kappa_{\mathrm{gd}}$ and $\kappa_{\mathrm{bd}}$,
respectively; for a schematic illustration compare Fig.\ref{fig:SAW-system-sketch}.
For the total quality factor $Q$, we can typically identify three
distinct regimes {[}cf. Fig.\ref{fig:SAW-cavity}{]}: For very small
groove depths $h/\lambda_{c}\lesssim2\%$, losses are dominated by
coupling to SAW modes outside of the cavity, dubbed as $Q_{r}$-regime
$\left(\kappa_{\mathrm{gd}}\gg\kappa_{\mathrm{bd}}\right)$, whereas
for very deep grooves losses due to conversion into bulk-modes become
excessive ($Q_{b}$-regime, $\kappa_{\mathrm{gd}}\ll\kappa_{\mathrm{bd}}$).
In between, for a sufficiently high number of grooves $N$, the quality
factor $Q$ can ultimately be limited by internal losses (surface
cracks etc.), referred to as $Q_{m}$-regime $\left(\kappa_{\mathrm{gd}}\ll\kappa_{\mathrm{bd}}\right)$.
For $N\approx300$, we find that the onset of the bulk-wave limit
occurs for $h/\lambda_{c}\gtrsim2.5\%$, in excellent agreement with
experimental findings \cite{bell76,li75}. With regard to applications
in quantum information schemes, the $Q_{r}$-regime plays a special
role in that resonator phonons leaking out through the acoustic mirrors
can be processed further by guiding them in acoustic SAW waveguides
(see below). To capture this behaviour quantitatively, we analyze
$\kappa_{\mathrm{gd}}/\kappa_{\mathrm{bd}}$ {[}cf. Fig.\ref{fig:SAW-cavity}{]}:
for $\kappa_{\mathrm{gd}}/\kappa_{\mathrm{bd}}\gg1$, leakage through
the mirrors is the strongest decay mechanism for the cavity phonon,
whereas the undesired decay channels are suppressed. Our analysis
shows that, for gigahertz frequencies $f_{c}\approx3\mathrm{GHz}$,
$N\approx100$ and $h/\lambda_{c}\approx2\%$, a quality factor of
$Q\approx10^{3}$ is achievable, together with an effective cavity
confinement $L_{c}\approx40\lambda_{c}$ (for $D\lesssim5\lambda_{c}$)
and $\kappa_{\mathrm{gd}}/\kappa_{\mathrm{bd}}\gtrsim20$ (illustrated
by the circle in Fig.\ref{fig:SAW-cavity}); accordingly,\textcolor{black}{{}
the probability for a cavity phonon to leak through the mirrors (rather
than into the bulk for example) is $\kappa_{\mathrm{gd}}/\left(\kappa_{\mathrm{gd}}+\kappa_{\mathrm{bd}}\right)\gtrsim95\%$.}
Note that the resulting total cavity linewidth of $\kappa/2\pi=f_{c}/Q\approx\left(1-3\right)\mathrm{MHz}$
is similar to the ones typically encountered in circuit QED \cite{sillanp=0000E4=0000E407}.
To compare this to the effective cavity-qubit coupling, we need to
fix the effective mode area of the SAW cavity. In addition to the
longitudinal confinement by the Bragg mirror (as discussed above)
a transverse confinement length $L_{\mathrm{trans}}$ (in direction
\textcolor{black}{$\hat{y}$}) can be provided, e.g., using waveguiding,
etching or (similar to cavity QED) focusing techniques\textcolor{black}{{}
\cite{morgan07,deLima04,deLima05}. For transverse confinement $L_{\mathrm{trans}}\approx\left(1-5\right)\mu\mathrm{m}$
and a typical resonant cavity wavelength $\lambda_{c}\approx1\mu\mathrm{m}$,
the effective mode area is then $A=L_{\mathrm{trans}}L_{c}\approx(40-200)\mu\mathrm{m^{2}}$.
In the desired regime $\kappa_{\mathrm{gd}}/\kappa_{\mathrm{bd}}\gg1$,
this is largely limited by the deliberately low reflectivity of a
single groove; accordingly, the cavity mode leaks strongly into the
grating such that $L_{c}\gg D$ {[}cf. Appendix \ref{sec:SAW-Cavities}
for details{]}. While we have focused on this standard Bragg design
(due to its experimentally validated frequency selectivity and quality
factors), let us shortly mention potential approaches to reduce $A$
and thus increase single-phonon effects even further: (i) The most
straightforward strategy (that is still compatible with the Bragg
mirror design) is to reduce $\lambda_{c}$ as much as possible, down
to the maximum frequency $f_{c}=v_{s}/\lambda_{c}$ that can still
be made resonant with the (typically highly tunable) qubit's transition
frequency $\omega_{\mathrm{q}}/2\pi$; note that fundamental Rayleigh
modes with $f_{c}\approx6\mathrm{GHz}$ have been demonstrated experimentally
\cite{b=0000FCy=0000FCkk=0000F6se12}. (ii) In order to increase the
reflectivity of a single groove, one could use deeper grooves. To
circumvent the resulting increased losses into the bulk {[}cf. Fig.\ref{fig:SAW-cavity}(b){]},
free-standing structures (where the effect of bulk phonon modes is
reduced) could be employed. (iii) Lastly, alternative cavity designs
such as so-called trapped energy resonators make it possible to strongly
confine acoustic resonances in the center of plate resonators \cite{auld73}. }

\textit{SAW waveguides.}---Not only can SAWs be confined in cavities,
but they can also be guided in acoustic waveguides (WGs) \cite{morgan07,oliner78}.
\textcolor{black}{Two dominant types of design are: (i) Topographic
WGs such as ridge-type WGs where the substrate is locally deformed
using etching techniques, or (ii) overlay WGs (such as strip- or slot-type
WGs) where one or two strips of one material are deposited on the
substrate of another to form core and clad regions with different
acoustic velocities. If the SAW velocity is slower (higher) in the
film than in the substrate, the film acts as a core (cladding) for
the guide whereas the unmodified substrate corresponds to the cladding
(core). }An attenuation coefficient of $\sim0.6\mathrm{dB/mm}$ has
been reported for a $10\mu\mathrm{m}$-wide slot-type WG, defined
by Al cladding layers on a GaAs substrate \cite{deLima04,deLima05}.
This shows that SAWs can propagate basically dissipation-free over
chip-scale distances exceeding several millimeters.\textcolor{blue}{{}
}\textcolor{black}{Typically, one-dimensional WG designs have been
investigated, but---to expand the design flexibility---one could use
multiple acoustic lenses in order guide SAWs around a bend \cite{white70}. }

\section{Universal Coupling\label{sec:Universal-Coupling}}

\begin{table*}
\begin{tabular}{l||c|c|c|c|}
 & charge qubit (DQD) & spin qubit (DQD) & trapped ion & NV-center\tabularnewline
\hline 
\hline 
coupling $g$ & $(200-450)\mathrm{MHz}$ & $(10-22.4)\mathrm{MHz}$ & $(1.8-4.0)\mathrm{kHz}$ & $(45-101)\mathrm{kHz}$\tabularnewline
\hline 
cooperativity $C$ & $11-55$ & $21-106$ & $7-36$ & $10-54$\tabularnewline
\hline 
\end{tabular}

\caption{\label{tab:g-cooperativity}Estimates for single-phonon coupling strength
$g$ and cooperativity $C$. We have used $A=(1-5)\mu\mathrm{m}\times40\lambda_{c}$,
$T=20\mathrm{mK}$ \cite{gustafsson14}, (conservative) quality factors
of $Q=\left(1,1,3,1\right)\times10^{3}$ and frequencies of $\omega_{\mathrm{c}}=2\pi\left(6,1.5,2\times10^{-3},3\right)\mathrm{GHz}$
for the four systems listed. \textcolor{black}{For the spin qubit
$T_{2}^{\star}=2\mu\mathrm{s}$ \cite{shulman14}, and for the trapped
ion scenario, $g_{\mathrm{ion}}(C_{\mathrm{ion}})$ is given for $d=150\mu\mathrm{m}$
due to the prolonged dephasing time further away from the surface
($C_{\mathrm{ion}}$ improves with increasing $d$, even though $g_{\mathrm{ion}}$
decreases, up to a point where other dephasing start to dominate).
}Further details can be found in the main text.}
\end{table*}

\textit{\textcolor{black}{Versatility.}}\textcolor{black}{---}To complete
the analogy with cavity QED, a non-linear element similar to an atom
needs to be introduced. In the following, we highlight three different
exemplary systems, illustrating the versatility of our SAW-based platform.\textcolor{black}{{}
We focus on quantum dots, trapped ions and NV-centers, but similar
considerations naturally apply to other promising quantum information
candidates such as superconducting qubits \cite{gustafsson14,kockum14,cleland04,oconnell10},
Rydberg atoms \cite{gao11} or electron spins bound to a phosphorus
donor atom in silicon \cite{b=0000FCy=0000FCkk=0000F6se12}. }In all
cases considered, a single cavity mode $a$, with frequency $\omega_{\mathrm{c}}$
close to the relevant transition frequency, is retained. We provide
estimates for the single-phonon coupling strength and cooperativity
{[}cf. Tab.\ref{tab:g-cooperativity}{]}, while more detailed analyses
go beyond the scope of this work and are subject to future research. 

(i) QD Charge qubit: A natural candidate for our scheme is a charge
qubit embedded in a lithographically defined GaAs double quantum dot
(DQD) containing a single electron. The DQD can well be described
by an effective two-level system, characterized by an energy offset
$\epsilon$ and interdot tunneling $t_{c}$ yielding a level splitting
$\Omega=\sqrt{\epsilon^{2}+4t_{c}^{2}}$ \cite{petersson10}. The
electron's charge $e$ couples to the piezoelectric potential; the
deformation coupling is much smaller than the piezoelectric coupling
and can therefore safely be neglected \cite{naber06}. \textcolor{black}{Since
the quantum dot is small compared to the SAW wavelength, we neglect
potential effects coming from the structure making up the dots (heterostructure
and metallic gates); for a detailed discussion see Appendix \ref{sec:QD-structure}.}
Performing a standard rotating-wave approximation (valid for $\delta,g_{\mathrm{ch}}\ll\omega_{\mathrm{c}}$),
we find that the system can be described by a Hamiltonian of Jaynes-Cummings
form, 
\begin{equation}
H_{\mathrm{dot}}=\delta S^{z}+g_{\mathrm{ch}}\frac{2t_{c}}{\Omega}\left(S^{+}a+S^{-}a^{\dagger}\right),\label{eq:Hamiltonian-charge-qubit-main-text-1}
\end{equation}
\label{eq:Hamiltonian-charge-qubit-main}where $\delta=\Omega-\omega_{\mathrm{c}}$
specifies the detuning between the qubit and the cavity mode, and
$S^{\pm}=\left|\pm\right\rangle \left\langle \mp\right|$ (and so
on) refer to pseudo-spin operators associated with the eigenstates
$\left|\pm\right\rangle $ of the DQD Hamiltonian (cf. Appendix \ref{sec:Charge-Qubit}).
The Hamiltonian $H_{\mathrm{dot}}$ describes the coherent exchange
of excitations between the qubit and the acoustic cavity mode. The
strength of this interaction $g_{\mathrm{ch}}=e\phi_{0}\mathcal{F}\left(kd\right)\sin\left(kl/2\right)$
is proportional to the charge $e$ and the piezoelectric potential
associated with a single phonon $\phi_{0}$. The decay of the SAW
mode into the bulk is captured by the function $\mathcal{F}\left(kd\right)$
{[}$d$ is the distance between the DQD and the surface; see Appendix
\ref{sec:Piezoelectric-SAW} for details{]}, while the factor $\sin\left(kl/2\right)$
reflects the assumed mode function along the axis connecting the two
dots, separated by a distance $l$. For (typical) values of $l\approx\lambda_{c}/2=250\mathrm{nm}$
and $d\approx50\mathrm{nm}\ll\lambda_{c}$, the geometrical factor
$\mathcal{F}\left(kd\right)\sin\left(kl/2\right)$ then leads to a
reduction in coupling strength compared to the bare value $e\phi_{0}$
(at the surface) by a factor of $\sim2$ only. In total, we then obtain
$g_{\mathrm{ch}}\approx2\mathrm{GHz}/\sqrt{A\left[\mu\mathrm{m}^{2}\right]}$.
\textcolor{black}{For lateral confinement $L_{\mathrm{trans}}\approx\left(1-5\right)\mu\mathrm{m}$,
the effective mode area is $A=L_{\mathrm{trans}}L_{c}\approx(20-100)\mu\mathrm{m^{2}}$.
}The resulting charge-resonator coupling strength $g_{\mathrm{ch}}\approx(200-450)\mathrm{MHz}$
compares well with values obtained using superconducting qubits coupled
to \textit{localized} nano-mechanical resonators made of piezoelectric
material where $g\approx(0.4-1.2)\mathrm{GHz}$ \cite{cleland04,oconnell10}
or\textcolor{black}{{} superconducting resonators coupled to }Cooper
pair box qubits $\left(g/2\pi\approx6\mathrm{MHz}\right)$ \cite{walraff04},
transmon qubits $\left(g/2\pi\approx100\mathrm{MHz}\right)$ \cite{schuster07}
and indium arsenide DQD qubits $\left(g/2\pi\approx30\mathrm{MHz}\right)$
\cite{petersson12}. Note that, in principle, the coupling strength
$g_{\mathrm{ch}}$ could be further enhanced by additionally depositing
a strongly piezoelectric material such as $\mathrm{LiNbO_{3}}$ on
the GaAs substrate \cite{gustaffson12}.\textcolor{black}{{} Moreover,
with a $\mathrm{LiNbO_{3}}$ film on top of the surface, also non-piezoelectric
materials such as Si or Ge could be used to host the quantum dots
\cite{zwanenburg13}.} The level splitting $\Omega\left(t\right)$
and interdot tunneling $t_{c}\left(t\right)$ can be tuned in-situ
via external gate voltages. By controlling $\delta$ one can rapidly
turn on and off the interaction between the qubit and the cavity:
For an effective interaction time $\tau=\pi/2g_{\mathrm{eff}}$ $\left(g_{\mathrm{eff}}=2g_{\mathrm{ch}}t_{c}/\Omega\right)$
on resonance $\left(\delta=0\right)$, an arbitrary state of the qubit
is swapped to the absence or presence of a cavity phonon, i.e., $\left(\alpha\left|-\right\rangle +\beta\left|+\right\rangle \right)\left|0\right\rangle \rightarrow\left|-\right\rangle \left(\alpha\left|0\right\rangle -i\beta\left|1\right\rangle \right)$,
where $\left|n\right\rangle $ labels the Fock states of the cavity
mode. Apart from this SWAP operation, further quantum control techniques
known from cavity QED may be accessible \cite{raimond01}. Note that
below we will generalize our results to \textit{spin} qubits embedded
in DQDs. 

(ii) Trapped ion: The electric field associated with the SAW mode
does not only extend into the solid, but, for a free surface, in general
there will also be an electrical potential decaying exponentially
into the vacuum above the surface $\sim\mathrm{exp}\left[-k\left|z\right|\right]$
\cite{simon96}; cf. Appendix \ref{sec:Piezoelectric-SAW}. This allows
for coupling to systems situated above the surface, without any mechanical
contact. For example, consider a single ion of charge $q$ and mass
$m$ trapped at a distance $d$ above the surface of a strongly piezoelectric
material such as $\mathrm{LiNbO}_{3}$ or AlN. The electric dipole
induced by the ion motion couples to the electric field of the SAW
phonon mode. The dynamics of this system are described by the Hamiltonian
\begin{equation}
H_{\mathrm{ion}}=\omega_{\mathrm{c}}a^{\dagger}a+\omega_{t}b^{\dagger}b+g_{\mathrm{ion}}\left(ab^{\dagger}+a^{\dagger}b\right),\label{eq:Hamiltonian-trapped-ion}
\end{equation}
where $b$ refers to the annihilation operator of the ion's motional
mode and $\omega_{t}$ is the (axial) trapping frequency. The single
phonon coupling strength is given by $g_{\mathrm{ion}}=qx_{0}\times k_{c}\phi_{0}F\left(k_{c}d\right)=q\phi_{0}F\left(k_{c}d\right)\eta_{\mathrm{LD}}$.
Apart from the exponential decay $F\left(kd\right)=\mathrm{exp}\left[-kd\right]$,
the effective coupling is reduced by the Lamb-Dicke parameter $\eta_{\mathrm{LD}}=2\pi x_{0}/\lambda_{c}$,
with $x_{0}=\sqrt{\hbar/2m\omega_{t}}$, since the the motion of the
ion is restricted to a region small compared with the SAW wavelength
$\lambda_{c}$. For $\mathrm{LiNbO}_{3}$, a surface mode area of
$A=(1-5)\mu\mathrm{m}\times40\lambda_{c}$, the commonly used $^{9}\mathrm{Be}^{+}$
ion and typical ion trap parameters with $d\approx30\mu\mathrm{m}$
and $\omega_{t}/2\pi\approx2\mathrm{MHz}$ \cite{labaziewicz08},
we obtain $g_{\mathrm{ion}}\approx(3-6.7)\mathrm{kHz}$. Here, $g_{\mathrm{ion}}$
refers to the coupling between the ion's motion and the cavity. However,
based on $H_{\mathrm{ion}}$, one can in principle generalize the
well-known protocols operating on the ion's spin and motion to operations
on the spin and the acoustic phonon mode \cite{kielpinski12}.

\begin{figure}
\includegraphics[width=1\columnwidth]{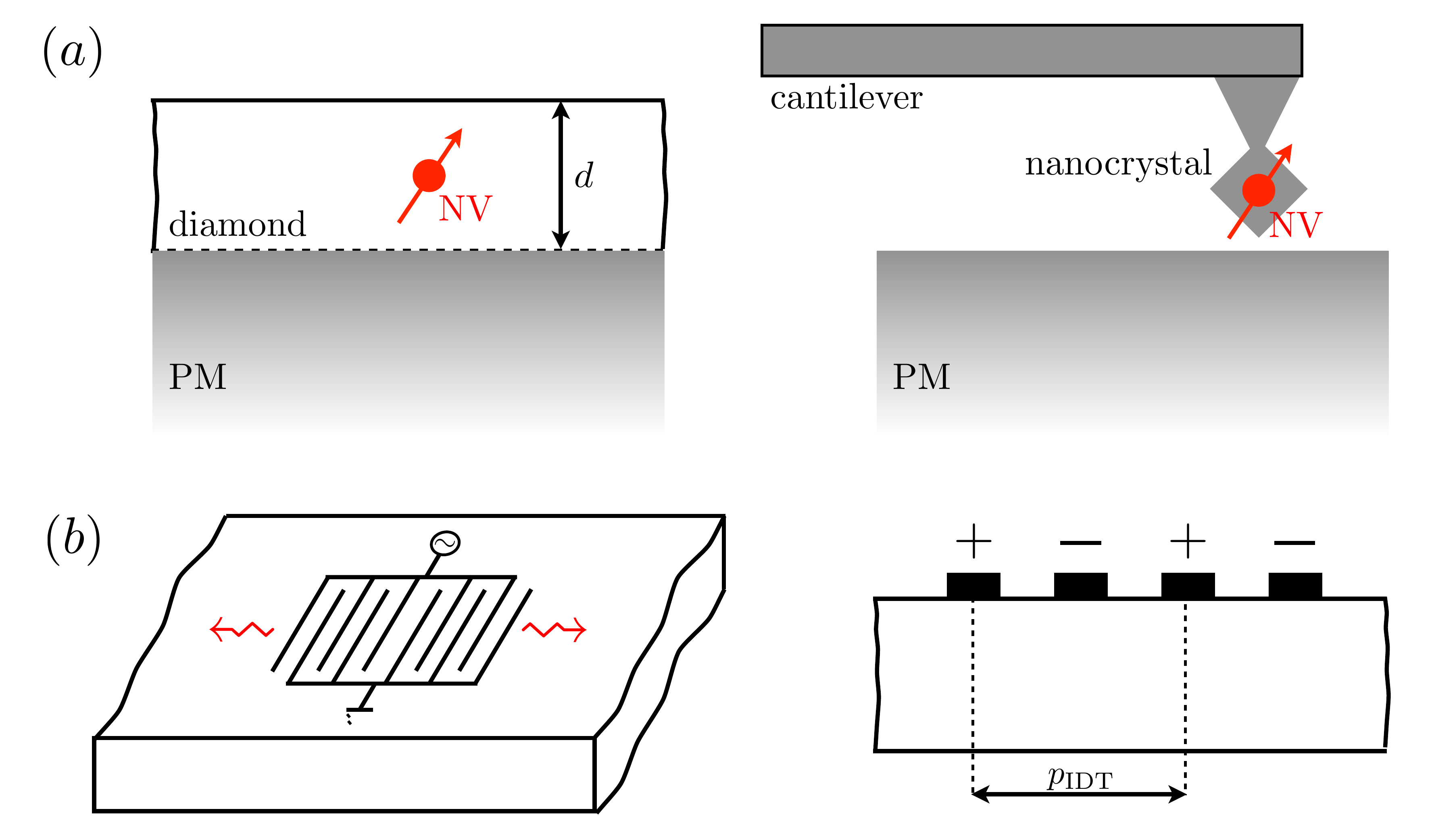}

\caption{\label{fig:NV-IDT}\textcolor{black}{(color online). (a) Schematic
illustration for coupling to a NV center via a piezo-magnetic (PM)
material (see text for details); surface grooves (not shown) can be
used to provide SAW phonon confinement. (b) SAWs can be generated
electrically based on standard interdigital transducers (IDTs) deposited
on the surface. Typically, an IDT consists of two thin-film electrodes
on a piezo-electric material, each formed by interdigitated fingers.
When an ac voltage is applied to the IDT on resonance (defined by
the periodicity of the fingers as $\omega_{\mathrm{IDT}}/2\pi=v_{s}/p_{\mathrm{IDT}}$,
where $v_{s}$ is the SAW propagation speed), it launches a SAW across
the substrate surface in the two directions perpendicular to IDT fingers\cite{gustafsson14,datta86,morgan07}.}}
\end{figure}

(iii) NV-center: Yet another system well suited for our scheme are
NV centers in diamond. Even though diamond itself is not piezoactive,
it has played a key role in the context of high-frequency SAW devices
due to its record-high sound velocity \cite{morgan07}; for example,
high-performance SAW resonators with a quality factor of $Q=12500$
at $\omega_{c}\gtrsim10\mathrm{GHz}$ were experimentally demonstrated
for AlN/diamond heterostructures \cite{benetti05,rodriguez-madrid12}.
To make use of the large magnetic coupling coefficient of the NV center
spin $\gamma_{\mathrm{NV}}=2\pi\times28\mathrm{GHz/T}$, here we consider
a hybrid device composed of a thin layer of diamond with a single
(negatively charged) NV center with ground-state spin ${\bf S}$ implanted
a distance $d\approx10\mathrm{nm}$ away from the interface with a
strongly piezo-magnetic material. Equivalently, building upon current
quantum sensing approaches \cite{rondin14,taylor08}, one could use
a diamond nanocrystal (typically $\sim10\mathrm{nm}$ in size) in
order to get the NV centre extremely close to the surface of the piezo-magnetic
material and thus maximize the coupling to the SAW cavity mode;\textcolor{black}{{}
compare Fig.\ref{fig:NV-IDT}(a) for a schematic illustration.} In
the presence of an external magnetic field ${\bf B}_{\mathrm{ext}}$
\cite{PM-background-field}, the system is described by 
\begin{eqnarray}
H_{\mathrm{NV}} & = & DS_{z}^{2}+\gamma_{\mathrm{NV}}{\bf B}_{\mathrm{ext}}\cdot{\bf S}+\omega_{\mathrm{c}}a^{\dagger}a\label{eq:Hamiltonian-NV-center}\\
 &  & +g_{\mathrm{NV}}\sum_{\alpha=x,y,z}\eta_{\mathrm{NV}}^{\alpha}S^{\alpha}\left(a+a^{\dagger}\right),\nonumber 
\end{eqnarray}
where $D=2\pi\times2.88\mathrm{GHz}$ is the zero-field splitting,
$g_{\mathrm{NV}}=\gamma_{\mathrm{NV}}B_{0}$ is the single phonon
coupling strength and $\eta_{\mathrm{NV}}^{\alpha}$ is a dimensionless
factor encoding the orientation of the NV spin with respect to the
magnetic stray field of the cavity mode. For $d\ll\lambda_{c}$, a
rough estimate shows that at least one component of $\eta_{\mathrm{NV}}^{\alpha}$
is of order unity \cite{cai14}. For a NV center close to a Terfenol-D
layer of thickness $h\gg\lambda_{c}$, we find $g_{\mathrm{NV}}\approx400\mathrm{kHz}/\sqrt{A\left[\mu\mathrm{m}^{2}\right]}$.
Thus, as compared to direct strain coupling $\lesssim200\mathrm{Hz}/\sqrt{A\left[\mu\mathrm{m}^{2}\right]}$,
the presence of the piezomagnetic layer is found to boost the single
phonon coupling strength by three orders of magnitude; this is in
agreement with previous theoretical results for a static setting \cite{cai14}. 

\textit{\textcolor{black}{Decoherence.}}\textcolor{black}{---}In the
analysis above, we have ignored the presence of decoherence which
in any realistic setting will degrade the effects of coherent qubit-phonon
interactions. In this context, the cooperativity parameter, defined
as $C=g^{2}T_{2}/\left[\kappa\left(\bar{n}_{\mathrm{th}}+1\right)\right]$,
is a key figure of merit. Here, $T_{2}$ refers to the corresponding
dephasing time, while $\bar{n}_{\mathrm{th}}=(\mathrm{exp}\left[\hbar\omega_{c}/k_{B}T\right]-1)^{-1}$
gives the thermal occupation number of the cavity mode at temperature
$T$. \textcolor{black}{The parameter $C$ compares the coherent single-phonon
coupling strength $g$ with the geometric mean of the qubit's decoherence
rate $\sim T_{2}^{-1}$ and the cavity's effective linewidth $\sim\kappa\left(\bar{n}_{\mathrm{th}}+1\right)$;
in direct analogy to cavity QED, $C>1$ marks the onset of coherent
quantum effects in a coupled spin-oscillator system, even in the presence
of noise; cf. Ref.\cite{kolkowitz12} and Appendix \ref{sec:cooperativity}
for a detailed discussion.} To estimate $C$, we take the following
parameters for the dephasing time $T_{2}$: For system (i) $T_{2}\approx10\mathrm{ns}$
has been measured close to the charge degeneracy point $\epsilon=0$
\cite{petersson10}. In scenario (ii) motional decoherence rates of
$0.5\mathrm{Hz}$ have been measured in a cryogenically cooled trap
for an ion height of $150\mu\mathrm{m}$ and $1\mathrm{MHz}$ motional
frequency \cite{labaziewicz08}. Since this rate scales as $\sim d^{-4}$
\cite{turchette00,kielpinski12}, we take $T_{2}\left[\mathrm{s}\right]\approx2(d\left[\mu\mathrm{m}\right]/150)^{4}$.
Lastly, for the NV-center (iii) $T_{2}\approx0.6\mathrm{s}$ has been
demonstrated for ensembles of NV spins \cite{bar-gill13} and we assume\textcolor{black}{{}
an optimistic value of $T_{2}=100\mathrm{ms}$, similarly to Ref.\cite{ovartchaiyapong14}.}
The results are summarized in Tab.\ref{tab:g-cooperativity}. We find
that $C>1$ should be experimentally feasible which is sufficient
to perform a quantum gate between two spins mediated by a \textit{thermal}
mechanical mode \cite{rabl10}. 

\textit{\textcolor{black}{Qubit-qubit coupling.}}\textcolor{black}{---When
placing a pair of qubits into the }\textit{\textcolor{black}{same}}\textcolor{black}{{}
cavity, the regime of large single spin cooperativity $C\gg1$ allows
for coherent cavity-phonon-mediated interactions and quantum gates
between the two spins via the effective interaction Hamiltonian $H_{\mathrm{int}}=g_{\mathrm{dr}}(S_{1}^{+}S_{2}^{-}+\mathrm{h.c.})$,
where $g_{\mathrm{dr}}=g^{2}/\delta\ll g$ in the so-called dispersive
regime \cite{blais04}. For the estimates given in Tab.\ref{tab:g-cooperativity},
we have restricted ourselves to the $Q_{r}$-regime with $Q\approx10^{3}$,
where leakage through the acoustic mirrors dominates over undesired
(non-scalable) phonon losses $\left(\kappa_{\mathrm{gd}}\gg\kappa_{\mathrm{bd}}\right)$.
However, note that small-scale experiments using a }\textit{\textcolor{black}{single
}}\textcolor{black}{cavity only (where there is no need for guiding
the SAW phonon into a waveguide for further quantum information processing)
can be operated in the $Q_{m}$-regime (which is limited only by internal
material losses), where the quality factor $Q\approx Q_{m}\gtrsim10^{5}$
is maximized (and thus overall phonon losses minimal). }

\textcolor{black}{As a specific example, consider two NV-centers,
both coupled with strength $g_{\mathrm{NV}}\approx100\mathrm{kHz}$
to the cavity and in resonance with each other, but detuned from the
resonator. Since for large detuning $\delta$ the cavity is only virtually
populated, the cavity decay rate is reduced to $\kappa_{\mathrm{dr}}=\left(g^{2}/\delta^{2}\right)\kappa\approx10^{-2}\kappa\approx1\mathrm{kHz}$
(for $f_{c}=3\mathrm{GHz}$, $Q=2\times10^{5}$), whereas the spin-spin
coupling is $g_{\mathrm{dr}}\approx0.1g_{\mathrm{NV}}\approx10\mathrm{kHz}$.
Therefore, $T_{2}=1\mathrm{ms}$ is already sufficient to approach
the strong-coupling regime where $g_{\mathrm{dr}}\gg\kappa_{\mathrm{dr}},T_{2}^{-1}$. }

\textcolor{black}{Finally, we note that, in all cases considered above,
one could implement a coherent, electrical control by pumping the
cavity mode using standard interdigital transducers (IDTs) \cite{morgan07,datta86,gustafsson14};
compare Fig.\ref{fig:NV-IDT}(b) for a schematic illustration. The
effect of the additional Hamiltonian $H_{\mathrm{drive}}=\Xi\cos\left(\omega_{\mathrm{IDT}}t\right)\left[a+a^{\dagger}\right]$
can be accounted for by replacing the cavity state by a coherent state,
that is $a\rightarrow\alpha$. For example, in the case of Eq.(\ref{eq:Hamiltonian-charge-qubit-main-text-1}),
one could then drive Rabi oscillations between the states $\left|+\right\rangle $
and $\left|-\right\rangle $ with the amplified Rabi frequency $\Omega_{R}=g\alpha$.}

\section{\textcolor{black}{State Transfer Protocol }}

\textcolor{black}{The possibility of quantum transduction between
SAWs and different realizations of stationary qubits enables a variety
of applications including quantum information achitectures that use
SAW phonons as a quantum bus to couple dissimilar and/or spatially
separated qubits. T}he most fundamental task in such a quantum network
is the implementation of a state transfer protocol between two remote
qubits $1$ and $2$, which achieves the mapping $\left(\alpha\left|0\right\rangle _{1}+\beta\left|1\right\rangle _{1}\right)\otimes\left|0\right\rangle _{2}\rightarrow\left|0\right\rangle _{1}\otimes\left(\alpha\left|0\right\rangle _{2}+\beta\left|1\right\rangle _{2}\right)$.
In analogy to optical networks, this can be accomplished via coherent
emission and reabsorption of a single phonon in a waveguide \cite{habraken12}.
As first shown in the context of atomic QED \cite{cirac97}, in principle
perfect, deterministic state transfer can be implemented by identifying
appropriate time-dependent control pulses. 

\textcolor{black}{Before we discuss a specific implementation of such
a transfer scheme in detail, we provide a general approximate result
for the state transfer fidelity $\mathcal{F}$. As demonstrated in
detail in Appendix \ref{sec:cooperativity}, for small infidelities
one can take 
\begin{equation}
\mathcal{F}\approx1-\varepsilon-\mathcal{C}C^{-1},\label{eq:fidelity-estimate}
\end{equation}
as a }\textit{\textcolor{black}{general}}\textcolor{black}{{} estimate
for the state transfer fidelity. Here, individual errors arise from
intrinsic phonon losses $\sim\varepsilon=\kappa_{\mathrm{bd}}/\kappa_{\mathrm{gd}}$
and qubit dephasing $\sim C^{-1}\sim T_{2}^{-1}$, respectively; the
numerical coefficient $\mathcal{C}\sim\mathcal{O}\left(1\right)$
depends on the specific control pulse and may be optimized for a given
set of experimental parameters \cite{stannigel10}. This simple, analytical
result holds for a Markovian noise model where qubit dephasing is
described by a standard pure dephasing term leading to an exponential
loss of coherence $\sim\mathrm{exp}\left(-t/T_{2}\right)$ and agrees
well with numerical results presented in Ref.\cite{stannigel10}.
For non-Markovian qubit dephasing an even better scaling with $C$
can be expected \cite{rabl10}. Using experimentally achievable parameters
$\varepsilon\approx5\%$ and $C\approx30$, we can then estimate $\mathcal{F}\approx90\%$,
showing that fidelities sufficiently high for quantum communication
should be feasible for all physical implementations listed in Tab.\ref{tab:g-cooperativity}. }

\begin{figure}
\includegraphics[width=1\columnwidth]{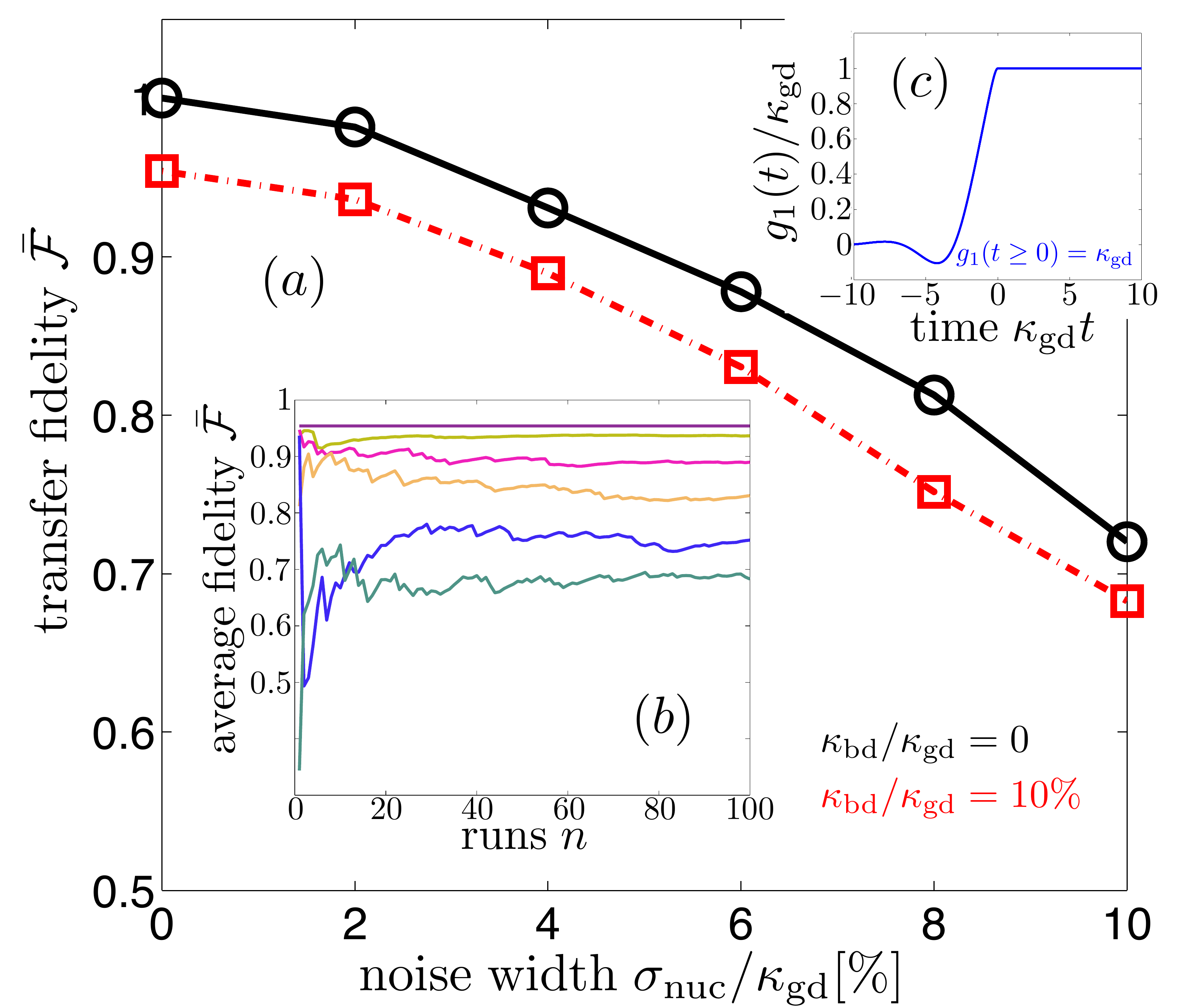}

\caption{\label{fig:state-transfer-simulation-OHnoise}(color online). (a)
Average fidelity $\bar{\mathcal{F}}$ of the state transfer protocol
for a coherent superposition $\left|\psi\right\rangle =\left(\left|0\right\rangle -\left|1\right\rangle \right)/\sqrt{2}$
in the presence of quasi-static (non-Markovian) Overhauser noise,
as a function of the root-mean-square fluctuations $\sigma_{\mathrm{nuc}}$
in the detuning parameters $\delta_{i}\left(i=1,2\right)$, for $\kappa_{\mathrm{bd}}/\kappa_{\mathrm{gd}}=0$
(solid line, circles) and $\kappa_{\mathrm{bd}}/\kappa_{\mathrm{gd}}=10\%$
(dash-dotted line, squares). (b) After $n=100$ runs with random values
for $\delta_{i}$, $\bar{\mathcal{F}}$ approximately reaches convergence.
The curves refer to $\sigma_{\mathrm{nuc}}/\kappa_{\mathrm{gd}}=(0,2,\dots,10)\%$
(from top to bottom) for $\kappa_{\mathrm{bd}}/\kappa_{\mathrm{gd}}=10\%$.
(c) Pulse shape $g_{1}\left(t\right)$ for first node. }
\end{figure}

In the following, we detail the implementation of a transfer scheme
based on \textit{spin} qubits implemented in gate-defined double quantum
dots (DQDs) \cite{spin-qubit-DQD}. In particular, we consider singlet-triplet-like
qubits encoded in lateral QDs, where two electrons are localized in
adjacent, tunnel-coupled dots. As compared to the charge qubits discussed
above, this system is known to feature superior coherence timescales
\cite{kloeffel13,bluhm11,foletti09,chekhovich13}; these are largely
limited by the relatively strong hyperfine interaction between the
electronic spin and the nuclei in the host environment \cite{chekhovich13},
resulting in a random, slowly evolving magnetic (Overhauser) field
for the electronic spin. To mitigate this decoherence mechanism, two
common approaches are (i) spin-echo techniques which allow to extend
spin-coherence from a time-ensemble-averaged dephasing time $T_{2}^{\star}\approx100\mathrm{ns}$
to $T_{2}\gtrsim250\mu\mathrm{s}$ \cite{bluhm11}, and (ii) narrowing
of the nuclear field distribution \cite{chekhovich13,vink09}. Recently,
real-time adaptive control and estimation methods (that are compatible
with arbitrary qubit operations) have allowed to narrow the nuclear
spin distribution to values that prolong $T_{2}^{\star}$ to $T_{2}^{\star}>2\mu\mathrm{s}$
\cite{shulman14}. For our purposes, the latter is particularly attractive
as it can be done simply before loading and transmitting the quantum
information, whereas spin-echo techniques can be employed as well,
however at the expense of more complex pulse sequences (see Appendix
for details). In order to couple the electric field associated with
the SAW cavity mode to the electron \textit{spin} states of such a
DQD, the essential idea is to make use of an effective electric dipole
moment associated with the exchange-coupled spin states of the DQD
\cite{taylor06,shulman12,burkard06,hu12}. As detailed in Appendix
\ref{sec:spin-qubit-QED}, we then find that in the usual singlet-triplet
subspace spanned by the two-electron states $\left\{ \left|\Uparrow\Downarrow\right\rangle ,\left|\Downarrow\Uparrow\right\rangle \right\} $,
a single node can well be described by the prototypical Jaynes-Cummings
Hamiltonian given in Eq.(\ref{eq:generic-Jaynes-Cummings-Hamiltonian}).
As compared to the direct charge coupling $g_{\mathrm{ch}}$, the
single phonon coupling strength $g$ is reduced since the qubit states
$\left|l\right\rangle $ have a small admixture of the localized singlet
$\left<S_{02}|l\right>$ $(l=0,1)$ only. Using typical parameters
values, we find $g\approx0.1g_{\mathrm{ch}}\approx200\mathrm{MHz}/\sqrt{A\left[\mu\mathrm{m}^{2}\right]}$
\cite{spin-coupling}. In this system, the coupling $g\left(t\right)$
can be tuned with great flexibility via both the tunnel-coupling $t_{c}$
and/or the detuning parameter $\epsilon$. 

The state transfer between two such singlet-triplet qubits connected
by a SAW waveguide can be adequately described within the theoretical
framework of cascaded quantum systems, as outlined in detail for example
in Refs.\cite{cirac97,habraken12,carmichael93,gardiner93}: The underlying
quantum Langevin equations describing the system can be converted
into an effective, cascaded Master equation for the system's density
matrix $\rho$. For the relevant case of two qubits, it can be written
as $\dot{\rho}=\mathcal{L}_{\mathrm{ideal}}\rho+\mathcal{L}_{\mathrm{noise}}\rho$,
where 
\begin{eqnarray}
\mathcal{L}_{\mathrm{ideal}}\rho & = & -i\left[H_{S}\left(t\right)+i\kappa_{\mathrm{gd}}\left(a_{1}^{\dagger}a_{2}-a_{2}^{\dagger}a_{1}\right),\rho\right]\nonumber \\
 &  & +2\kappa_{\mathrm{gd}}\mathcal{D}\left[a_{1}+a_{2}\right]\rho,\label{eq:Liouviilian-ideal}\\
\mathcal{L}_{\mathrm{noise}}\rho & = & 2\kappa_{\mathrm{bd}}\sum_{i=1,2}\mathcal{D}\left[a_{i}\right]\rho-i\sum_{i}\delta_{i}\left[S_{i}^{z},\rho\right].\label{eq:Liovillian-noise}
\end{eqnarray}
Here, $\mathcal{D}\left[a\right]\rho=a\rho a^{\dagger}-\frac{1}{2}\left\{ a^{\dagger}a,\rho\right\} $
is a Lindblad term with jump operator $a$ and $H_{S}\left(t\right)=\sum_{i}H_{i}\left(t\right)$,
with $H_{i}\left(t\right)=g_{i}\left(t\right)[S_{i}^{+}a_{i}+S_{i}^{-}a_{i}^{\dagger}]$
describes the coherent Jaynes-Cummings dynamics of the two nodes.
The ideal cascaded interaction is captured by $\mathcal{L}_{\mathrm{ideal}}$
which contains the \textit{non-local} coherent environment-mediated
coupling transferring excitations from qubit 1 to qubit 2 \cite{unidirectional-propagation},
while $\mathcal{L}_{\mathrm{noise}}$ summarizes undesired decoherence
processes: We account for intrinsic phonon losses (bulk-mode conversion
etc.) with a rate $\kappa_{\mathrm{bd}}$ and (non-exponential) qubit
dephasing. Since the nuclear spins evolve on relatively long time-scales,
the electronic spins in quantum dots typically experience non-Markovian
noise leading to a non-exponential loss of coherence on a characteristic
time-scale $T_{2}^{\star}$ given by the width of the nuclear field
distribution $\sigma_{\mathrm{nuc}}$ as $T_{2}^{\star}=\sqrt{2}/\sigma_{\mathrm{nuc}}$
\cite{chekhovich13,shulman14}. Recently, a record-low value of $\sigma_{\mathrm{nuc}}/2\pi=80\mathrm{kHz}$
has been reported \cite{shulman14}, yielding an extended time-ensemble-averaged
electron dephasing time of $T_{2}^{\star}=2.8\mu\mathrm{s}$. In our
model, to realistically account for the dephasing induced by the quasi-static,
yet unknown Overhauser field, the detuning parameters $\delta_{i}$
are sampled independently from a normal distribution $p\left(\delta_{i}\right)$
with zero mean (since nominal resonance can be achieved via the electronic
control parameters) and standard deviation $\sigma_{\mathrm{nuc}}$
\cite{vink09}; see Appendix \ref{sec:spin-qubit-QED} for details.
In Appendix \ref{sec:State-Transfer-Protocol} we also provide numerical
results for standard Markovian dephasing, showing that non-Markovian
noise is beneficial in terms of faithful state transfer. 

Under ideal conditions where $\mathcal{L}_{\mathrm{noise}}=0$, the
setup is analogous to the one studied in Ref.\cite{cirac97} and the
same time-symmetry arguments can be employed to determine the optimal
control pulses $g_{i}\left(t\right)$ for faithful state transfer:
if a phonon is emitted by the first node, then, upon reversing the
direction of time, one would observe perfect reabsorption. By engineering
the emitted phonon wavepacket such that it is invariant under time
reversal and using a time-reversed control pulse for the second node
$g_{2}\left(t\right)=g_{1}\left(-t\right)$, the absorption process
in the second node is a time-reversed copy of the emission in the
first and therefore in principle perfect. Based on this reasoning
(for details see Ref.\cite{cirac97}), we find the explicit,\textcolor{black}{{}
optimal control pulse shown in Fig.\ref{fig:state-transfer-simulation-OHnoise}(c).} 

To account for noise, we simulate the full master equation numerically.
The results are displayed in Fig.\textcolor{black}{\ref{fig:state-transfer-simulation-OHnoise}}(a),
where for every random pair $\delta=\left(\delta_{1},\delta_{2}\right)$
the fidelity of the protocol is defined as the overlap between the
target state $\left|\psi_{\mathrm{tar}}\right\rangle $ and the actual
state after the transfer $\rho\left(t_{f}\right)$, that is $\mathcal{F}_{\delta}=\left<\psi_{\mathrm{tar}}|\rho\left(t_{f}\right)|\psi_{\mathrm{tar}}\right>$.
The average fidelity $\bar{\mathcal{F}}$ of the protocol is determined
by averaging over the classical noise in $\delta$, that is $\bar{\mathcal{F}}=\int d\delta_{1}d\delta_{2}p\left(\delta_{1}\right)p\left(\delta_{2}\right)\mathcal{F}_{\delta}$.
Taking an effective mode area $A\approx100\mu\mathrm{m^{2}}$ as above
and $Q\approx10^{3}$ to be well within the $Q_{r}$-regime where
$\kappa_{\mathrm{bd}}/\kappa_{\mathrm{gd}}\approx5\%$, we have $g\approx\kappa_{\mathrm{gd}}\approx20\mathrm{MHz}$.
For two nodes separated by millimeter distances, propagation losses
are negligible and $\kappa_{\mathrm{bd}}/\kappa_{\mathrm{gd}}\approx5\%$
captures well all intrinsic phonon losses during the transfer.\textcolor{black}{{}
We then find that for realistic undesired phonon losses $\kappa_{\mathrm{bd}}/\kappa_{\mathrm{gd}}\approx5\%$
and $\sigma_{\mathrm{nuc}}/2\pi=80\mathrm{kHz}$ (such that $\sigma_{\mathrm{nuc}}/\kappa_{\mathrm{gd}}\approx2.5\%$)
\cite{shulman14}, transfer fidelities close to $95\%$ seem feasible.
Notably, this could be improved even further using spin-echo techniques
such that $T_{2}\approx10^{2}T_{2}^{\star}$ \cite{bluhm11}. }Therefore,
state transfer fidelities $\mathcal{F}>2/3$ as required for quantum
communication \cite{massar95} seem feasible with present technology.
Near unit fidelities might be approached from further optimizations
of the system's parameters, the cavity design, the control pulses
and/or from communication protocols that correct for errors such as
phonon losses \cite{vanEnk97,vanEnk98,vanEnk99}. Once the transfer
is complete, the system can be tuned adiabatically into a storage
regime which immunizes the qubit against electronic noise and dominant
errors from hyperfine interaction with ambient nuclear spins can be
mitigated by standard, occasional refocusing of the spins \cite{taylor06,bluhm11}.\textcolor{black}{{}
Alternatively, one could also investigate silicon dots: while this
setup requires a more sophisticated hetero-structure including some
piezo-electric layer (as studied experimentally in Ref.\cite{b=0000FCy=0000FCkk=0000F6se12}),
it potentially benefits from prolonged dephasing times $T_{2}^{\star}>100\mu\mathrm{s}$
\cite{veldhorst14}, since nuclear spins are largely absent in isotopically
purified $^{28}\mathrm{Si}$.}

\section{Summary \& Outlook}

In summary, we have proposed and analyzed SAW phonons in piezo-active
materials (such as GaAs) as a universal quantum transducer that allows
to convert quantum information between stationary and propagating
realizations. We have shown that a \textcolor{black}{sound-based}
quantum information architecture based on SAW cavities and waveguides
\textcolor{black}{is very versatile,} \textcolor{black}{bears striking
similarities to cavity QED} and can serve as a scalable mediator of
long-range spin-spin interactions between a variety of qubit implementations\textcolor{black}{,
allowing for faithful quantum state transfer between remote qubits
with existing experimental technology.} The proposed combination of
techniques and concepts known from quantum optics and quantum information,
in conjunction with the technological expertise for SAW devices, is
likely to lead to further, rapid theoretical and experimental progress. 

Finally, we highlight possible directions of research going beyond
our present work: First, since our scheme is not specific to any particular
qubit realization, novel hybrid systems could be developed by embedding
dissimilar systems such as quantum dots and superconducting qubits
into a common SAW architecture. Second, our setup could also be used
as a transducer between \textit{different} propagating quantum systems
such as phonons and photons. Light can be coupled into the SAW circuit
via (for example) NV-centers or self-assembled quantum dots and structures
guiding both photons and SAW phonons have already been fabricated
experimentally \cite{deLima04,deLima05}. Finally, the SAW architecture
opens up a novel, on-chip test-bed for investigations reminiscent
of quantum optics, bringing the highly developed toolbox of quantum
optics and cavity-QED to the widely anticipated field of \textit{quantum
acoustics} \cite{ruskov14,soykal11,gustaffson12,gustafsson14}. Potential
applications include quantum simulation of many-body dynamics \cite{soykal13},
quantum state engineering (yielding for example squeezed states of
sound), quantum-enhanced sensing, sound detection, and sound-based
material analysis. 
\begin{acknowledgments}
We would like to thank the Benasque Center for Science for great working
conditions to start this project. M.J.A.S., G.G. and J.I.C. acknowledge
support by the DFG within SFB 631, and the Cluster of Excellence NIM.
E.M.K. acknowledges support by the Harvard Quantum Optics Center and
the Institute for Theoretical Atomic and Molecular Physics. L.M.K.V.
acknowledges support by a European Research Council Synergy grant.
Work at Harvard was supported by NSF, CUA, CIQM, and AFOSR MURI. M.J.A.S.
would like to thank M. Endres, A. Gonzalez-Tudela and O. Romero-Isart
for fruitful discussions. 
\end{acknowledgments}
\appendix

\section{Classical Description of Nonpiezoelectric Surface Acoustic Waves
\label{sec:Nonpiezoelectric-SAW}}

In this Appendix, we review the general (classical) theoretical framework
describing SAW in cubic lattices, such as diamond or GaAs. We derive
an analytical solution for propagation in the {[}110{]} direction.
The latter is of particular interest in piezoelectric systems. The
classical description of SAW is explicitly shown here to make our
work self-contained, but follows standard references such as Refs.\cite{simon96}.

\textit{Wave equation.}---The propagation of acoustic waves (bulk
and surface waves) in a solid is described by the equation
\begin{equation}
\rho\ddot{u}_{i}({\bf x},t)=\frac{\partial T_{ij}}{\partial x_{j}},\label{eq:EOM}
\end{equation}
where ${\bf u}$ denotes the displacement vector with $u_{i}$ being
the displacement along the cartesian coordinate $\hat{x}_{i}$ $(\hat{x}_{1}=\hat{x},\hat{x}_{2}=\hat{y},\hat{x}_{3}=\hat{z})$,
$\rho$ gives the mass density and $T$ is the stress tensor; $T_{ij}$
is the $i$th component of force per unit area perpendicular to the
$\hat{x}_{j}$-axis. Moreover, ${\bf x}$ is the cartesian coordinate
vector, where in the following we assume a material with infinite
dimensions in $\hat{x},\hat{y}$, and a surface perpendicular to the
$\hat{z}$-direction at $z=0$. The stress tensor obeys a generalized
Hooke's law (stress is linearly proportional to strain)
\begin{equation}
T_{ij}=c_{ijkl}u_{kl},
\end{equation}
where the strain tensor is defined as 
\begin{equation}
u_{kl}=\frac{1}{2}\left(\frac{\partial u_{k}}{\partial x_{l}}+\frac{\partial u_{l}}{\partial x_{k}}\right).\label{eq:def-strain-tensor}
\end{equation}
Using the symmetry $c_{ijkl}=c_{ijlk}$, in terms of displacements
we find 
\begin{equation}
T_{ij}=c_{ijkl}\frac{\partial u_{k}}{\partial x_{l}},
\end{equation}
such that Eq.(\ref{eq:EOM}) takes the form of a set of three coupled
wave equations
\begin{equation}
\rho\ddot{u}_{i}({\bf x},t)-c_{ijkl}\frac{\partial^{2}u_{k}}{\partial x_{j}\partial x_{l}}=0.
\end{equation}
The elasticity tensor $\underline{c}$ obeys the symmetries $c_{ijkl}=c_{jikl}=c_{ijlk}=c_{klij}$
and is largely defined by the crystal symmetry. 

\begin{table}
\begin{tabular}{c||c|c|c|c|c|}
 & $c_{11}$  & $c_{12}$  & $c_{44}$  & $\rho[\mathrm{kg}/\mathrm{m}^{3}]$ & $e_{14}$ \tabularnewline
\hline 
\hline 
Diamond & 107.9 & 12.4 & 57.8 & 3515 & 0\tabularnewline
\hline 
GaAs & $12.26$ & $5.71$ & $6.00$ & 5307 & 0.157\tabularnewline
\hline 
\end{tabular}

\caption{\label{tab:Material-properties}Material properties \cite{simon96}
for both diamond and GaAs. The elastic tensor $\underline{c}$ has
three independent parameters, given in units of $[10^{10}\mathrm{N}/\mathrm{m}^{2}]$,
while the piezoelectric tensor $\underline{e}$ has a single independent
parameter $e_{14}$ for cubic materials (units of $\mathrm{C}/\mathrm{m}^{2}$). }
\end{table}

\textit{Mechanical boundary condition.}---The free surface at $z=0$
is stress free (no external forces are acting upon it), such that
the three components of stress across $z=0$ shall vanish, that is
$T_{13}=T_{23}=T_{33}=0$. This results in the boundary conditions
\begin{equation}
T_{i\hat{z}}=c_{i\hat{z}kl}\frac{\partial u_{k}}{\partial x_{l}}=0\,\,\,\,\,\,\,\mathrm{at}\,\, z=0.\label{eq:mech-boundary-condition}
\end{equation}

\begin{figure}
\includegraphics[width=1\columnwidth]{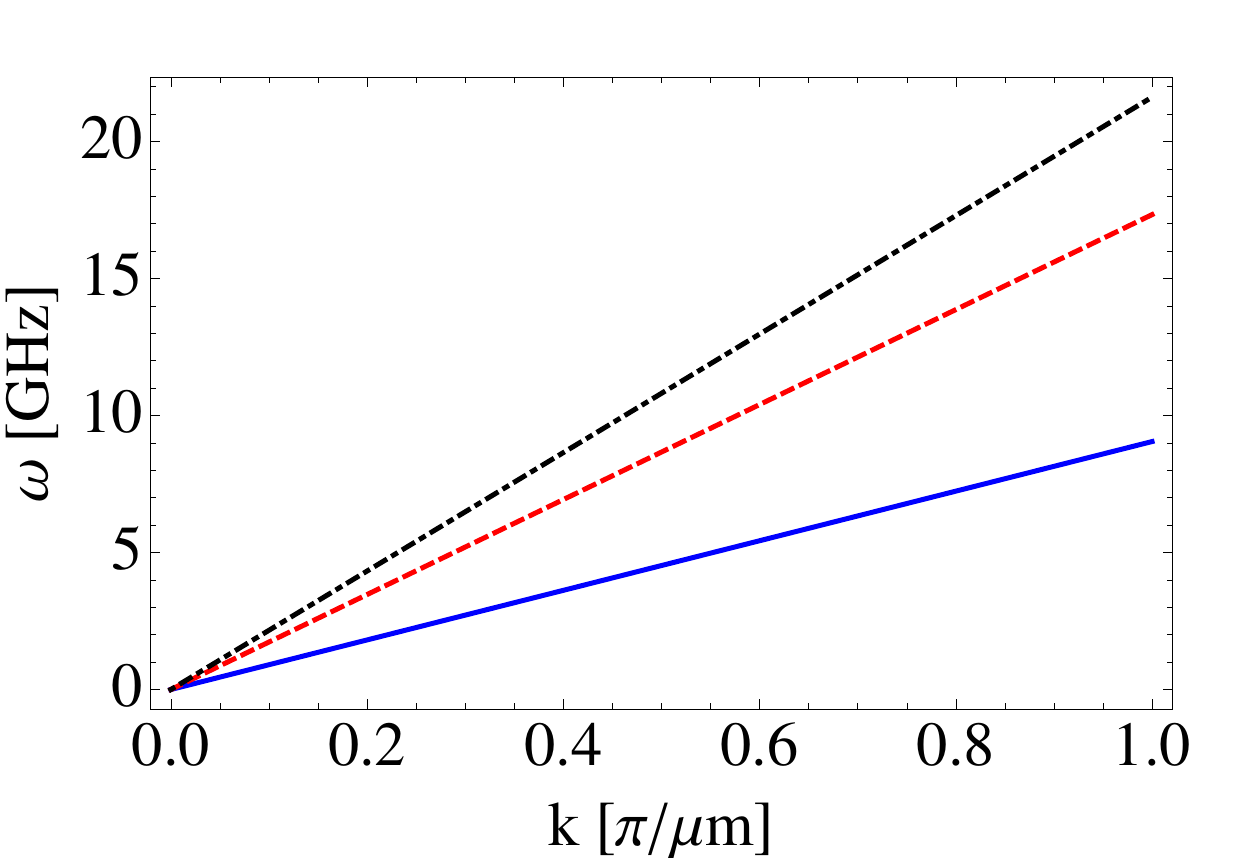}

\caption{\label{fig:dispersion-SAW-110-GaAs}(color online). Dispersion relation
$\omega_{n}=v_{n}k$ of the three $(n=1,2,3)$ Rayleigh-type SAW modes
for propagation along $\hat{x}'||\left[110\right]$. If not stated
otherwise, we refer to the the lowest frequency solution as the SAW
mode (solid line). }
\end{figure}

\textit{Cubic lattice.}---For a cubic lattice (such as GaAs or diamond)
the elastic tensor $c_{ijkl}$ has three independent elastic constants,
generally denoted by $c_{11},c_{12}$, and $c_{44}$; compare Tab.\ref{tab:Material-properties}.
Taking the three direct two-fold axes as the coordinate axes, the
wave equations then read
\begin{eqnarray}
\rho\frac{\partial^{2}u_{x}}{\partial t^{2}} & = & c_{11}\frac{\partial^{2}u_{x}}{\partial x^{2}}+c_{44}\left[\frac{\partial^{2}u_{x}}{\partial y^{2}}+\frac{\partial^{2}u_{x}}{\partial z^{2}}\right]\nonumber \\
 &  & +\left(c_{12}+c_{44}\right)\left[\frac{\partial^{2}u_{y}}{\partial x\partial y}+\frac{\partial^{2}u_{z}}{\partial x\partial z}\right],
\end{eqnarray}
(and cyclic permutations) while the mechanical boundary conditions
can be written as 
\begin{eqnarray}
T_{13} & = & c_{44}\left(\frac{\partial u_{z}}{\partial x}+\frac{\partial u_{x}}{\partial z}\right)=0,\\
T_{23} & = & c_{44}\left(\frac{\partial u_{z}}{\partial y}+\frac{\partial u_{y}}{\partial z}\right)=0,\\
T_{33} & = & c_{11}\frac{\partial u_{z}}{\partial z}+c_{12}\left(\frac{\partial u_{x}}{\partial x}+\frac{\partial u_{y}}{\partial y}\right)=0,
\end{eqnarray}
at $z=0$. In the following we seek for solutions which propagate
along the surface with a wavevector ${\bf k}=k\left(l\hat{x}+m\hat{y}\right)$,
where $l=\mathrm{cos}\left(\theta\right)$, $m=\mathrm{sin}\left(\theta\right)$
and $\theta$ is the angle between the $\hat{x}$-axis and ${\bf k}$.
Following Ref.\cite{stoneley55}, we make the ansatz
\begin{equation}
\left(\begin{array}{c}
u_{x}\\
u_{y}\\
u_{z}
\end{array}\right)=\left(\begin{array}{c}
U\\
V\\
W
\end{array}\right)e^{-kqz}e^{ik\left(lx+my-ct\right)},
\end{equation}
where the decay constant $q$ describes the exponential decay of the
surface wave into the bulk and $c$ is the phase velocity. Plugging
this ansatz into the mechanical wave equations can be rewritten as
$\mathcal{M}{\bf A}=0$, where \begin{widetext} 
\begin{equation}
\mathcal{M}=\left(\begin{array}{ccc}
c_{11}l^{2}+c_{44}\left(m^{2}-q^{2}\right)-\rho c^{2} & lm\left(c_{12}+c_{44}\right) & lq\left(c_{12}+c_{44}\right)\\
lm\left(c_{12}+c_{44}\right) & c_{11}m^{2}+c_{44}\left(l^{2}-q^{2}\right)-\rho c^{2} & mq\left(c_{12}+c_{44}\right)\\
lq\left(c_{12}+c_{44}\right) & mq\left(c_{12}+c_{44}\right) & c_{11}q^{2}-c_{44}+\rho c^{2}
\end{array}\right),
\end{equation}
\end{widetext}and ${\bf A}=\left(U,V,iW\right)$. Nontrivial solutions
for this homogeneous set of equations can be found if the determinant
of $\mathcal{M}$ vanishes, resulting in the so-called \textit{secular
equation} $\det\left(\mathcal{M}\right)=0$. The secular equation
is of sixth order in $q$; as all coefficients in the secular equation
are real, there are, in general, three complex-conjugate roots $q_{1}^{2},q_{2}^{2},q_{3}^{2}$,
with the phase velocity $c$ and propagation direction $\theta$ as
parameters. If the medium lies in the half space $z>0$, the roots
with negative real part will lead to a solution which does not converge
as $z\rightarrow\infty$. Thus, only the roots which lead to vanishing
displacements deep in the bulk are kept. Then, the most general solution
can be written as a superposition of surface waves with allowed $q_{r}$
values as 
\begin{equation}
\left(u_{x},u_{y},iu_{z}\right)=\sum_{r=1,2,3}\left(\xi_{r},\eta_{r},\zeta_{r}\right)K_{r}e^{-kq_{r}z}e^{ik\left(lx+my-ct\right)},
\end{equation}
where, for any $q_{r}=q_{r}\left(c,\theta\right)$, the ratios of
the amplitudes can be calculated according to 
\begin{equation}
K_{r}=\frac{U_{r}}{\xi_{r}}=\frac{V_{r}}{\eta_{r}}=\frac{iW_{r}}{\zeta_{r}},\label{eq:amplitude-ratios}
\end{equation}
where we have introduced the quantities 
\begin{eqnarray}
\xi_{r} & = & \left|\begin{array}{cc}
c_{11}m^{2}+c_{44}\left(l^{2}-q_{r}^{2}\right)-\rho c^{2} & mq_{r}\left(c_{12}+c_{44}\right)\\
mq_{r}\left(c_{12}+c_{44}\right) & c_{11}q_{r}^{2}-c_{44}+\rho c^{2}
\end{array}\right|,\nonumber \\
\eta_{r} & = & \left|\begin{array}{cc}
mq_{r}\left(c_{12}+c_{44}\right) & lm\left(c_{12}+c_{44}\right)\\
c_{11}q_{r}^{2}-c_{44}+\rho c^{2} & lq_{r}\left(c_{12}+c_{44}\right)
\end{array}\right|,
\end{eqnarray}
and 
\begin{equation}
\zeta_{r}=\left|\begin{array}{cc}
lm\left(c_{12}+c_{44}\right) & c_{11}m^{2}+c_{44}\left(l^{2}-q_{r}^{2}\right)-\rho c^{2}\\
lq_{r}\left(c_{12}+c_{44}\right) & mq_{r}\left(c_{12}+c_{44}\right)
\end{array}\right|.
\end{equation}
Note that for each root $q_{r}$ and displacement $u_{i}$ there is
an associated amplitude. The phase velocity $c$, however, is the
same for every root $q_{r}$, and needs to be determined from the
mechanical boundary conditions as described below. Similarly to the
acoustic wave equations, the boundary conditions can be rewritten
as $\mathcal{B}\left(K_{1},K_{2},K_{3}\right)=0$, where the boundary
condition matrix $\mathcal{B}$ is\begin{widetext} 
\begin{equation}
\mathcal{B}=\left(\begin{array}{ccc}
l\zeta_{1}-q_{1}\xi_{1} & l\zeta_{2}-q_{2}\xi_{2} & l\zeta_{3}-q_{3}\xi_{3}\\
m\zeta_{1}-q_{1}\eta_{1} & m\zeta_{2}-q_{2}\eta_{2} & m\zeta_{3}-q_{3}\eta_{3}\\
l\xi_{1}+m\eta_{1}+aq_{1}\zeta_{1} & l\xi_{2}+m\eta_{2}+aq_{2}\zeta_{2} & l\xi_{3}+m\eta_{3}+aq_{3}\zeta_{3}
\end{array}\right),
\end{equation}
\end{widetext}with $a=c_{11}/c_{12}$. Again, nontrivial solutions
are found for $\det\left(\mathcal{B}\right)=0$. The requirements
$\det\left(\mathcal{M}\right)=0$, $\det\left(\mathcal{B}\right)=0$
together with Eq.(\ref{eq:amplitude-ratios}) constitute the formal
solution of the problem \cite{stoneley55}; $\det\left(\mathcal{M}\right)=0$
and $\det\left(\mathcal{B}\right)=0$ may be seen as determining $c^{2}$
and $q^{2}$, and Eq.(\ref{eq:amplitude-ratios}) then gives the the
ratios of the components of the displacement. In the following, we
discuss a special case where one can eliminate the $q$-dependence
in $\det\left(\mathcal{B}\right)=0$, leading to an explicit, analytically
simple equation for the phase velocity $c$, which depends only on
the material properties. 

\textit{Propagation in {[}110{]} direction.}---The wave equations
simplify for propagation in high-symmetry directions. Here, we consider
propagation in the $\left[110\right]$-direction, for which $l=m=1/\sqrt{2}$;
we define the diagonal as $\hat{x}'=\left(\hat{x}+\hat{y}\right)/\sqrt{2}$.
Subtracting the second row from the first in $\mathcal{M}$, one finds
that the common factor $\left(c_{11}-c_{12}\right)/2-c_{44}q^{2}-\rho c^{2}$
divides through the first row, which then becomes $\left(1,-1,0\right)$.
Thus, $U=V$ and the wave equations can be simplified to $\mathcal{M}_{110}\left(U,iW\right)=0$,
where
\begin{equation}
\mathcal{M}_{110}=\left(\begin{array}{cc}
c'_{11}-\rho c^{2}-c_{44}q^{2} & \frac{q}{\sqrt{2}}\left(c_{12}+c_{44}\right)\\
\sqrt{2}q\left(c_{12}+c_{44}\right) & c_{11}q^{2}-c_{44}+\rho c^{2}
\end{array}\right),
\end{equation}
with $c'_{11}=\left(c_{11}+c_{12}+2c_{44}\right)/2$. Then, the secular
equation $\det\left(\mathcal{M}_{110}\right)=0$ is found to be 
\begin{eqnarray}
\left(c'_{11}-\rho c^{2}-c_{44}q^{2}\right)\left(c_{44}-\rho c^{2}-c_{11}q^{2}\right)\nonumber \\
+\left(c_{12}+c_{44}\right)^{2}q^{2} & = & 0,\label{eq:secular-equation-110}
\end{eqnarray}
yielding the roots $q_{1}^{2},q_{2}^{2}$. We choose the roots commensurate
with the convergence condition yielding the general ansatz 
\begin{equation}
\left(\begin{array}{c}
u_{x'}\\
iu_{z}
\end{array}\right)=\sum_{r=1,2}\left(\begin{array}{c}
U'_{r}\\
iW_{r}
\end{array}\right)e^{-kq_{r}z}e^{ik\left(x'-ct\right)}.\label{eq:ansatz-110-propagation}
\end{equation}
with $u_{x}=u_{y}=u_{x'}/\sqrt{2}$. The amplitude ratios $\gamma'_{r}=iW_{r}/U'_{r}$
can be obtained from the kernel of $\mathcal{M}$ as 
\begin{equation}
\gamma'_{r}=q_{r}\frac{c_{12}+c_{44}}{c_{44}-c_{11}\left(X+q_{r}^{2}\right)},\label{eq:amplitude-ratios-110}
\end{equation}
where $X=\rho c^{2}/c_{11}$. In the coordinate system $\left\{ \hat{x}',\hat{z}\right\} $,
the mechanical boundary conditions read 
\begin{eqnarray}
\frac{\partial u_{z}}{\partial x'}+\frac{\partial u_{x'}}{\partial z} & = & 0,\,\,\,\,\,\left(z=0\right)\\
c_{12}\frac{\partial u_{x'}}{\partial x'}+c_{11}\frac{\partial u_{z}}{\partial z} & = & 0.\,\,\,\,\,\left(z=0\right)
\end{eqnarray}
For the ansatz given in Eq.(\ref{eq:ansatz-110-propagation}), they
can be reformulated as $\mathcal{B}_{110}\left(U'_{1},U'_{2}\right)=0$
with 
\begin{equation}
\mathcal{B}_{110}=\left(\begin{array}{cc}
\gamma'_{1}-q_{1} & \gamma'_{2}-q_{2}\\
1+\frac{c_{11}}{c_{12}}q_{1}\gamma'_{1} & 1+\frac{c_{11}}{c_{12}}q_{2}\gamma'_{2}
\end{array}\right).\label{eq:boundary-condition-matrix-110}
\end{equation}
The requirement $\det\left(\mathcal{B}_{110}\right)=0$ can be written
as 
\begin{eqnarray*}
q_{1}\left[c_{12}+\rho c^{2}+c_{11}q_{1}^{2}\right]\left[c_{12}\left(c_{44}-\rho c^{2}\right)+c_{11}c_{44}q_{2}^{2}\right]-\\
q_{2}\left[c_{12}+\rho c^{2}+c_{11}q_{2}^{2}\right]\left[c_{12}\left(c_{44}-\rho c^{2}\right)+c_{11}c_{44}q_{1}^{2}\right] & = & 0.
\end{eqnarray*}
From the symmetry of this equation it is clear that one can remove
a factor $\left(q_{1}-q_{2}\right)$ leading to 
\begin{eqnarray*}
c_{12}\left(\frac{c_{12}}{c_{11}}+X\right)\left(\frac{c_{44}}{c_{11}}-X\right)+c_{44}q_{1}^{2}q_{2}^{2}+c_{12}\times\\
\left(\frac{c_{44}}{c_{11}}-X\right)\left(q_{1}^{2}+q_{2}^{2}+q_{1}q_{2}\right)-c_{44}q_{1}q_{2}\left(\frac{c_{12}}{c_{11}}+X\right) & = & 0.
\end{eqnarray*}
Using simple expressions for $q_{1}^{2}q_{2}^{2}$ and $q_{1}^{2}+q_{2}^{2}$
obtained from Eq.(\ref{eq:secular-equation-110}), one arrives at
the following explicit equation for the wave velocity $c$ \cite{stoneley55,simon96}
\begin{equation}
\left(1-\frac{c_{11}}{c_{44}}X\right)\left(\frac{c_{11}c'_{11}-c_{12}^{2}}{c_{11}^{2}}-X\right)^{2}=X^{2}\left(\frac{c'_{11}}{c_{11}}-X\right),\label{eq:phase-velocity-eq-110}
\end{equation}
which is cubic in $X=\rho c^{2}/c_{11}$. If not stated otherwise,
we consider the mode with the lowest sound velocity, referred to as
Rayleigh mode; compare Fig.\ref{fig:dispersion-SAW-110-GaAs}. 

\begin{figure}
\includegraphics[width=1\columnwidth]{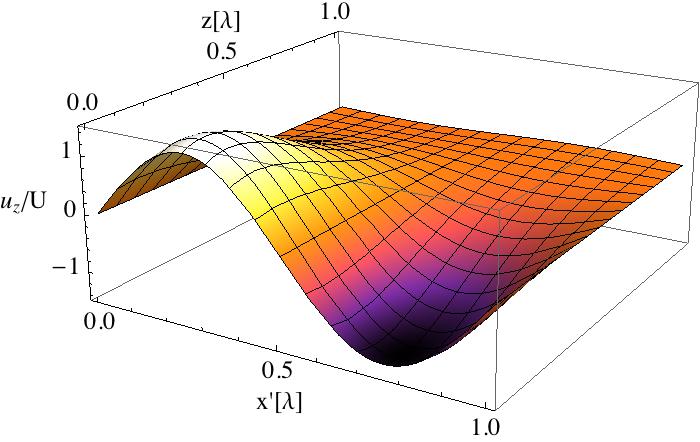}

\caption{\label{fig:displacement-uz-2d-GaAs}(color online). Depth dependence
of the (normalized) vertical displacement $u_{z}/U$ along $\hat{x}'||\left[110\right]$
for a Rayleigh surface acoustic wave propagating on a $(001)$ GaAs
crystal. The acoustic amplitude decays away from the surface into
the bulk on a characteristic length scale approximately given by the
SAW wavelength $\lambda=2\pi/k\approx1\mu\mathrm{m}$. }
\end{figure}

Using the secular equation given in Eq.(\ref{eq:secular-equation-110})
and the mechanical boundary conditions, the ansatz given in Eq.(\ref{eq:ansatz-110-propagation})
can be simplified as follows: The roots compatible with convergence
in the bulk are complex conjugate, i.e. $q\equiv q_{1}=q_{2}^{*}$,
and therefore $\gamma\equiv\gamma'_{1}=\gamma_{2}^{*}$. Then, using
the first row in the boundary condition matrix {[}compare Eq.(\ref{eq:boundary-condition-matrix-110}){]},
we can deduce 
\begin{equation}
U'_{1}=Ue^{-i\varphi},\,\,\,\,\, U'_{2}=Ue^{i\varphi},
\end{equation}
where 
\begin{equation}
e^{-2i\varphi}=-\frac{\gamma^{*}-q^{*}}{\gamma-q}.\label{eq:phase-eq-110}
\end{equation}

In summary, we find the following solution \cite{simon96} 
\begin{eqnarray}
u_{x'} & = & U\left(e^{-qkz-i\varphi}+\mathrm{h.c.}\right)e^{ik\left(x'-ct\right)},\nonumber \\
iu_{z} & = & U\left(\gamma e^{-qkz-i\varphi}+\mathrm{h.c.}\right)e^{ik\left(x'-ct\right)},\label{eq:nonpiezo-solution-110-propagation}
\end{eqnarray}
where the material-dependent parameters $c$, $q$, $\gamma$ and
$\varphi$ are determined by Eq.(\ref{eq:phase-velocity-eq-110}),
Eq.(\ref{eq:secular-equation-110}), Eq.(\ref{eq:amplitude-ratios-110})
and Eq.(\ref{eq:phase-velocity-eq-110}), respectively. For the GaAs
parameters given in Tab.\ref{tab:Material-properties}, we get $c=2878\mathrm{m}/\mathrm{s}$,
$q=0.5+0.48i$, $\gamma=-0.68+1.16i$ and $\varphi=1.05$, respectively.
The corresponding (normalized) transversal displacement is displayed
in Fig.\ref{fig:displacement-uz-2d-GaAs}.

\section{Surface Acoustic Waves In Piezoelectric Materials \label{sec:Piezoelectric-SAW}}

In a piezoelectric material, elastic and electromagnetic waves are
coupled. In principle, the field distribution can be found only by
solving simultaneously the equations of both Newton and Maxwell. The
corresponding solutions are hybrid elasto-electromagnetic waves, i.e.,
elastic waves with velocity $v_{s}$ accompanied by electric fields,
and electromagnetic waves with velocity $c\approx10^{5}v_{s}$ accompanied
by mechanical strains. For the first type of wave, the magnetic field
is negligible, because it is due to an electric field traveling with
a velocity $v_{s}$ much slower than the speed of light $c$; therefore,
one can approximate Maxwell's equations as $\nabla\times{\bf E}=-\partial{\bf B}/\partial t\approx0$,
giving ${\bf E}=-\nabla\phi$. Thus, the propagation of elastic waves
in a piezoelectric material can be described within the quasi-static
approximation, where the electric field is essentially static compared
to electromagnetic fields \cite{royer00}. The potential $\phi$ and
the associated electric field are not electromagnetic in nature but
rather a component of the predominantly mechanical wave propagating
with velocity $v_{s}$.

\subsection{General Analysis}

\textit{Wave equation.}---The basic equations that govern the propagation
of acoustic waves in a piezoelectrical material connect the mechanical
stress $T$ and the electrical displacement ${\bf D}$ with the mechanical
strain and the electrical field. The coupled constitutive equations
are 
\begin{eqnarray}
T_{ij} & = & c_{ijkl}\frac{\partial u_{k}}{\partial x_{l}}+e_{kij}\frac{\partial\phi}{\partial x_{k}},\nonumber \\
D_{i} & = & -\epsilon_{ij}\frac{\partial\phi}{\partial x_{j}}+e_{ijk}\frac{\partial u_{j}}{\partial x_{k}},\label{eq:constituive-equations}
\end{eqnarray}
where $e$ with $\left(e_{ijk}=e_{ikj}\right)$ and $\epsilon$ are
the piezoelectric and permittivity tensor, respectively. Here, Hooke's
law is extended by the additional stress term due to the piezoelectric
effect, while the equation for the displacement $D_{i}$ includes
the polarization produced by the strain. Therefore, Newton's law becomes
\begin{equation}
\rho\ddot{u}_{i}=c_{ijkl}\frac{\partial^{2}u_{k}}{\partial x_{j}\partial x_{l}}+e_{kij}\frac{\partial^{2}\phi}{\partial x_{j}\partial x_{k}}.
\end{equation}
For an insulating solid, the electric displacement $D_{i}$ must satisfy
Poisson's equation $\partial D_{i}/\partial x_{i}=0$ which yields
\begin{eqnarray}
e_{ijk}\frac{\partial^{2}u_{j}}{\partial x_{i}\partial x_{k}}-\epsilon_{ij}\frac{\partial^{2}\phi}{\partial x_{i}\partial x_{j}} & = & 0,\,\,\,\,\, z>0\\
\triangle\phi & = & 0,\,\,\,\,\, z>0.
\end{eqnarray}

\textit{Mechanical boundary conditions.}---In the presence of piezoelectric
coupling the mechanical boundary conditions {[}compare Eq.(\ref{eq:mech-boundary-condition}){]}
generalize to 
\begin{equation}
T_{i\hat{z}}=c_{i\hat{z}kl}\frac{\partial u_{k}}{\partial x_{l}}+e_{ki\hat{z}}\frac{\partial\phi}{\partial x_{k}}=0\,\,\,\,\,\,\,\mathrm{at}\,\, z=0.\label{eq:mech-bounday-condition-piezo}
\end{equation}
Using the symmetries $c_{ijkl}=c_{jikl}$ and $e_{kij}=e_{kji}$ it
is easy to check that this is equivalent to Eq.(41) in Ref.\cite{simon96}. 

\textit{Electric boundary condition.}---In addition to the stress-free
boundary conditions, piezoelectricity introduces an electric boundary
condition: The normal component of the electric displacement needs
to be continuous across the surface \cite{deLima05}, that is 
\begin{equation}
D_{z}\left(z=0^{+}\right)=D_{z}\left(z=0^{-}\right),\label{eq:normal-component-elec-displacement}
\end{equation}
where by definition $D_{z}=e_{\hat{z}jk}\partial u_{j}/\partial x_{k}-\epsilon_{\hat{z}j}\partial\phi/\partial x_{j}$.
Outside of the medium $\left(z<0\right)$, we assume vacuum; thus,
$D_{z}=\epsilon_{0}E_{z}=-\epsilon_{0}\partial\phi_{\mathrm{out}}/\partial z$,
where the electrical potential has to satisfy Poisson's equation $\triangle\phi_{\mathrm{out}}=0$.
The ansatz 
\begin{equation}
\phi_{\mathrm{out}}=A_{\mathrm{out}}e^{ik\left(x'-ct\right)}e^{\Omega kz}
\end{equation}
gives $\triangle\phi_{\mathrm{out}}=\left(-k^{2}+\Omega^{2}k^{2}\right)\phi_{\mathrm{out}}=0$.
Thus, for proper convergence far away from the surface $z\rightarrow-\infty$,
we take the decay constant $\Omega=1$; accordingly, the electrical
potential decays exponentially into the vacuum above the surface on
a typical length scale given by the SAW wavelength $\lambda=2\pi/k\approx1\mu\mathrm{m}$.
Therefore, for the electrical displacement outside of the medium,
we find $D_{z}=-\epsilon_{0}k\phi$. Lastly, the electrical potential
has to be continuous across the surface \cite{simon96}, i.e., 
\begin{equation}
\phi\left(z=0^{+}\right)=\phi_{\mathrm{out}}\left(z=0^{-}\right),
\end{equation}
which allows us to determine the amplitude $A_{\mathrm{out}}$. In
summary, Eq.(\ref{eq:normal-component-elec-displacement}) ca be rewritten
as 
\begin{equation}
\left.\left(e_{\hat{z}jk}\partial u_{j}/\partial x_{k}-\epsilon_{\hat{z}j}\partial\phi/\partial x_{j}+\epsilon_{0}k\phi\right)\right|_{z=0}=0.\label{eq:elec-boundary-condition-piezo}
\end{equation}

\subsection{Cubic lattice}

\textit{Cubic lattice.}---For a cubic, piezoelectric system there
is only one independent nonzero component of the piezoelectric tensor
called $e_{14}$ \cite{simon96,royer00}. With this piezoelectric
coupling, the wave equations are given by four coupled partial differential
equations
\begin{eqnarray}
\rho\frac{\partial^{2}u_{x}}{\partial t^{2}} & = & c_{11}\frac{\partial^{2}u_{x}}{\partial x^{2}}+c_{44}\left[\frac{\partial^{2}u_{x}}{\partial y^{2}}+\frac{\partial^{2}u_{x}}{\partial z^{2}}\right]\label{eq:wave-eqn-piezo-ux}\\
 &  & +\left(c_{12}+c_{44}\right)\left[\frac{\partial^{2}u_{y}}{\partial x\partial y}+\frac{\partial^{2}u_{z}}{\partial x\partial z}\right]+2e_{14}\frac{\partial^{2}\phi}{\partial y\partial z},\nonumber \\
\epsilon\triangle\phi & = & 2e_{14}\left[\frac{\partial^{2}u_{x}}{\partial y\partial z}+\frac{\partial^{2}u_{y}}{\partial x\partial z}+\frac{\partial^{2}u_{z}}{\partial x\partial y}\right],\label{eq:wave-eqn-piezo-phi}
\end{eqnarray}
and cyclic for $u_{y}$ and $u_{z}$. Here, $\triangle$ is the Laplacian
and $\epsilon$ is the dielectric constant of the medium. For a cubic
lattice, the \textit{mechanical boundary} \textit{conditions} at $z=0$
explicitly read 
\begin{eqnarray}
T_{13} & = & c_{44}\left(\frac{\partial u_{z}}{\partial x}+\frac{\partial u_{x}}{\partial z}\right)+e_{14}\frac{\partial\phi}{\partial y}=0,\\
T_{23} & = & c_{44}\left(\frac{\partial u_{z}}{\partial y}+\frac{\partial u_{y}}{\partial z}\right)+e_{14}\frac{\partial\phi}{\partial x}=0,\\
T_{33} & = & c_{11}\frac{\partial u_{z}}{\partial z}+c_{12}\left(\frac{\partial u_{x}}{\partial x}+\frac{\partial u_{y}}{\partial y}\right)=0,
\end{eqnarray}
while the \textit{electrical boundary condition} {[}compare the general
relation in Eq.(\ref{eq:elec-boundary-condition-piezo}){]} leads
to 
\begin{equation}
\left[e_{14}\left(\frac{\partial u_{x}}{\partial y}+\frac{\partial u_{y}}{\partial x}\right)-\epsilon\frac{\partial\phi}{\partial z}+\epsilon_{0}k\phi\right]_{z=0}=0.\label{eq:elec-boundary-condition-cubic}
\end{equation}
In general, the wave equations can be formulated into a $4\times4$
matrix $\mathcal{M}$; the condition $\det\mathcal{M}=0$ can then
used to find the four decay constants. In addition, the mechanical
and electrical boundary conditions can be recast to a $4\times4$
boundary condition matrix $\mathcal{B}$, from which one can deduce
the allowed phase velocities of the piezoelectric SAW by solving $\det\mathcal{B}=0$. 

\begin{figure}
\includegraphics[width=1\columnwidth]{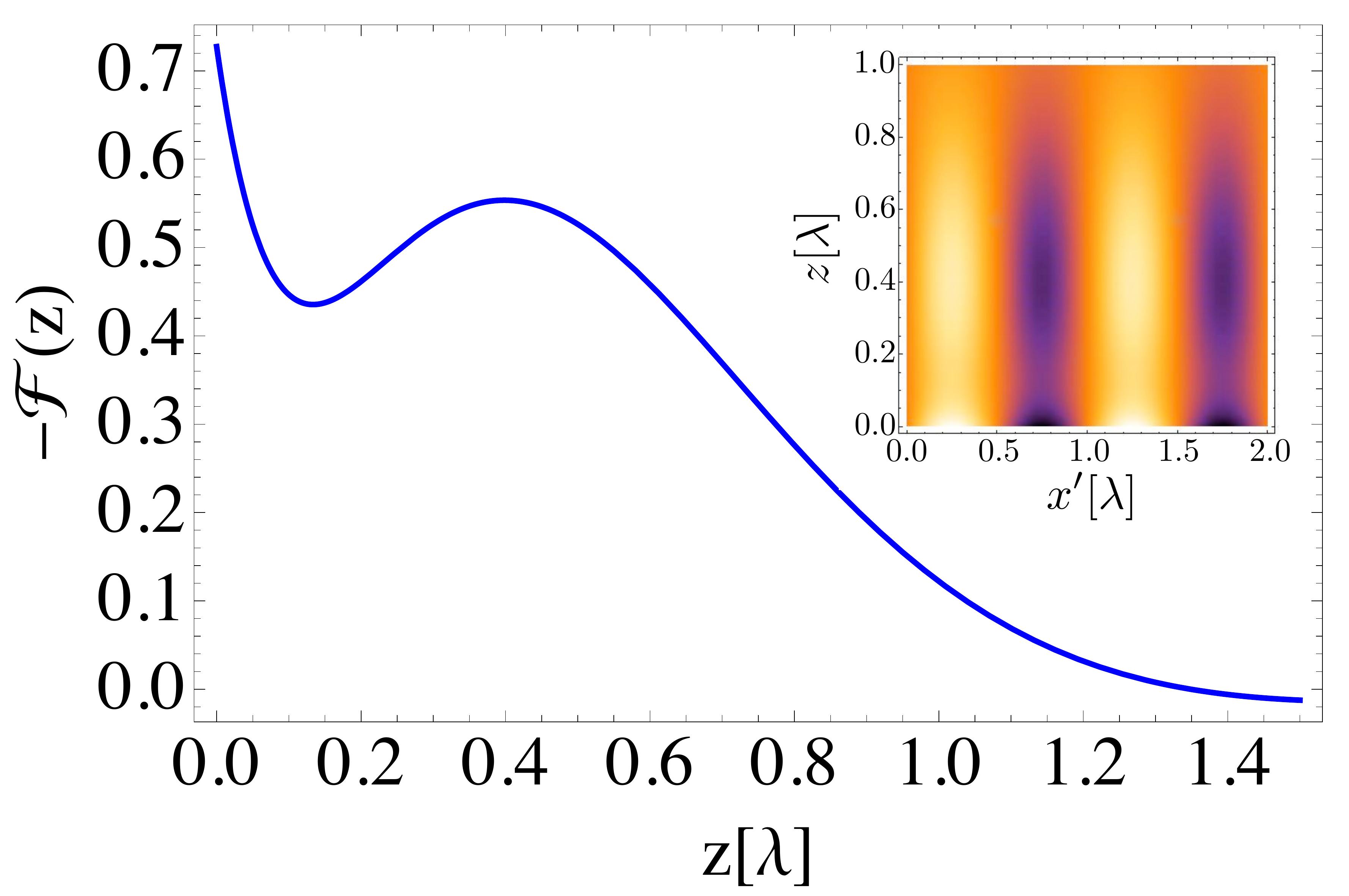}

\caption{\label{fig:elec-potential-phi}(color online). The dimensionless function
$\mathcal{F}\left(z\right)$ determines the decay of the electrical
potential away from the surface into the bulk; the characterisitc
length scale is approximately set by the SAW wavelength $\lambda=2\pi/k\approx1\mu\mathrm{m}$.
Inset: Density plot of the (normalized) electric potential $\mathrm{Re}\left[\phi\right]/\phi_{0}=-\mathcal{F}\left(kz\right)\sin\left(kx'-\omega t\right)$
along $\hat{x}'||\left[110\right]$ for a Rayleigh surface acoustic
wave propagating on a $(001)$ GaAs crystal at $t=0$.}
\end{figure}

\textit{Perturbative treatment.}---For materials with weak piezoelectric
coupling (such as GaAs), the properties of surface acoustic waves
are primarily determined by the elastic constants and density of the
medium. Then, within a perturbative treatment of the piezoelectric
coupling, one can obtain analytical expressions for the strain and
piezoelectric fields. Here, we summarize the results for SAWs propagating
along $\hat{x}'||[110]$ of the $\hat{z}||[001]$ surface following
Refs.\cite{simon96,aizin98}. Since the piezoelectric coupling $e_{14}$
is small, it follows from Eq.(\ref{eq:wave-eqn-piezo-phi}) that $\phi$
will be order $e_{14}$ smaller than the mechanical displacements
$u$, that is
\begin{equation}
\phi\sim\frac{e_{14}}{\epsilon}u.
\end{equation}
This results in additional terms in the wave equations that are of
order $\sim e_{14}^{2}/\epsilon\approx10^{8}\mathrm{N/m^{2}}$. Since
the elastic constants are 2 to 3 orders of magnitude bigger than this
piezoelectric term, the wave equations Eqs.(\ref{eq:wave-eqn-piezo-ux}),
and (cyclic versions for $u_{y},u_{z}$) will be solved by the nonpiezoelectric
solution with corrections only at order $e_{14}^{2}$. The nonpiezoelectric
solution derived in detail in Sec.~\ref{sec:Nonpiezoelectric-SAW}
can be summarized as 
\begin{eqnarray}
u_{x'} & = & 2U\mathrm{Re}\left[e^{-qkz-i\varphi}\right]e^{ik\left(x'-vt\right)},\nonumber \\
u_{y'} & = & 0,\nonumber \\
u_{z} & = & -2iU\mathrm{Re}\left[\gamma e^{-qkz-i\varphi}\right]e^{ik\left(x'-vt\right)},\label{eq:nonpiezo-solution-110-ux-uz}
\end{eqnarray}
where the sound velocity $v$ for the Rayleigh-mode follows from the
smallest solution of 
\begin{equation}
\left(c_{44}-\rho v^{2}\right)\left(c_{11}c'_{11}-c_{12}^{2}-c_{11}\rho v^{2}\right)^{2}=c_{11}c_{44}\rho^{2}v^{4}\left(c'_{11}-\rho v^{2}\right),
\end{equation}
with $c'_{11}=c_{44}+\left(c_{11}+c_{12}\right)/2$. The decay constant
$q$ is a solution of 
\begin{equation}
\left(c'_{11}-\rho v^{2}-c_{44}q^{2}\right)\left(c_{44}-\rho v^{2}-c_{11}q^{2}\right)+q^{2}\left(c_{12}+c_{44}\right)^{2}=0.
\end{equation}
Lastly, the parameters $\gamma$, $\varphi$ can be obtained from
\begin{equation}
\gamma=\frac{\left(c_{12}+c_{44}\right)q}{c_{44}-c_{11}q^{2}-\rho v^{2}},\,\,\,\,\,\,\,\,\,\, e^{-2i\varphi}=-\frac{\gamma^{*}-q^{*}}{\gamma-q}.
\end{equation}
Now, based on the nonpiezoelectric solution given in Eq.(\ref{eq:nonpiezo-solution-110-ux-uz}),
the potential $\phi$ is constructed such that both the wave equation
in Eq.(\ref{eq:wave-eqn-piezo-phi}) and the electrical boundary condition
in Eq.(\ref{eq:elec-boundary-condition-cubic}) are solved. In the
$\left\{ \hat{x}',\hat{y}',\hat{z}\right\} $ coordinate system they
read explicitly 
\begin{eqnarray}
\epsilon\triangle\phi & = & e_{14}\left(2\frac{\partial^{2}u_{x'}}{\partial x'\partial z}+\frac{\partial^{2}u_{z}}{\partial x'\partial x'}\right),\\
0 & = & \left.\left(\epsilon_{0}k\phi+e_{14}\frac{\partial u_{x'}}{\partial x'}-\epsilon\frac{\partial\phi}{\partial z}\right)\right|_{z=0}.
\end{eqnarray}
One can readily check that this is achieved by the form proposed in
Ref.\cite{simon96,aizin98}
\begin{equation}
\phi=\begin{cases}
i\phi_{0}\mathcal{F}\left(kz\right)e^{ik\left(x'-vt\right)}, & z>0\\
A_{\mathrm{out}}e^{kz}e^{ik\left(x'-ct\right)}, & z<0,
\end{cases}
\end{equation}
where $\phi_{0}=\left(e_{14}/\epsilon\right)U$ and $A_{\mathrm{out}}=i\phi_{0}\mathcal{F}\left(0\right)$.
Here, we have introduced the dimensionless function $\mathcal{F}\left(kz\right)$
which determines the length scale on which the electrical potential
generated by the SAW decays into the bulk. It is given by 
\begin{equation}
\mathcal{F}\left(kz\right)=2\left|A_{1}\right|e^{-\alpha kz}\cos\left(\beta kz+\varphi+\xi\right)+A_{3}e^{-kz},
\end{equation}
with $A_{1}=\left|A_{1}\right|e^{-i\xi}$, $q=\alpha+\beta i$, and
\begin{eqnarray}
A_{1} & = & \frac{\gamma-2q}{q^{2}-1},\\
A_{3} & = & -\frac{2}{\epsilon+\epsilon_{0}}\left[\epsilon\cos\varphi+\epsilon\mathrm{Re}\left[A_{1}qe^{-i\varphi}\right]+\epsilon_{0}\mathrm{Re}\left[A_{1}e^{-i\varphi}\right]\right].\nonumber 
\end{eqnarray}
For $\mathrm{Al}_{x}\mathrm{Ga}_{1-x}\mathrm{As}$ we obtain the following
parameter values {[}compare Ref.\cite{simon96}{]}: $\left|A_{1}\right|\approx1.59$,
$A_{3}=-3.1$, $\alpha\approx0.501$, $\beta\approx0.472$, $\varphi=1.06$,
and $\xi=-0.33$. The electric potential for this parameter set is
shown in Fig.\ref{fig:elec-potential-phi}.

\section{Mechanical Zero-Point Fluctuation\label{sec:Mechanical-Zero-Point-Fluctuation}}

In this Appendix we provide more detailed calculations and estimates
for the mechanical zero-point motion $U_{0}$ of a SAW. We show that
they agree very well with the simple estimate given in the main text.
Finally, we provide details on the material parameters used to obtain
the numerical estimates. 

Our first approach follows closely the one presented in Ref.\cite{aspelmeyer14}.
The analysis starts out from the mechanical displacement operator
in the Heisenberg picture 
\begin{equation}
\hat{{\bf u}}\left({\bf x},t\right)=\sum_{n}\left[{\bf v}_{n}\left({\bf x}\right)a_{n}e^{-i\omega_{n}t}+\mathrm{h.c.}\right].
\end{equation}
To obtain the proper normalization of the displacement profiles, let
us assume a single phonon Fock state, that is $\left|\Psi\right\rangle =a_{n}^{\dagger}\left|\mathrm{vac}\right\rangle =\left|0,\dots,0,1_{n},0.\dots\right\rangle $,
where $\left|\mathrm{vac}\right\rangle =\prod_{n}\left|0\right\rangle _{n}$
is the phonon vacuum and compute the expectation value of additional
field energy above the vacuum $E_{\mathrm{mech}}$, defined as twice
the kinetic energy, since for a mechanical mode half of the energy
is kinetic, the other one potential \cite{aspelmeyer14}. We find
\begin{eqnarray}
E_{\mathrm{mech}} & = & 2\omega_{n}^{2}\int d^{3}{\bf r}\rho\left({\bf r}\right){\bf v}_{n}^{*}\left({\bf r}\right)\cdot{\bf v}_{n}\left({\bf r}\right)\\
 & = & 2\rho V\omega_{n}^{2}\mathrm{max}\left[\left|{\bf v}_{n}\left({\bf r}\right)\right|^{2}\right],
\end{eqnarray}
where the last equality defines the effective mode-volume for mode
$n$. Setting $U_{0}=\mathrm{max}\left[\left|{\bf v}_{n}\left({\bf r}\right)\right|\right]$,
and assuming the phonon energy as $E_{\mathrm{mech}}=\hbar\omega_{n}$,
we arrive at the general result for a phonon mode, $U_{0}=\sqrt{\hbar/2\rho V\omega_{n}}$;
this confirms the simple estimate given in the main text. 

\begin{table}
\begin{tabular}{c||c|c|c|c|c}
 & decay const. $\Omega$  & $\gamma=\left|\gamma\right|e^{-i\Theta}$  & $\varphi$  & $v_{s}[\mathrm{m/s}]$ & $\delta$\tabularnewline
\hline 
\hline 
GaAs & $0.50+0.48i$ & $-0.68+1.16i$ & 1.05 & 2878 & 1.2\tabularnewline
\hline 
Diamond & $0.60+0.22i$ & $-1.05+0.75i$ & 1.26 & 11135 & 0.44\tabularnewline
\hline 
\end{tabular}

\caption{\label{tab:derived-Rayleigh-wave-properties}Derived properties for
Rayleigh surface waves, for both GaAs and diamond. }
\end{table}

\textit{Explicit example.}---Next, we provide a calculation based
on the exact analytical results derived in Appendix \ref{sec:Nonpiezoelectric-SAW}
and \ref{sec:Piezoelectric-SAW}. In what follows, we assume that,
in analogy to cavity QED, cavity confinement leads to the quantization
$k_{n}=n\pi/L_{c}$, where $A=L_{c}^{2}$ is the effective quantization
area. In a full 3D model, $A=L_{x}L_{y}$ where $L_{y}$ is related
to the spread of the transverse mode function as discussed (for example)
in Refs.\cite{morgan07}. For simplicity, here we take $L_{x}=L_{y}$.
Surface wave resonators can routinely be designed to show only one
resonance $k_{0}$ \cite{morgan07}. Within this single-mode approximation,
based on results derived in Appendix \ref{sec:Nonpiezoelectric-SAW}
for a SAW traveling wave, we take the quantized mechanical displacement
describing a SAW standing wave along the axis $\hat{x}'=\left(110\right)$
as 
\begin{equation}
\hat{u}\left(x',z\right)=U_{0}\left(\begin{array}{c}
\chi_{0}\left(z\right)\cos\left(k_{0}x'\right)\\
0\\
\zeta_{0}\left(z\right)\sin\left(k_{0}x'\right)
\end{array}\right)\left[a+a^{\dagger}\right],\label{eq:mechanical-displacement-perfect-SAW-cavity}
\end{equation}
Here, the functions $\chi_{0}\left(z\right)$ and $\zeta_{0}\left(z\right)$
describe how the SAW decays into the bulk, 
\begin{eqnarray}
\chi_{0}\left(z\right) & = & 2e^{-\Omega_{r}k_{0}z}\cos\left(\Omega_{i}k_{0}z+\varphi\right),\\
\zeta_{0}\left(z\right) & = & 2\left|\gamma\right|e^{-\Omega_{r}k_{0}z}\cos\left(\Omega_{i}k_{0}z+\varphi+\theta\right),
\end{eqnarray}
with material-dependent parameters $\Omega=\Omega_{r}+i\Omega_{i}$,
$\gamma=\left|\gamma\right|\exp\left[-i\theta\right]$ and $\varphi$;
numerical values are presented in Table \ref{tab:derived-Rayleigh-wave-properties}.
We note that for GaAs we find $\zeta\left(0\right)/\chi\left(0\right)\approx1.33$.
This is in very good agreement with the numerical values of $c_{x}=|u_{x}/\phi|=0.98\mathrm{nm/V}$
and $c_{z}=|u_{z}/\phi|=1.31\mathrm{nm/V}$ as given in Ref.\cite{datta86}.
Normalization of the mode-function allows us to determine the parameter
$U_{0}$. Performing the integration, we find 
\begin{equation}
U_{0}=\sqrt{\frac{\Omega_{r}}{\delta}}\sqrt{\frac{\hbar}{2\rho v_{s}A}},\label{eq:single-phonon-displacement-derived-formula}
\end{equation}
\textcolor{black}{where the parameter $\delta$ depends on the material
parameters; see Table \ref{tab:derived-Rayleigh-wave-properties}.
}Using typical material parameters, we obtain for GaAs (diamond) $\sqrt{\Omega_{r}/\delta}=0.64(1.17)$
and $U_{0}\approx1.2\left(1.36\right)\mathrm{fm}/\sqrt{A[\mu\mathrm{m}^{2}]}$.
This is in very good agreement with the numerical values presented
in the main text. 

\textit{Estimates derived from literature.}---In Ref.\cite{simon96},
it is shown that the SAW Rayleigh mode studied in Appendix \ref{sec:Nonpiezoelectric-SAW}
has a classical energy density $\mathcal{E}$ (energy per unit surface
area) given by
\begin{equation}
\mathcal{E}=kU^{2}H,
\end{equation}
where $U$ is the amplitude of the wave, $k$ the wave vector and
$H$ a material-dependent factor which is given as $H\approx28.2\times10^{10}\mathrm{N/m^{2}}$
for GaAs. By equating the classical energy of the SAW given by $\mathcal{E}A$,
where $A$ is the quantization area, with its quantum-mechanical analog
$N_{\mathrm{ph}}\hbar\omega$ ($N_{\mathrm{ph}}$ is the number of
phonons), we can estimate the single phonon displacement $U_{0}$
as 
\begin{equation}
U_{0}=\sqrt{\frac{\hbar\omega}{kHA}}=\sqrt{\frac{\hbar v_{s}}{HA}}\approx1.05\times10^{-21}\frac{m^{2}}{\sqrt{A}},\label{eq:estimate-zero-point-motion-amplitude-Simon}
\end{equation}
with $U=U_{0}\sqrt{N_{\mathrm{ph}}}$. This estimate is also found
to be in very good agreement with a result given in Ref.\cite{deLima05}
as 
\begin{equation}
U_{0}=C\sqrt{\frac{2\hbar}{\rho v_{s}A}}\approx1.7\times10^{-21}\frac{m^{2}}{\sqrt{A}},\label{eq:single-phonon-displacement-U0-deLima}
\end{equation}
where $C$ is a normalization constant with numerical value $C\approx0.45$
for GaAs \cite{deLima05}. Therefore, for an effective mode area of
$L_{c}=\sqrt{A}=1\mu\mathrm{m}$ we find a single phonon displacement
of $U_{0}\approx1\mathrm{fm}$. This confirms the estimates given
in the main text.

\section{Zero-Point Estimates\label{sec:Zero-Point-Estimates}}

\begin{table}
\begin{tabular}{c||c|c|c|c|c|}
 & $U_{0}\left[\mathrm{fm}\right]$  & $s_{0}[10^{-9}]$ & $\phi_{0}\left[\mu\mathrm{V}\right]$ & $\xi_{0}\left[\mathrm{V/m}\right]$ & $B_{0}\left[\mu\mathrm{T}\right]$\tabularnewline
\hline 
\hline 
GaAs & 1.9 & 11.7 & 3.1 & 19.2 & ---\tabularnewline
\hline 
$\mathrm{LiNbO_{3}}$ & 1.8 & 11.3 & $0.9-25.8$ & $5.8-162.2$ & ---\tabularnewline
\hline 
Quartz & 2.75 & 17.3 & $2.8-12.0$ & $17.3-75.4$ & ---\tabularnewline
\hline 
Terfenol-D & 2.2 & 13.8 & --- & --- & 2.3\tabularnewline
\hline 
$\mathrm{CoFe_{2}O_{4}}$ & 1.8 & 11.4 & --- & --- & 6.3\tabularnewline
\hline 
Diamond & 1.17 & 7.4 & --- & --- & ---\tabularnewline
\hline 
\end{tabular}

\caption{\label{tab:zero-point-fluctuations}\textcolor{black}{Estimates for
zero-point fluctuations (mechanical amplitude $U_{0}$, strain $s_{0}$,
electrical potential $\phi_{0}$, electric field $\xi_{0}$ and magentic
field $B_{0}$) close to the surface $\left(d\ll\lambda\right)$ for
typical piezo-electric and piezo-magnetic (magnetostrictive) materials.
All values must be multiplied by the universal scaling factor $1/\sqrt{A\left[\mu\mathrm{m}^{2}\right]}$;
thus, they refer to an effective surface mode area of size $A=1\mu\mathrm{m}^{2}$.
Lower (upper) bounds for $\phi_{0}$ and $\xi_{0}$ comprise minimum
(maximum) non-zero element of $\underline{e}$ with maximum (minimum)
non-zero element of $\underline{\epsilon}$. We have set $k=2\pi/\mu\mathrm{m}$.
Details on cut-directions and material parameters are given in the
text.}}
\end{table}

In this Appendix we provide details on piezo-magnetic materials and
numerical estimates of the zero-point quantities for several relevant
materials. The results are summarized in Tab.\ref{tab:zero-point-fluctuations}.
The underlying input parameters are given below. 

\textcolor{black}{Theoretically, piezo-magnetic materials with a large
magneto-strictive effect are typically described in a 1:1 correspondence
to Eqs.(\ref{eq:Newton-third-law-PE-material}) and (\ref{eq:Poisson-PE-material}),
with the appropriate replacements (using standard notation) ${\bf E}\rightarrow{\bf H}$,
${\bf D}\rightarrow{\bf B}$, $\epsilon_{ij}\rightarrow\mu_{ij}$
and $e_{ijk}\rightarrow h_{ijk}$ \cite{nowacki06}. Coupling between
mechanical and magnetic degrees of freedom is described by the piezomagnetic
tensor $\underline{h}$ which can reach values as high as $\sim700\mathrm{T/strain}$
\cite{pang08}; for our estimates we have referred to Terfenol-D,
where $h_{15}\approx167\mathrm{T/strain}$. The magnetic field associated
with a single phonon can then be estimated as $B_{0}\approx h_{15}s_{0}$,
where $h_{15}$ refers to a typical (non-zero) element of $\underline{h}$.}

For the piezoelectric materials GaAs, $\mathrm{LiNbO_{3}}$ and Quartz
all material parameters have been obtained from Ref.\cite{royer00}.
Phase velocities for typical cut directions have been used, that is
$(100)[001]$ GaAs, Y-Z $\mathrm{LiNbO_{3}}$ and ST Quartz. For the
piezomagnetic (magnetostrictive) materials $\mathrm{CoFe_{2}O_{4}}$
and Terfenol-D all material parameters have been taken from Ref.\cite{liu08}.
We have used the phase velocities of the bulk shear waves given in
there as $v_{\mathrm{sh}}=3.02\times10^{3}\mathrm{m/s}$ and $v_{\mathrm{sh}}=1.19\times10^{3}\mathrm{m/s}$
for $\mathrm{CoFe_{2}O_{4}}$ and Terfenol-D, respectively. This gives
an conservative estimate for $U_{0}$, since Rayleigh modes have phase
velocities that are lower than the ones of bulk modes \cite{morgan07}.
For example, in the case of $\mathrm{CoFe_{2}O_{4}}$ in Ref.\cite{pang08}
wave velocities for Rayleigh-type surface waves in a piezoelectric-piezomagnetic
layered half space are found to be $v_{s}\approx2840\mathrm{m/s}<v_{\mathrm{sh}}$.

\section{SAW Cavities\label{sec:SAW-Cavities}}

\begin{figure}
\includegraphics[width=1\columnwidth]{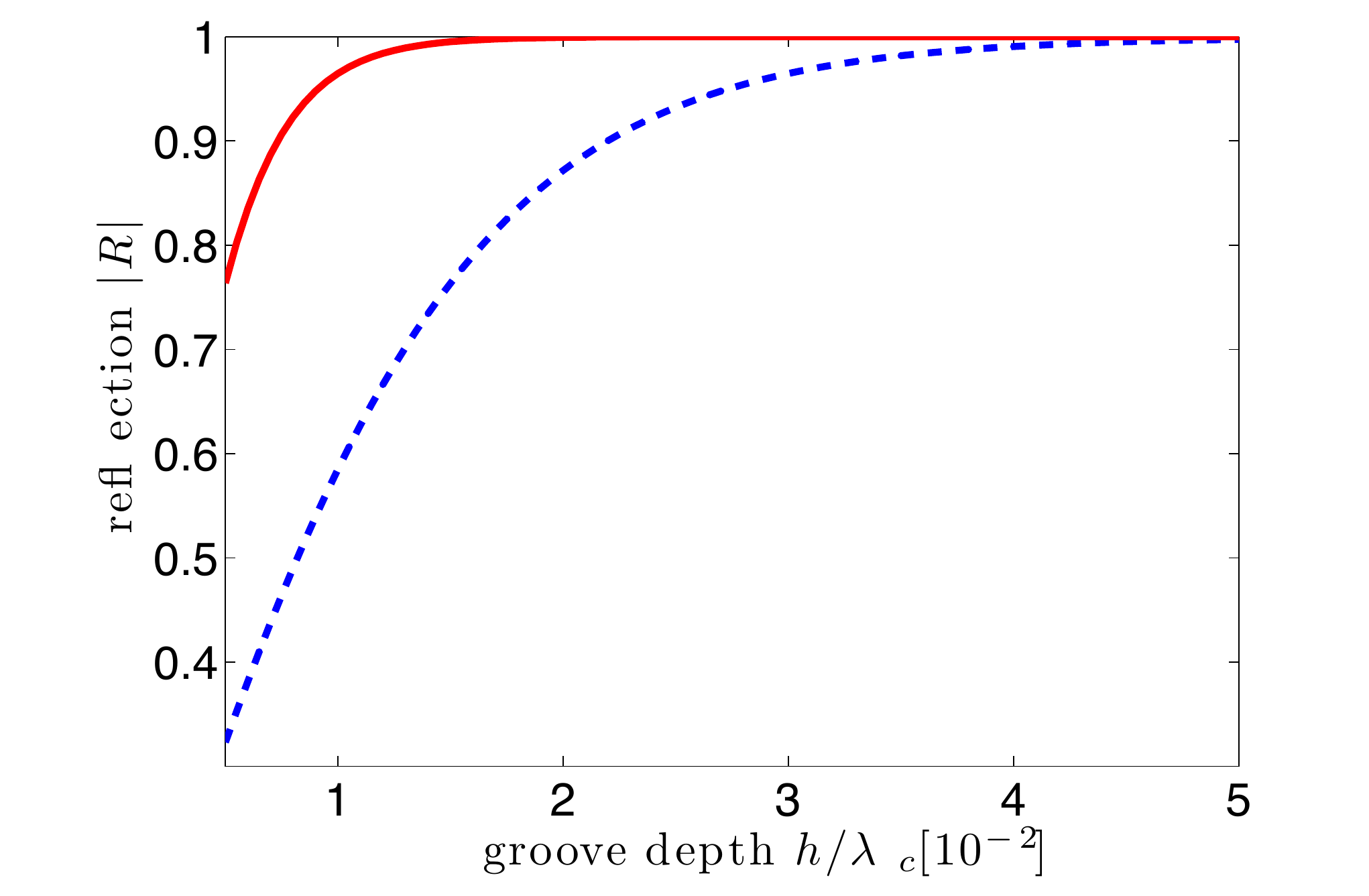}

\caption{\label{fig:reflection-coefficient}(color online). Total reflection
coefficient $\left|R\right|$ as a function of the normalized groove-depth
$h/\lambda_{c}$ for $N=100$ (blue dashed) and $N=300$ (red solid).
Here, $w/p=0.5$ and material parameters for $\mathrm{LiNbO_{3}}$
have been used (see text). }
\end{figure}

In this Appendix, we present a detailed discussion of the theoretical
model describing the SAW resonator. 

\textcolor{black}{Typically, a SAW cavity is based on an on-chip distributed
Bragg reflector formed by a periodic array of either metal electrodes
or grooves etched into the surface; see Fig.\ref{fig:SAW-system-sketch}.
In such a grating, each strip reflects only weakly, but, for many
strips $N\gg1$, the total reflection $\left|R\right|$ can approach
unity if the pitch $p$ equals half the wavelength, $p=\lambda_{c}/2$.
This Bragg condition defines the center frequency
\begin{equation}
f_{c}=v_{s}/2p.
\end{equation}
At $f=f_{c}$, the total reflection coefficient is given by 
\begin{equation}
\left|R\right|=\tanh\left(N\left|r_{s}\right|\right),
\end{equation}
where $N$ is the number of strips and $r_{s}$ is the reflection
coefficient associated with a single strip \cite{datta86,morgan07}.
The total reflection coefficient $\left|R\right|$ goes to unity in
the limit $N\left|r_{s}\right|\gg1$; see Fig.(\ref{fig:reflection-coefficient}).
Typically, $N\gtrsim200$ and $|r_{s}|\approx\left(1-2\right)\%$
\cite{morgan07}. For $f\approx f_{c}$, $|r_{s}|$ increases with
the normalized groove depth as $|r_{s}|=C_{1}h/\lambda_{c}\mathrm{sin}\left(\pi w/p\right)+C_{2}\left(h/\lambda_{c}\right)^{2}\cos\left(\pi w/p\right)$,
with material-dependent pre-factors \cite{li75}. For $\mathrm{LiNbO}_{3}$,
$C_{1}=0.67$ and $C_{2}=42$ \cite{li75}. As argued in Ref.\cite{li75},
the first term $\sim C_{1}$ is due to a impedance mismatch, while
the second one $\sim C_{2}$ is due to the stored energy effect. }

\textcolor{black}{Due to the distributed nature of the mirror, strong
reflection occurs over a fractional bandwidth only, given by $\delta f/f_{c}\approx2|r_{s}|/\pi$.
In practice, the cavity formed by two reflective gratings can be viewed
as an acoustic Fabry-Perot resonator with effective reflection centers,
sketched by localized mirrors in Fig.\ref{fig:SAW-system-sketch},
situated at some effective penetration distance into the grating,
given by $L_{p}=\tanh\left[\left(N-1\right)\left|r_{s}\right|\right]\lambda_{c}/\left(4\left|r_{s}\right|\right)\approx\lambda_{c}/4|r_{s}|$
\cite{morgan07,bell76,datta86}. Therefore, the total effective cavity
size along the mirror axis is $L_{c}\approx D+2L_{p}$, where $D$
is the physical gap between the gratings; compare Fig.\ref{fig:SAW-system-sketch}.
For $N\approx100-300$, $h/\lambda_{c}\approx2\%$, we then obtain
$L_{c}\approx38\lambda_{c}$ and $L_{c}\approx42\lambda_{c}$ for
$D=0.75\lambda_{c}$ and $D=5.25\lambda_{c}$, respectively. In analogy
to an optical Fabry-Perot resonator, the mode spacing can then by
estimated as $\Delta f/f_{c}=\lambda_{c}/2L_{c}\approx|r_{s}|$. Since
this is larger than $\delta f/f_{c}$, SAW resonators can be designed
to host a single resonance only \cite{morgan07}. }

\textcolor{black}{The total decay rate of this resonance $\kappa$
can be decomposed into four relevant contributions \cite{li75}, $\kappa=\kappa_{bk}+\kappa_{d}+\kappa_{m}+\kappa_{r}$,
which inludes conversion into bulk modes $\sim\kappa_{bk}$, diffraction
losses $\sim\kappa_{d}$, internal losses due to material imperfections
$\sim\kappa_{m}$, and leakage (radiation) losses due to imperfect
mirrors $\sim\kappa_{r}$. The associated $Q$-factors given by $Q_{i}=\omega_{c}/\kappa_{i}$.
The desired decay rate is $\kappa_{\mathrm{gd}}=\kappa_{r}$, whereas
the undesired one is $\kappa_{\mathrm{bd}}=\kappa_{bk}+\kappa_{m}+\kappa_{d}$.
Here, $\kappa_{d}$ is associated with diffraction losses due to spill-over
beyond the aperture of the reflector. It can be made negligible by
lateral confinement using for example waveguide structures, focusing
or etching techniques \cite{morgan07,deLima05,deLima04}. $Q_{m}$
refers to losses due to interaction with thermal phonons, losses due
to defects in the material and propagation losses due to contamination
\cite{li75,bell76}. These losses ultimately limit $Q$: Low temperature
experiments on quartz have demonstrated SAW resonators with $Q_{m}\times f\left[\mathrm{GHz}\right]>10^{5}$
\cite{elHabti96,magnusson14}. Another source of losses is due to
mode-conversion into bulk-modes. Measurements show that $Q_{bk}=2\pi N_{\mathrm{eff}}/[C_{b}\left(h/\lambda_{c}\right)^{2}]$
with $N_{\mathrm{eff}}=L_{c}/\lambda_{c}$ and a material-dependent
pre-factor $C_{b}$ \cite{tanski79}; for $\mathrm{LiNbO}_{3}$ (Quartz),
$C_{b}=8.7(10)$, respectively \cite{parekh78,tanski79}. Typically,
$\kappa_{bk}$ is found to be negligible for small groove depths,
$h/\lambda_{c}<2\%$ \cite{li75}. Finally, $\kappa_{r}$ arises from
leakage through imperfectly reflecting gratings $\left(|R|<1\right)$;
in direct analogy to optical Fabry-Perot resonators, the associated
$Q$-factor is given by $Q_{r}=2\pi N_{\mathrm{eff}}/\left(1-|R|^{2}\right)$.
Assuming negligible diffraction losses (that can be minimized via
waveguide-like confinement \cite{morgan07,oliner78}) and cryostat
temperatures, the total $Q$-factor is then given by} 
\begin{equation}
Q^{-1}=Q_{m}^{-1}+Q_{bk}^{-1}+Q_{r}^{-1}.
\end{equation}

\section{Charge Qubit Coupled to SAW Cavity Mode\label{sec:Charge-Qubit}}

We consider a GaAs charge qubit embedded in a tunnel-coupled double
quantum dot (DQD) containing a single electron. In the one-electron
regime the single-particle orbital level spacing is on the order of
$\sim1\mathrm{meV}$. Therefore, the system is well described by an
effective two-level system: The state of the qubit is set by the position
of the electron in the double-well potential, with the logical basis
$\left|L\right\rangle ,\left|R\right\rangle $ corresponding to the
electron localized in the left (right) orbital. The Hamiltonian describing
this system reads 
\begin{equation}
H_{\mathrm{ch}}=\frac{\epsilon}{2}\sigma^{z}+t_{c}\sigma^{x},\label{eq:Hamiltonian-charge-qubit}
\end{equation}
with the (orbital) Pauli-operators defined as $\sigma^{z}=\left|L\right\rangle \left\langle L\right|-\left|R\right\rangle \left\langle R\right|$
and $\sigma^{x}=\left|L\right\rangle \left\langle R\right|+\left|R\right\rangle \left\langle L\right|$,
respectively. In Eq.(\ref{eq:Hamiltonian-charge-qubit-main-text-1}),
$\epsilon$ refers to the level detuning between the dots, while $t_{c}$
gives the tunnel coupling. The level splitting between the eigenstates
of $H_{\mathrm{ch}}$ is given by $\Omega=\sqrt{\epsilon^{2}+4t_{c}^{2}}$,
with a pure tunnel-splitting of $\Omega=2t_{c}$ at the charge degeneracy
point $\left(\epsilon=0\right)$; typical parameter values are $t_{c}\sim\mu\mathrm{eV}$
and $\epsilon\sim\mu\mathrm{eV}$, such that the level splitting $\Omega\sim\mathrm{GHz}$
lies in the microwave regime. At the charge degeneracy point, where
to first order the qubit is insensitive to charge fluctuations $\left(d\Omega/d\epsilon=0\right)$,
the coherence time has been found to be $T_{2}\approx10\mathrm{ns}$
\cite{petersson10}. 

We now consider a charge qubit as described above inside a SAW resonator
with a single resonance frequency $\omega_{c}$ close to the qubit's
transition frequency, $\omega_{c}\approx\Omega$, that is the regime
of small detuning $\delta=\Omega-\omega_{c}\approx0$; note that single-resonance
SAW cavities can be realized routinely with today's standard techniques
\cite{morgan07}. Within this single-mode approximation, the Hamiltonian
describing the SAW cavity simply reads 
\begin{equation}
H_{\mathrm{cav}}=\omega_{c}a^{\dagger}a,\label{eq:cavity-Hamiltonian}
\end{equation}
where $a^{\dagger}\left(a\right)$ creates (annihilates) a phonon
inside the cavity.\textcolor{black}{{} The electrostatic potential associated
with this mode is given by $\hat{\phi}\left({\bf x}\right)=\phi\left({\bf x}\right)\left[a+a^{\dagger}\right],$
where the mode-function $\phi\left({\bf x}\right)$ can be obtained
from the corresponding mechanical mode-function ${\bf w}\left({\bf x}\right)$
via the relation $\epsilon\triangle\phi\left({\bf x}\right)=e_{kij}\partial_{j}\partial_{k}w_{i}\left({\bf x}\right)$;
here, $\triangle$ is the Laplacian, $e_{kij}$ the piezoelectric
tensor and $\epsilon$ the permittivity of the material. The electron's
charge $e$ couples to the phonon induced electrical potential $\hat{\phi}$.
In second quantization, the piezoelectric interaction reads $H_{\mathrm{int}}=e\int d{\bf x}\,\hat{\phi}\left({\bf x}\right)\hat{n}\left({\bf x}\right)$,
where $e$ is the electron's charge, $\hat{n}\left({\bf x}\right)=\sum_{\sigma}\psi_{\sigma}^{\dagger}\left({\bf x}\right)\psi_{\sigma}\left({\bf x}\right)$
is the electron number density operator and $\psi_{\sigma}^{\dagger}\left({\bf x}\right)$
creates an electron with spin $\sigma$ at position ${\bf x}$ \cite{kornich14}.
Since $\hat{\phi}\left({\bf x}\right)$ varies on a micron length-scale
which is large compared to the spatial extension $\sim40\mathrm{nm}$
of the electron's wavefunction in a QD \cite{chekhovich13}, the electron
density is approximately given by a delta-function at the center of
the corresponding dots. For the DQD system under consideration $H_{\mathrm{int}}$
is then approximately given by $H_{\mathrm{int}}=e\sum_{i}\hat{\phi}\left({\bf x}_{i}\right)n_{i}$;
here, ${\bf x}_{i}$ refers to the center of the electronic orbital
wavefunction $\psi_{i}\left({\bf x}\right)$ of dot $i=L,R$. Note
that this form of $H_{\mathrm{int}}$ becomes exact if the overlap
integral vanishes, that is if $\int d{\bf x}\,\phi\left({\bf x}\right)\psi_{L}^{*}\left({\bf x}\right)\psi_{R}\left({\bf x}\right)=0$
is satisfied. As shown below, for a mode-function $\phi\left({\bf x}\right)$
of sine-form, this condition maximizes the piezoelectric coupling
strength between the electronic DQD system and the phonon mode. }For
the \textit{charge} qubit system under consideration coupling to the
cavity mode is then described by $H_{\mathrm{int}}=e\left(a+a^{\dagger}\right)\left[\phi\left(x_{L}\right)\left|L\right\rangle \left\langle L\right|+\phi\left(x_{R}\right)\left|R\right\rangle \left\langle R\right|\right]$;
here, $x_{i}$ refers to the center of the electronic orbital wavefunction
$\varphi_{i}\left({\bf x}\right)$ of dot $i=L,R$ and the transverse
direction $\hat{y}$ has been integrated out already. To obtain strong
coupling between the qubit and the cavity, we assume a mode profile
$\phi\left(x\right)=\varphi_{0}\mathrm{sin}\left(kx\right)$, with
a node tuned between the two dots, such that $\phi\left(x_{L}\right)=\varphi_{0}\mathrm{sin}\left(kl/2\right)=-\phi\left(x_{R}\right)$;
here, $l$ gives the distance between the two dots. Note that the
single phonon amplitude, defined as $\varphi_{0}=\phi_{0}\mathcal{F}\left(kd\right)$,
with $\mathcal{F}\left(kd\right)\approx0$ for $d\gg\lambda$, accounts
for the decay of the SAW into the bulk. For $\lambda=\left(0.5-1\right)\mu\mathrm{m}$
and a 2DEG (where the DQD is embedded) situated a distance $d=50\mathrm{nm}$
below the surface, however, the single phonon amplitude is reduced
by a factor of $\sim2$ only, $\varphi_{0}\approx\left(0.45-0.52\right)\phi_{0}$;
see Appendix \ref{sec:Piezoelectric-SAW} for details. Then, the coupling
between qubit and cavity reads 
\begin{equation}
H_{\mathrm{int}}=g_{\mathrm{ch}}\left(a+a^{\dagger}\right)\otimes\sigma^{z},\label{eq:Hamiltonian-coupling-charge-qubit-SAW-mode}
\end{equation}
where the single-phonon coupling strength is 
\begin{equation}
g_{\mathrm{ch}}=e\phi_{0}\mathcal{F}\left(kd\right)\mathrm{sin}\left(kl/2\right)\approx\frac{1.5\mu\mathrm{eV}}{\sqrt{A\left[\mu\mathrm{m}^{2}\right]}}.
\end{equation}
Here, we have assumed $\lambda\approx2l$ such that the geometrical
factor $\mathrm{sin}\left(\pi l/\lambda\right)\approx1$ \cite{naber06}.
In principle, the coupling strength $g_{\mathrm{ch}}$ could be further
enhanced by additionally depositing a strongly piezoelectric material
such as $\mathrm{LiNbO_{3}}$ on the GaAs substrate. Moreover, comparison
with standard literature shows that the piezoelectric electron-phonon
coupling strength can be expressed as $g_{\mathrm{pe}}=\sqrt{P}U_{0}\approx e\left(e_{14}/\epsilon\right)U_{0}=e\phi_{0}$,
where $P=\left(ee_{14}/\epsilon\right)^{2}$ is a material parameter
quantifying the piezoelectric coupling strength in zinc-blend structures
\cite{bruus93,brandes99}. Using $P=5.4\times10^{-20}\mathrm{J^{2}m^{-2}}$
for GaAs, the single phonon Rabi frequency can be estimated as $g_{\mathrm{pe}}\approx2.87\mu\mathrm{eV}/\sqrt{A[\mu\mathrm{m}^{2}]}$.
This corroborates our estimate for $g_{\mathrm{ch}}$. 

In summary, the total system can be described by the Hamiltonian $H=H_{\mathrm{ch}}+H_{\mathrm{cav}}+H_{\mathrm{int}}$,
\begin{equation}
H=\frac{\epsilon}{2}\sigma^{z}+t_{c}\sigma^{x}+\omega_{c}a^{\dagger}a+g_{\mathrm{ch}}\left(a+a^{\dagger}\right)\otimes\sigma^{z}.
\end{equation}
This corresponds to the generic Hamiltonian for a qubit-resonator
system \cite{aspelmeyer14}. It is instructive to rewrite $H$ in
the eigenbasis of $H_{\mathrm{ch}}$, given by 
\begin{eqnarray}
\left|+\right\rangle  & = & \sin\theta\left|L\right\rangle +\cos\theta\left|R\right\rangle ,\\
\left|-\right\rangle  & = & \cos\theta\left|L\right\rangle -\sin\theta\left|R\right\rangle ,
\end{eqnarray}
where the mixing angle $\theta$ is defined via $\tan\theta=2t_{c}/\left(\epsilon+\Omega\right)$.
In a rotating wave approximation $\left(\delta,g_{\mathrm{ch}}\ll\omega_{c}\right)$,
$H$ then reduces to the well-known Hamiltonian of Jaynes-Cummings
form
\begin{equation}
H\approx\delta S^{z}+g_{\mathrm{ch}}\frac{2t_{c}}{\Omega}\left(S^{+}a+S^{-}a^{\dagger}\right),
\end{equation}
where $\delta=\Omega-\omega_{\mathrm{c}}$, $S^{z}=\left(\left|+\right\rangle \left\langle +\right|-\left|-\right\rangle \left\langle -\right|\right)/2$
and $S^{\pm}=\left|\pm\right\rangle \left\langle \mp\right|$.

\section{SAW-Based Cavity QED with Spin Qubits in Double Quantum Dots \label{sec:spin-qubit-QED}}

In this Appendix, we show in detail how to realize the prototypical
Jaynes-Cummings dynamics based on a spin qubit encoded in a double
quantum dot (DQD) inside a SAW resonator. 

We consider a double quantum dot (in the two-electron regime) coupled
to the electrostatic potential generated by a SAW. The system is described
by the Hamiltonian
\begin{equation}
H_{\mathrm{DQD}}=H_{0}+H_{\mathrm{cav}}+H_{\mathrm{int}},
\end{equation}
where $H_{0}$, $H_{\mathrm{cav}}$ and $H_{\mathrm{int}}$ describe
the DQD, the cavity and the electrostatically mediated coupling between
them, respectively. In the following, the different contributions
are discussed in detail. 

\textit{Double Quantum Dot.}---The DQD is modeled by the standard
Hamiltonian
\begin{equation}
H_{0}=H_{C}+H_{t}+H_{Z}.
\end{equation}
Here, $H_{C}$ gives the electrostatic energy 
\begin{equation}
H_{C}=\sum_{i,\sigma}\epsilon_{i}n_{i\sigma}+U\sum_{i=L,R}n_{i\uparrow}n_{i\downarrow}+U_{LR}n_{L}n_{R},
\end{equation}
where (due to strong confinement) both the left and right dot are
assumed to support a single orbital level with energy $\epsilon_{i}\left(i=L,R\right)$
only; $U$ and $U_{LR}$ refer to the on-site and interdot Coulomb
repulsion, respectively. As usual, $n_{i\sigma}=d_{i\sigma}^{\dagger}d_{i\sigma}$
and $n_{i}=n_{i\uparrow}+n_{i\downarrow}$ refer to the spin-resolved
and total electron number operators, respectively, with the fermionic
creation (annihilation) operators $d_{i\sigma}^{\dagger}\left(d_{i\sigma}\right)$
creating (annihilating) an electron with spin $\sigma=\uparrow,\downarrow$
in the orbital $i=L,R$. We focus on a setting where an applied bias
between the two dots approximately compensates the Coulomb energy
of two electrons occupying the right dot; that is, $\epsilon_{L}\approx\epsilon_{R}+U-U_{LR}$.
Thus, we restrict ourselves to a region in the stability diagram where
only $\left(1,1\right)$ and $\left(0,2\right)$ charge states are
of interest. All levels with $\left(1,1\right)$ charge configuration
have an electrostatic energy of $E_{\left(1,1\right)}=\epsilon_{L}+\epsilon_{R}+U_{LR}$,
while the $\left(0,2\right)$ configuration has $E_{\left(0,2\right)}=2\epsilon_{R}+U$.
As usual, we introduce the detuning parameter $\epsilon=\epsilon_{L}-\epsilon_{R}+U_{LR}-U$.
In this regime, the relevant electronic levels are defined as $\left|T_{+}\right\rangle =\left|\Uparrow\Uparrow\right\rangle $,
$\left|T_{-}\right\rangle =\left|\Downarrow\Downarrow\right\rangle $,
$\left|T_{0}\right\rangle =(\left|\Uparrow\Downarrow\right\rangle +\left|\Downarrow\Uparrow\right\rangle )/\sqrt{2}$,
$\left|S_{11}\right\rangle =(\left|\Uparrow\Downarrow\right\rangle -\left|\Downarrow\Uparrow\right\rangle )/\sqrt{2}$
and $\left|S_{02}\right\rangle =d_{R\uparrow}^{\dagger}d_{R\downarrow}^{\dagger}\left|0\right\rangle $
with $\left|\sigma\sigma'\right\rangle =d_{L\sigma}^{\dagger}d_{R\sigma'}^{\dagger}\left|0\right\rangle $. 

Next, $H_{t}$ describes coherent, spin-preserving interdot tunneling
\begin{eqnarray}
H_{t} & = & t_{c}\sum_{\sigma}d_{L\sigma}^{\dagger}d_{R\sigma}+\mathrm{h.c.},
\end{eqnarray}
where $t_{\mathrm{c}}$ is the interdot tunneling amplitude. Lastly,
$H_{Z}$ accounts for the Zeeman energies, 
\begin{equation}
H_{Z}=g\mu_{B}\sum_{i=L,R}\mathbf{B}_{i}\cdot\mathbf{S}_{i},\label{eq:Zeeman-Hamiltonian}
\end{equation}
where $g$ is the electron $g$-factor and $\mu_{B}$ the Bohr magneton,
respectively. In the presence of a micro-/nanomagnet, the two local
magnetic fields ${\bf B}_{i}$ are inhomogeneous, ${\bf B}_{L}\neq{\bf B}_{R}$.
We can then write ${\bf B}_{i}={\bf B}_{0}+{\bf B}_{\mathrm{m}}\left({\bf x}_{i}\right)$,
where ${\bf B}_{0}$ is the external homogeneous magnetic field, while
${\bf B}_{\mathrm{m}}\left({\bf x}_{i}\right)$ is the micromagnet
slanting field at the location of dot ${\bf x}_{i}$. In practice,
$B_{0}$ is a few Tesla, at least larger than the saturation field
of the micromagnet $B_{0}\gtrsim0.5\mathrm{T}$, while the magnetic
gradient $\Delta B=\left\Vert {\bf B}_{\mathrm{m}}\left({\bf x}_{R}\right)-{\bf B}_{\mathrm{m}}\left({\bf x}_{L}\right)\right\Vert $
can reach $\Delta B\approx100\mathrm{mT}$, corresponding to an electronic
energy scale of $\left|g\mu_{B}\Delta B\right|\approx2\mu\mathrm{eV}$
\cite{chesi14}. Field derivatives realized experimentally are $\partial B_{\mathrm{m},z}/\partial x\approx1.5\mathrm{mT}/\mathrm{nm}$.
Alternatively, the magnetic gradient can be realized via the Overhauser
field, as experimentally demonstrated for example in Ref.\cite{bluhm11}. 

Note that the Fermi contact hyperfine interaction between electron
and nuclear spins reads $H_{\mathrm{HF}}=\sum_{i}{\bf h}_{i}\cdot{\bf S}_{i}$.
Here, ${\bf h}_{i}$ is the Overhauser field in QD $i=L,R$. When
treating ${\bf h}_{i}$ as a classical (random) variable, $H_{\mathrm{HF}}$
is equivalent to $H_{Z}$ and thus one can absorb ${\bf h}_{i}$ into
the definition of the magnetic field ${\bf B}_{i}$ in Eq.(\ref{eq:Zeeman-Hamiltonian});
also see Ref.\cite{chesi14}. 

\begin{figure}
\includegraphics[width=1\columnwidth]{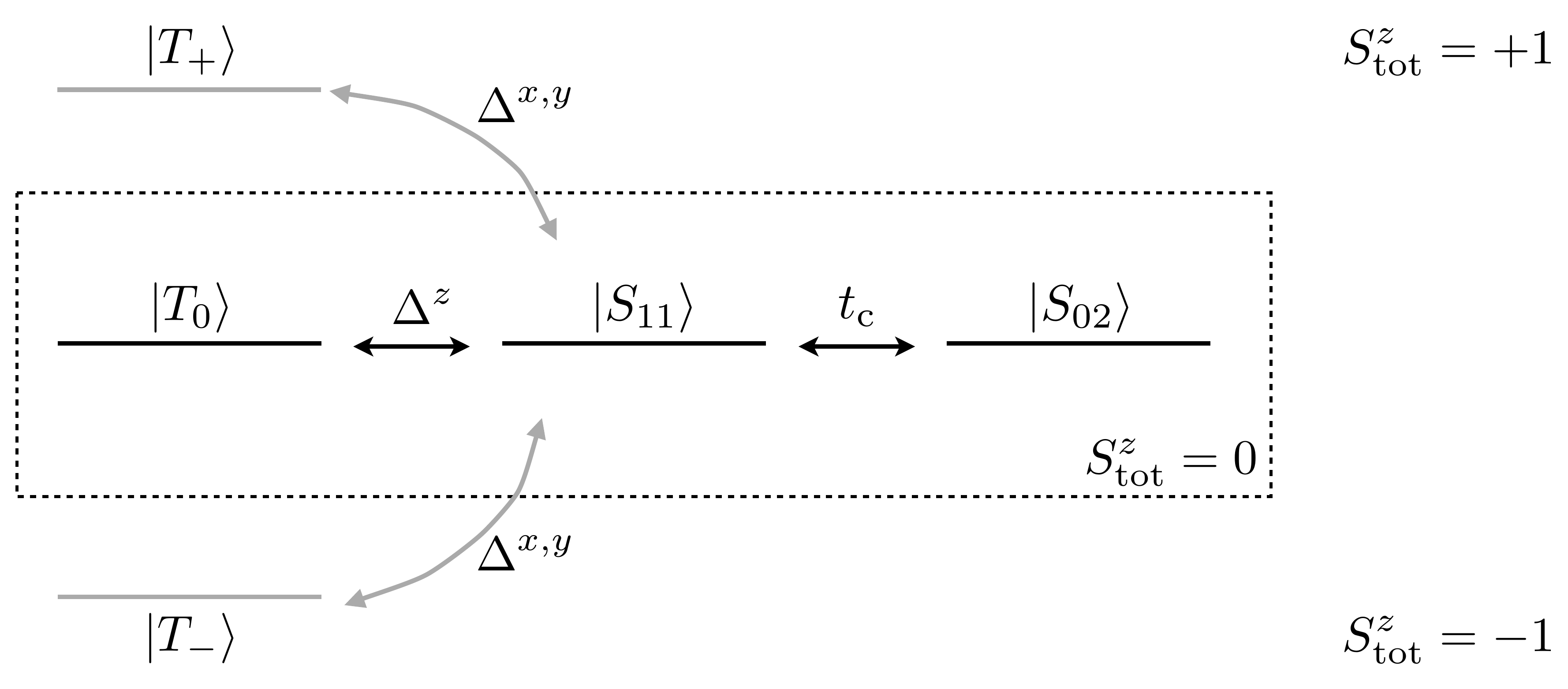}

\caption{\label{fig:DQD-suspace}(color online). Illustration of the relevant
electronic levels under consideration. The triplet levels with $S_{\mathrm{tot}}^{z}\neq1$
can be tuned off-resonance by applying a sufficiently large homogeneous
magnetic field. }
\end{figure}

To facilitate the discussion, we introduce the magnetic sum field
${\bf B}=\left({\bf B}_{L}+{\bf B}_{R}\right)/2$ and the difference
field $\Delta{\bf B}=\left({\bf B}_{R}-{\bf B}_{L}\right)/2$. While
${\bf B}$ conserves the total spin, that is $\left[{\bf B}\left({\bf S}_{L}+{\bf S}_{R}\right),({\bf S}_{L}+{\bf S}_{R})^{2}\right]=0$,
the gradient field $\Delta{\bf B}$ does not. We set the quantization
axis $\hat{z}$ along ${\bf B}=B\hat{z}$. For sufficiently large
magnetic field $B$ the electronic levels with $S_{\mathrm{tot}}^{z}=S_{L}^{z}+S_{R}^{z}\neq0$
are far detuned and can be neglected for the remainder of the discussion.
Therefore, in the following, we restrict ourselves to the $S_{\mathrm{tot}}^{z}=0$
subspace. The components $\Delta{\bf B}^{x,y}$ give rise to transitions
out of the (logical) subspace $S_{\mathrm{tot}}^{z}=0$. Since these
processes are assumed to be far off-resonance, they are neglected
leaving us with the only relevant magnetic gradient $\Delta=g\mu_{B}\Delta{\bf B}^{z}/2$;
compare also Refs.\cite{bluhm11,foletti09}. For a schematic illustration,
compare Fig.\ref{fig:DQD-suspace}.

\begin{figure}
\includegraphics[width=0.95\columnwidth]{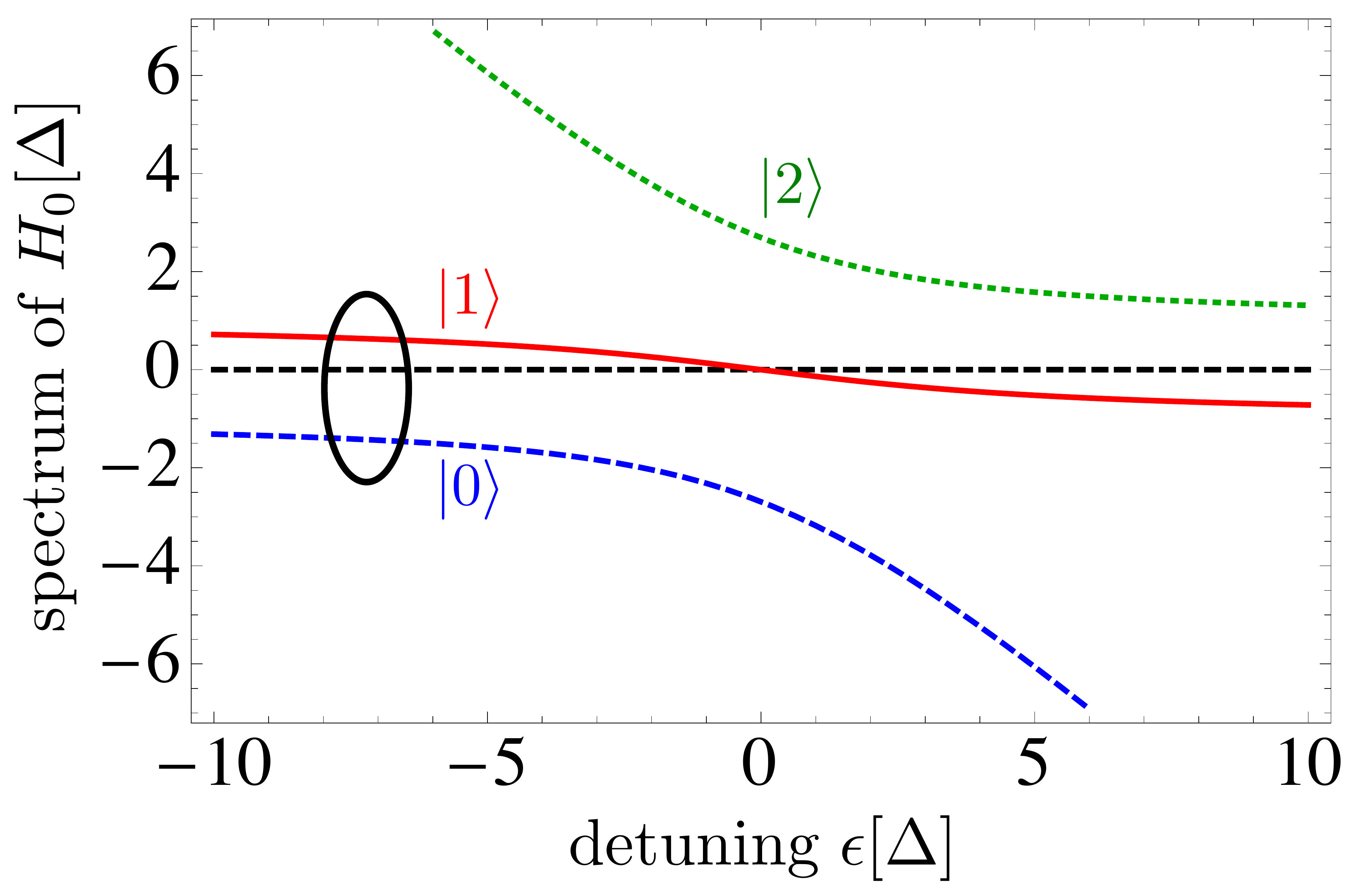}

\caption{\label{fig:SAW-DQD}(color online). Spectrum of $H_{0}$ for $t_{c}=5\Delta$.
The three eigenstates $\left|l\right\rangle $ are displayed in green
dotted $\left(l=2\right)$, red solid $\left(l=1\right)$ and blue
dashed $\left(l=0\right)$, respectively. The triplets $\left|T_{\pm}\right\rangle $
are assumed to be far detuned by a large external field and not shown.
For large negative detuning $\epsilon>-t_{c}$, the hybridized levels
$\left\{ \left|0\right\rangle ,\left|1\right\rangle \right\} $ can
be used as qubit. }
\end{figure}

In summary, in the regime of interest $H_{0}$ simplifies to 
\begin{eqnarray}
H_{0} & = & \frac{t_{c}}{2}\left(\left|S_{02}\right\rangle \left\langle S_{11}\right|+\mathrm{h.c.}\right)-\epsilon\left|S_{02}\right\rangle \left\langle S_{02}\right|\label{eq:H0-DQD-three-bare-levels}\\
 &  & -\Delta\left(\left|T_{0}\right\rangle \left\langle S_{11}\right|+\mathrm{h.c.}\right).\nonumber 
\end{eqnarray}
The eigenstates of $H_{0}$ within the relevant $S_{\mathrm{tot}}^{z}=S_{L}^{z}+S_{R}^{z}=0$
subspace can be expressed as 
\begin{equation}
\left|l\right\rangle =\alpha_{l}\left|T_{0}\right\rangle +\beta_{l}\left|S_{11}\right\rangle +\kappa_{l}\left|S_{02}\right\rangle ,
\end{equation}
with corresponding eigenenergies $\epsilon_{l}\left(l=0,1,2\right)$.
The spectrum is displayed in Fig.\ref{fig:SAW-DQD}. For large negative
detuning $-\epsilon\gg t_{c}$, the level $\left|2\right\rangle $
is far detuned, and the electronic subsystem can be simplified to
an effective two-level system comprising the levels $\left\{ \left|0\right\rangle ,\left|1\right\rangle \right\} $,
that is 
\begin{equation}
H_{0}\approx\frac{\omega_{0}}{2}\left(\left|1\right\rangle \left\langle 1\right|-\left|0\right\rangle \left\langle 0\right|\right),
\end{equation}
which can be identified with a 'singlet-triplet'-like logical qubit
subspace. Here, $\omega_{0}=\epsilon_{1}-\epsilon_{0}$ refers to
the qubit's transition frequency. Note that the magnetic gradient
causes efficient mixing between $\left|T_{0}\right\rangle $ and $\left|S_{11}\right\rangle $
for $\Delta\gtrsim\left|t_{c}^{2}/\epsilon\right|$. In the regime
of interest, the dominant character of the qubit's levels is $\left|1\right\rangle \approx\left|\Downarrow\Uparrow\right\rangle $,
$\left|0\right\rangle \approx\left|\Uparrow\Downarrow\right\rangle $
(or vice versa) \cite{foletti09} and the transition frequency is
approximately $\omega_{0}\approx2\Delta$. For $\Delta\approx1\mu\mathrm{eV}$,
the transition frequency $\omega_{0}=\epsilon_{1}-\epsilon_{0}\approx2\mu\mathrm{eV}\approx3\mathrm{GHz}$
matches typical SAW frequencies $\sim\mathrm{GHz}$. 

\textit{Coupling to SAW phonon mode.}---Along the lines of Appendix
\ref{sec:Charge-Qubit}, again we consider a SAW resonator with a
single relevant confined phonon mode of frequency $\omega_{c}$ close
to the qubit's transition frequency $\omega_{0}$. For a DQD in the\textit{
two}-electron regime, in the basis of Eq.(\ref{eq:H0-DQD-three-bare-levels})
coupling to the resoantor mode can be written as \cite{naber06} 
\begin{eqnarray}
H_{\mathrm{int}} & = & g_{0}\left[a+a^{\dagger}\right]\otimes\left|S_{02}\right\rangle \left\langle S_{02}\right|,\label{eq:bare-interaction-Hamiltonian-cavity-DQD}
\end{eqnarray}
where $g_{0}=e\varphi_{0}\eta_{\mathrm{geo}}$. Here, $\eta_{\mathrm{geo}}$
is a geometrical factor accounting for the DQD's position with respect
to the mode-function $\phi\left({\bf x}\right)$; it is defined according
to $\varphi_{0}\eta_{\mathrm{geo}}=\phi\left({\bf x}_{R}\right)-\phi\left({\bf x}_{L}\right)$.
For example, taking a standing wave pattern along $\hat{x}$ as demonstrated
experimentally in Ref.\cite{li75a}, together with a transverse mode
function restricting the spread in the $\hat{y}$-direction \cite{oliner78,deLima04},
we obtain $\eta_{\mathrm{geo}}=\mathrm{sin}\left(2\pi x_{R}/\lambda\right)-\mathrm{sin}\left(2\pi x_{L}/\lambda\right)$.
It takes on its maximum value $\eta_{\mathrm{opt}}$, when tuning
a node of the standing wave at the center between the two dots, that
is $x_{R}=l/2$, $x_{L}=-l/2$; this gives $\eta_{\mathrm{opt}}=2\mathrm{sin}\left(\pi l/\lambda\right)$,
where $l$ is the distance between the two dots \cite{naber06};\textcolor{black}{{}
as compared to the charge qubit described in Appendix \ref{sec:Charge-Qubit},
there is an additional factor of two, since here we consider a DQD
in the }\textit{\textcolor{black}{two}}\textcolor{black}{-electron
regime, whereas the charge qubit consists of }\textit{\textcolor{black}{one}}\textcolor{black}{{}
electron only.} For typical parameters ($l=220\mathrm{nm}$, $\lambda\approx1.4\mu\mathrm{m}$)
as used in Ref.\cite{naber06}, we get $\eta_{\mathrm{opt}}\approx0.95$,
while $l=220\mathrm{nm}$, $\lambda\approx0.5\mu\mathrm{m}$ leads
to the largest possible value of $\eta_{\mathrm{opt}}\approx2$. 

\begin{figure}
\includegraphics[width=1\columnwidth]{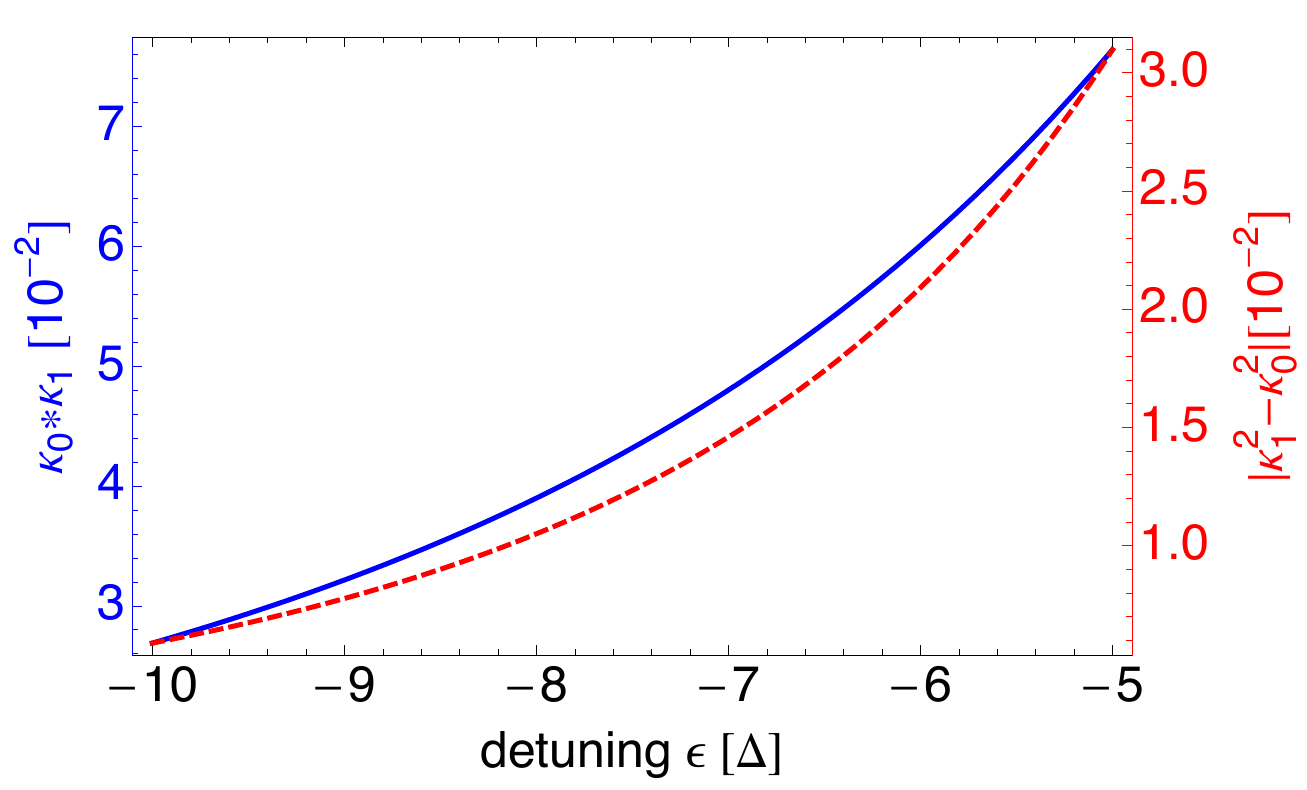}

\caption{\label{fig:DQD-effective-single-phonon-Rabi-frequency}(color online).
The product $\kappa_{0}\kappa_{1}$ directly affects the effective
single-phonon Rabi frequency $g_{\mathrm{QD}}/g_{0}=\kappa_{0}\kappa_{1}$
\cite{taylor06}, while the difference $\left|\kappa_{1}^{2}-\kappa_{0}^{2}\right|$
determines the robustness of the qubit against charge noise. Here,
$t_{c}=5\Delta$.}
\end{figure}

In summary, within the effective electronic two-level subspace $\left\{ \left|0\right\rangle ,\left|1\right\rangle \right\} $,
the system is described by the Hamiltonian 
\begin{eqnarray}
H_{\mathrm{DQD}} & = & \sum_{l=0,1}\left(\epsilon_{l}+\kappa_{l}^{2}\hat{V}_{\mathrm{pe}}\right)\left|l\right\rangle \left\langle l\right|\label{eq:effective-Jaynes-Cummings-Hamiltonian-before-RWA}\\
 &  & +\kappa_{0}\kappa_{1}\hat{V}_{\mathrm{pe}}\left(\left|0\right\rangle \left\langle 1\right|+\mathrm{h.c.}\right)+\omega_{c}a^{\dagger}a,\nonumber 
\end{eqnarray}
where $\hat{V}_{\mathrm{pe}}=g_{0}\left[a+a^{\dagger}\right]$. Applying
a unitary transformation to a frame rotating at the cavity frequency
$\omega_{c}$ according to $\tilde{H}_{\mathrm{DQD}}=UH_{\mathrm{DQD}}U^{\dagger}+i\dot{U}U^{\dagger}$,
with $U=\exp\left[i\omega_{c}t\left(a^{\dagger}a+\frac{1}{2}\left|1\right\rangle \left\langle 1\right|-\frac{1}{2}\left|0\right\rangle \left\langle 0\right|\right)\right]$,
performing a rotating-wave approximation (RWA), and dropping a global
energy shift $\tilde{\epsilon}=\left(\epsilon_{0}+\epsilon_{1}\right)/2$,
we arrive at the effective (time-independent) Hamiltonian of Jaynes-Cummings
form 
\begin{equation}
\tilde{H}_{\mathrm{DQD}}=\bar{\delta}S^{z}+g_{\mathrm{QD}}\left[S^{+}a+S^{-}a^{\dagger}\right],\label{eq:effective-Jaynes-Cummings-Hamiltonian}
\end{equation}
where we have introduced the spin operators $S^{+}=\left|1\right\rangle \left\langle 0\right|$
and $S^{z}=\left(\left|1\right\rangle \left\langle 1\right|-\left|0\right\rangle \left\langle 0\right|\right)/2$.
Moreover, $\bar{\delta}=\omega_{0}-\omega_{c}$ is the detuning between
the qubit's transition frequency $\omega_{0}$ and the cavity frequency
$\omega_{c}$, and the effective single-phonon Rabi frequency is defined
as 
\begin{equation}
g_{\mathrm{QD}}=\kappa_{0}\kappa_{1}\eta_{\mathrm{geo}}e\phi_{0}\mathcal{F}\left(kd\right)\approx2\kappa_{0}\kappa_{1}g_{\mathrm{ch}}.
\end{equation}
The coupling between the qubit and the cavity mode is mediated by
the piezoelectric potential; therefore, it is proportional to the
electron's charge $e$ and the single-phonon electric potential $\phi_{0}$.
Due to the prolonged decoherence timescales, here we consider an effective
(singlet-triplet like) spin-qubit rather than a charge qubit, such
that the coupling $g_{\mathrm{QD}}$ is reduced by the (small) admixtures
with the localized singlet $\kappa_{l}=\left<S_{02}|l\right>$. Increasing
$\kappa_{0}\kappa_{1}$ leads to a stronger Rabi frequency $g_{\mathrm{QD}}$,
but an increased difference in charge configuration $\left|\kappa_{1}^{2}-\kappa_{0}^{2}\right|$
makes the qubit more susceptible to charge noise. For typical numbers
$\left(t_{c}\approx5\mu\mathrm{eV},\,\epsilon\approx-7\mu\mathrm{eV},\,\Delta\approx1\mu\mathrm{eV}\right)$,
we get $\kappa_{0}\kappa_{1}\approx5\times10^{-2}$, $\left|\kappa_{1}^{2}-\kappa_{0}^{2}\right|\approx2\times10^{-2}$;
see Fig.\ref{fig:DQD-effective-single-phonon-Rabi-frequency}. For
$l\approx250\mathrm{nm}$, $\lambda\approx0.5\mu\mathrm{m}$, and
$d\approx50\mathrm{nm}$ we can then estimate 
\begin{equation}
g_{\mathrm{QD}}/\hbar\approx\frac{200\mathrm{MHz}}{\sqrt{A[\mu\mathrm{m}^{2}]}}.
\end{equation}
We take this coupling strength as a conservative estimate, since optimization
against the relevant noise sources as done in Ref.\cite{taylor06}
yields an optimal point with $\kappa_{0}\kappa_{1}\approx0.3$ and
$\omega_{0}/2\pi\approx1.5\mathrm{GHz}$. Resonance $\left(\delta=0\right)$
yields a SAW wavelength of $\lambda\approx2\mu\mathrm{m}$; accordingly,
for a fixed dot-to-dot distance $l=250\mathrm{nm}$, $\eta_{\mathrm{geo}}\approx0.76$
(whereas a larger DQD size $ $$l=400\mathrm{nm}$ as used in Ref.\cite{kornich14}
gives $\eta_{\mathrm{geo}}\approx1.18$). For $l=250\mathrm{nm}$,
$d=50\mathrm{nm}$, we then obtain $g_{\mathrm{QD}}/\hbar\approx600\mathrm{MHz}/\sqrt{A[\mu\mathrm{m}^{2}]}$,
which is a factor of three larger than the estimate quoted above. 

\begin{table}
\begin{tabular}{c||c|c|c|c}
 & cavity size $A[\mu\mathrm{m}^{2}]$  & $g_{\mathrm{QD}}[\mathrm{MHz}]$  & $Q$  & $C$\tabularnewline
\hline 
\hline 
small cavity & 1 & 200 & 10 & 4.25\tabularnewline
\hline 
large cavity & 500 & 9 & $10^{4}$ & 8.5\tabularnewline
\hline 
\end{tabular}

\caption{\label{tab:cooperativity-DQD}Estimates of the single spin cooperativity
$C$ for a DQD singlet-triplet qubit with $T_{2}^{\star}\approx100\mathrm{ns}$,
in a SAW cavity at gigahertz frequencies $\omega_{c}/2\pi\approx1.5\mathrm{GHz}$
and cryostat temperatures where $\bar{n}_{\mathrm{th}}\approx0$ for
both a small, low $Q$ and large, high $Q$ SAW-resonator. The coupling
strength $g_{\mathrm{QD}}$ could be further increased by additionally
depositing a strongly piezoelectric material such as $\mathrm{LiNbO_{3}}$
on the GaAs substrate and spin-echo (and/or narrowing) techniques
allow for dephasing times extended by up to three orders of magnitude
\cite{kloeffel13,shulman14}. }
\end{table}

\textit{Cooperativity.}---In this context, an important figure of
merit is the single spin cooperativity \cite{ovartchaiyapong14},
$C=g_{\mathrm{QD}}^{2}T_{2}/\kappa\left(\bar{n}_{\mathrm{th}}+1\right)$,
where $\kappa=\omega_{c}/Q$ is the mechanical damping rate and $\bar{n}_{\mathrm{th}}=1/(e^{\hbar\omega_{c}/k_{B}T}-1)$
is the equilibrium phonon occupation number at temperature $T$; here,
since $\hbar\omega_{c}\gg k_{B}T$ for cryostatic temperatures, $\bar{n}_{\mathrm{th}}\approx0$.
For singlet-triplet qubits in lateral QDs, $T_{2}^{\star}\approx100\mathrm{ns}$
\cite{kloeffel13}; using spin-echo techniques, experimentally this
has even been extended to $T_{2}=276\mu\mathrm{s}$. Even in the absence
of spin-echo pulses, with a far-from-optimistic dephasing time $T_{2}^{\star}\approx100\mathrm{ns}$
\cite{shulman14}, for a moderately small cavity size $A\approx100\mu\mathrm{m}^{2}$,
a quality factor of $Q=900$ is sufficient to reach $C\approx g_{\mathrm{QD}}^{2}T_{2}^{\star}Q/\omega_{c}\approx3.8$.
Note that $C>1$ allows to perform a quantum gate between two spins
mediated by a thermal mechanical mode \cite{rabl10}. 

\begin{figure}
\includegraphics[width=1\columnwidth]{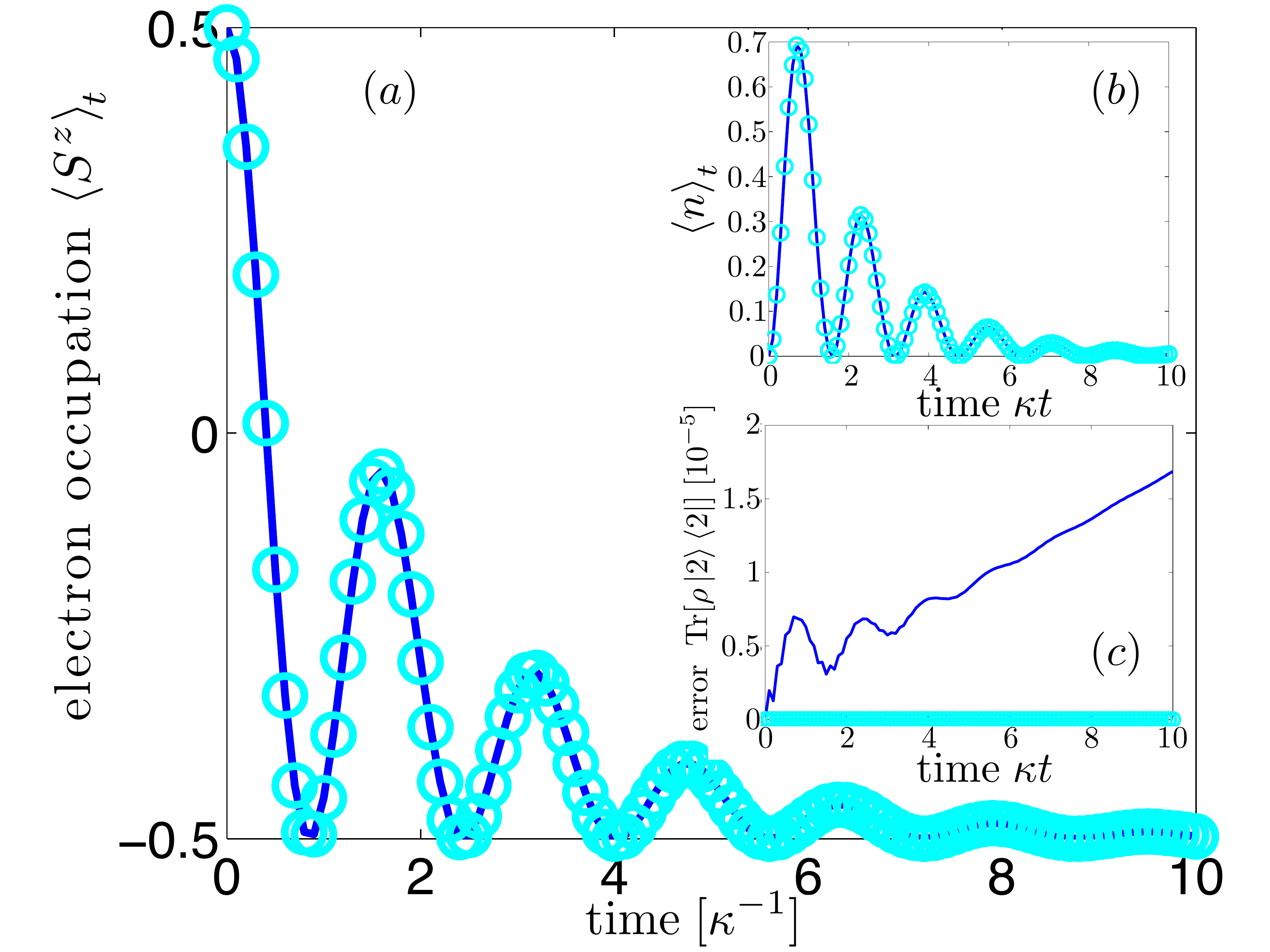}

\caption{\label{fig:SAW-DQD-numerical-simulation-approximations}(color online).
Exact numerical simulations of the full (blue, solid line) and approximate
(cyan circles) Master equations as given in Eqs.(\ref{eq:Master-equation-DQD-cavity-full})
and (\ref{eq:Master-equation-DQD-cavity-approx}) respectively. Plots
are shown for (a) the electronic inversion $\left\langle S^{z}\right\rangle _{t}$,
(b) the cavity occupation $\left\langle n\right\rangle _{t}$ and
(c) the error $\mathrm{Tr}\left[\rho\left|2\right\rangle \left\langle 2\right|\right]$
quantifying the leakage to $\left|2\right\rangle $. The latter is
found to be negligibly small $\sim10^{-5}$. We have set $\bar{\delta}=\omega_{0}-\omega_{c}=0$.
Numerical parameters: $t_{\mathrm{c}}=10\mu\mathrm{eV}$, $\epsilon=-7\mu\mathrm{eV}$,
$\Delta=1\mu\mathrm{eV}$, $\eta_{\mathrm{geo}}e\varphi_{0}=5.2\times10^{-2}\mu\mathrm{eV}$
such that $g_{\mathrm{QD}}=4\times10^{-3}\mu\mathrm{eV}\approx6\mathrm{MHz}$.
The cavity decay rate is $\kappa=g_{\mathrm{QD}}/2$, corresponding
to $Q\approx10^{3}$. }
\end{figure}

\textit{Discussion of approximations.}---To arrive at the effective
Hamiltonian given in Eq.(\ref{eq:effective-Jaynes-Cummings-Hamiltonian}),
we have made two essential approximations: (i) first, we have neglected
the electronic level $\left|2\right\rangle $ yielding an effective
two-level system (TLS), and (ii) second, we have applied a RWA leading
to a major simplification of $H_{\mathrm{DQD}}$; see Eq.(\ref{eq:effective-Jaynes-Cummings-Hamiltonian})
as compared to Eq.(\ref{eq:effective-Jaynes-Cummings-Hamiltonian-before-RWA}).
In order to corroborate these approximations, we now compare exact
numerical simulations of the full system where none of the approximations
have been applied to the simplified, approximate description described
above. While the dynamics of the former is described by the Master
equation 
\begin{equation}
\dot{\rho}=-i\left[H_{0}+H_{\mathrm{cav}}+H_{\mathrm{int}},\rho\right]+\kappa\mathcal{D}\left[a\right]\rho,\label{eq:Master-equation-DQD-cavity-full}
\end{equation}
with $H_{0}$, $H_{\mathrm{cav}}$ and $H_{\mathrm{int}}$ given in
Eqs.(\ref{eq:H0-DQD-three-bare-levels}), (\ref{eq:cavity-Hamiltonian})
and (\ref{eq:bare-interaction-Hamiltonian-cavity-DQD}), respectively,
the latter is described by a similar Master equation with the coherent
Hamiltonian term replaced by the prototypical Jaynes-Cummings Hamiltonian,
\begin{equation}
\dot{\rho}=-i\left[\bar{\delta}S^{z}+g_{\mathrm{QD}}\left(S^{+}a+S^{-}a^{\dagger}\right),\rho\right]+\kappa\mathcal{D}\left[a\right]\rho,\label{eq:Master-equation-DQD-cavity-approx}
\end{equation}
with $\rho$ referring to the density matrix of the combined system
comprising the DQD and the cavity mode. Here, we have also accounted
for decay of cavity phonons out of the resonator with a rate $\kappa$,
described by the Lindblad term $\mathcal{D}\left[a\right]\rho=a\rho a^{\dagger}-\frac{1}{2}\left\{ a^{\dagger}a,\rho\right\} $.
As a figure of merit to validate approximation (i) we determine the
population of the level level $\left|2\right\rangle $, that is $\mathrm{Tr}\left[\rho\left|2\right\rangle \left\langle 2\right|\right]$,
describing the undesired leakage out of the logical subspace; ideally,
this should be zero. Note that leakage into the triplet levels $\left|T_{\pm}\right\rangle $
could be accounted for along the lines, but they can be tuned far
off-resonance by another, independent experimental knob, the external
homogenous magnetic field. The results are summarized in Fig.\ref{fig:SAW-DQD-numerical-simulation-approximations}:
We find very good agreement between the exact and the approximate
model, with a negligibly small error $\mathrm{Tr}\left[\rho\left|2\right\rangle \left\langle 2\right|\right]\sim\mathcal{O}\left(10^{-5}\right)$.
This justifies the approximations made above and shows that (in the
regime of interest) the system can simply be described by Eq.(\ref{eq:Master-equation-DQD-cavity-approx}).

\subsection*{Noise Sources for the DQD-based System}

\begin{figure}
\includegraphics[width=1\columnwidth]{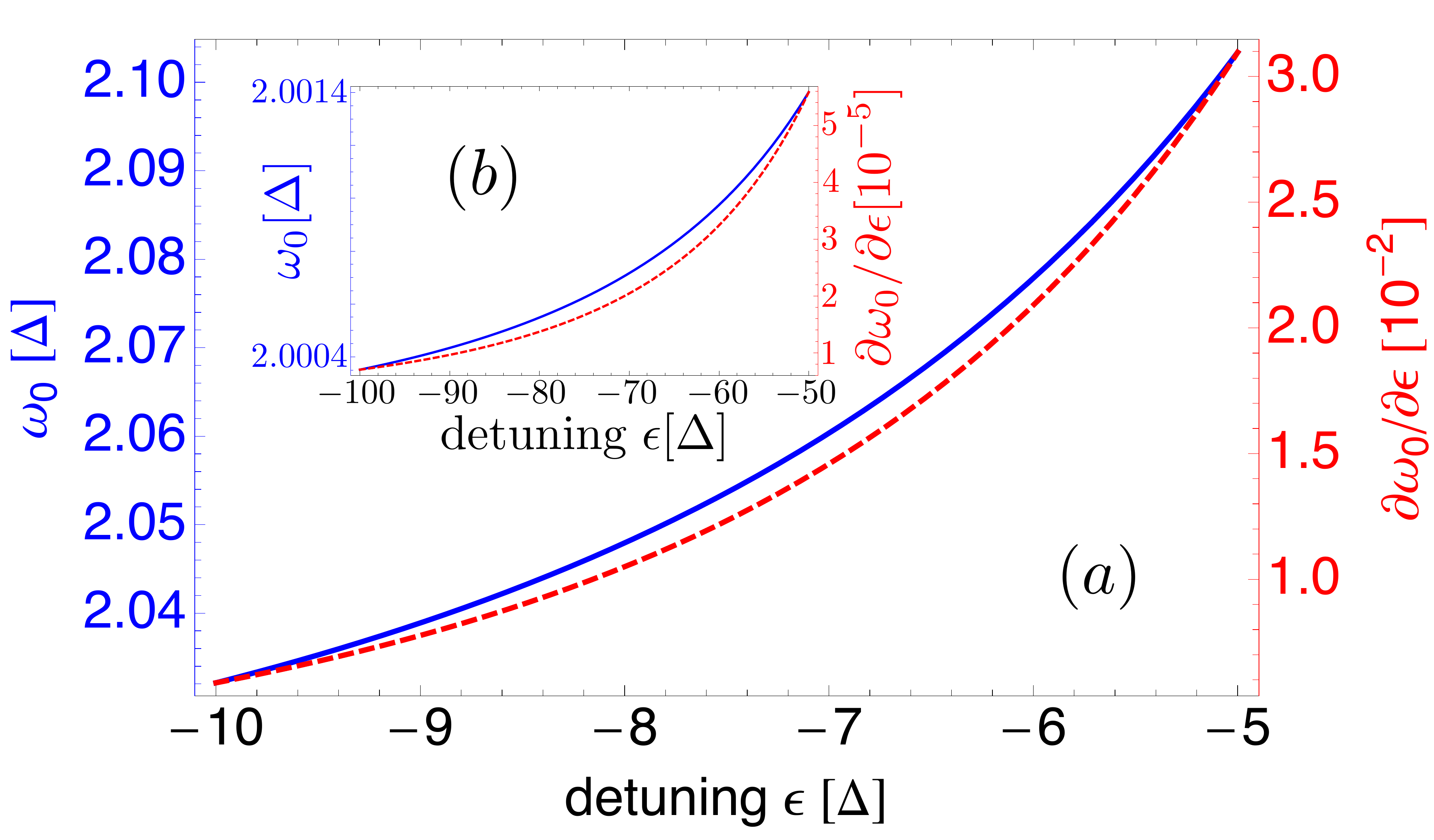}

\caption{\label{fig:SAW-DQD-charge-noise}(color online). Transition frequency
$\omega_{0}$ (blue solid) and its sensitivity against charge noise
induced fluctuations in $\epsilon$ for intermediate (a) and large
(b) negative detuning; for large negative detuning $\omega_{0}\approx2\Delta$,
and the sensitivity $\partial\omega_{0}/\partial\epsilon$ practically
vanishes leaving nuclear noise as the dominant dephasing process.
By occasional refocusing of the spin states, this regime can be used
for long-term storage of quantum information \cite{taylor06}. Other
numerical parameters: $t_{\mathrm{c}}=5\Delta$. }
\end{figure}

\textit{Charge noise.}---In a DQD device background charge fluctuations
and noise in the gate voltages may cause undesired dephasing processes.
In a recent experimental study \cite{dial13}, voltage fluctuations
in the intedot detuning parameter $\epsilon$ have been identified
as the main source of charge noise in a singlet-triplet qubit. Charge
noise can be treated by introducing a Gaussian distribution in $\epsilon$,
with a variance $\sigma_{\epsilon}$; typically $\sigma_{\epsilon}\approx\left(1-3\right)\mu\mathrm{eV}$
\cite{chesi14}. The qubit's transition frequency $\omega_{0}$, however,
turns out to be rather insensitive to fluctuations in $\epsilon$,
with a (tunable) sensitivity of approximately $\partial\omega_{0}/\partial\epsilon\lesssim10^{-2}$;
see Fig.\ref{fig:SAW-DQD-charge-noise}. In agreement with experimental
results presented in Ref.\cite{dial13}, we find $\partial\omega_{0}/\partial\epsilon\sim\omega_{0}$,
indicating $\omega_{0}$ to be an exponential function of $\epsilon$.
At very negative detuning $\epsilon$, dephasing due to charge noise
is practically absent, and $T_{2}^{\star}$ will be limited by nuclear
noise \cite{dial13}. Fluctuations in the tunneling amplitude $t_{c}$
can be treated along the lines: we find $\omega_{0}$ to be similarly
insensitive to noise in $t_{c}$, $\partial\omega_{0}/\partial t_{c}\approx10^{-2}$. 

\textit{Nuclear noise: Spin echo.}---The electronic qubit introduced
above has been defined for a fixed set of parameters $\left(t_{c},\epsilon,\Delta\right)$;
compare Eq.(\ref{eq:H0-DQD-three-bare-levels}). Now, let us consider
the effect of deviations from this fixed parameters, $H_{0}\rightarrow H_{0}+\delta H$,
where $\delta H$ can be decomposed as $\delta H=\delta H_{\mathrm{el}}+\delta H_{\mathrm{nuc}}$
with 
\begin{eqnarray}
\delta H_{\mathrm{el}} & = & \frac{\delta t_{c}}{2}\left(\left|S_{02}\right\rangle \left\langle S_{11}\right|+\mathrm{h.c.}\right)-\delta\epsilon\left|S_{02}\right\rangle \left\langle S_{02}\right|,\\
\delta H_{\mathrm{nuc}} & = & -\delta\Delta\left(\left|T_{0}\right\rangle \left\langle S_{11}\right|+\mathrm{h.c.}\right),
\end{eqnarray}
where $\delta t_{c}$ and $\delta\epsilon$ can be tuned electrostatically
and basically in-situ. In most practical situations this does not
hold for $\delta\Delta$: The primary source of decoherence in this
system has been found to come from (slow) fluctuations in the Overhauser
field generated by the nuclear spins \cite{bluhm11,foletti09,shulman14}.
In our model, this can directly be identified with a random, slowly
time-dependent parameter $\delta\Delta=\delta\Delta\left(t\right)$.
In the relevant subspace $\left\{ \left|0\right\rangle ,\left|1\right\rangle \right\} $,
$\delta H_{\mathrm{nuc}}$ is given by $\delta H_{\mathrm{nuc}}=-\delta\Delta\sum_{k,l}\alpha_{k}\beta_{l}\left[\left|k\right\rangle \left\langle l\right|+\mathrm{h.c.}\right]$.
Typically, $\delta\Delta\approx0.1\mu\mathrm{eV}\ll\omega_{c}$ is
fulfilled, such that we can apply a RWA yielding $\delta H_{\mathrm{nuc}}\approx-2\delta\Delta\sum_{l}\alpha_{l}\beta_{l}\left|l\right\rangle \left\langle l\right|$;
physically, $\delta H_{\mathrm{nuc}}$ is too weak to drive transitions
between the electronic levels $\left|0\right\rangle $ and $\left|1\right\rangle $
which are energetically separated by $\omega_{0}\approx\omega_{c}$.
Then, in the spin basis used in Eq.(\ref{eq:effective-Jaynes-Cummings-Hamiltonian}),
we find 
\begin{equation}
\delta H_{\mathrm{nuc}}=\delta\left(t\right)S^{z},\label{eq:nuclear-noise-Sz}
\end{equation}
where the gradient noise is given by $\delta\left(t\right)=2\delta\Delta\left(t\right)\left(\alpha_{0}\beta_{0}-\alpha_{1}\beta_{1}\right)$.
For $\left(t_{c},\epsilon,\Delta\right)\approx\left(5,-7,1\right)\mu\mathrm{eV}$,
$\alpha_{0}\beta_{0}-\alpha_{1}\beta_{1}\approx0.9$. Therefore, when
also accounting for nuclear noise as described by Eq.(\ref{eq:nuclear-noise-Sz}),
the full model {[}compare Eq.(\ref{eq:effective-Jaynes-Cummings-Hamiltonian}){]}
reads
\begin{equation}
\tilde{H}_{\mathrm{DQD}}=\left[\bar{\delta}+\delta\left(t\right)\right]S^{z}+g_{\mathrm{QD}}\left[S^{+}a+S^{-}a^{\dagger}\right].
\end{equation}
Since the nuclear spins evolve on timescales much longer than all
other relevant timescales $\sim\kappa^{-1},g^{-1}$ \cite{chekhovich13},
the Overhauser noise term can be approximated as quasi-static, that
is $\delta\left(t\right)=\delta$. As experimentally demonstrated
in Ref.\cite{bluhm11}, the slow (nuclear) noise term $\sim\delta\left(t\right)S^{z}$
can be neutralized by Hahn-echo techniques. Here, the dephasing time
of the electron spin qubit was extended by more than three orders
of magnitude from $T_{2}^{\star}\approx\left(10-100\right)\mathrm{ns}$
to $T_{2}=276\mu\mathrm{s}$. 

In the following, assuming nominal resonance $\bar{\delta}=0$, we
detail a sequence of Hahn-echo pulses that cancels the undesired noise
term and restores the pure, resonant Jaynes-Cummings dynamics: We
consider four short time intervals of length $\tau$, for which the
unitary evolution is approximately given by $U_{i}\approx\mathds1-iH_{i}\tau$,
interspersed by three $\pi$-pulses. First, we let the system evolve
with $H_{1}=\delta S^{z}+g_{\mathrm{QD}}\left[S^{+}a+S^{-}a^{\dagger}\right]$,
then we apply a $\pi$-pulse along $\hat{x}$ $\left(S^{z}\rightarrow-S^{z},S^{x}\rightarrow S^{x},S^{y}\rightarrow-S^{y}\right)$
such that $S^{\pm}\rightarrow S^{\mp}$ and the system evolves in
the second time interval with $H_{2}=-\delta S^{z}+g_{\mathrm{QD}}\left[S^{-}a+S^{+}a^{\dagger}\right]$.
Next, we apply a $\pi$-pulse along $\hat{z}$ $\left(S^{z}\rightarrow S^{z},S^{x}\rightarrow-S^{x},S^{y}\rightarrow-S^{y}\right)$
such that $S^{\pm}\rightarrow-S^{\pm}$ leading to $H_{3}=-\delta S^{z}-g_{\mathrm{QD}}\left[S^{-}a+S^{+}a^{\dagger}\right]$.
Finally, a $\pi$-pulse along $\hat{y}$ $\left(S^{z}\rightarrow-S^{z},S^{x}\rightarrow-S^{x},S^{y}\rightarrow S^{y}\right)$
is applied such that $S^{\pm}\rightarrow-S^{\mp}$ giving $H_{4}=\delta S^{z}+g_{\mathrm{QD}}\left[S^{+}a+S^{-}a^{\dagger}\right]$.
In summary, the system evolves over a time interval of $4\tau$ according
to $U_{\mathrm{eff}}=U_{4}U_{3}U_{2}U_{1}\approx\mathds1-i\tau\sum_{i}H_{i}=\mathds1-iH_{\mathrm{eff}}4\tau$
with the effective Hamiltonian 
\begin{equation}
H_{\mathrm{eff}}=\frac{g_{\mathrm{QD}}}{2}\left[S^{+}a+S^{-}a^{\dagger}\right].
\end{equation}
Thus, in order to cancel the noise term, the effective single-phonon
coupling strength is only lowered by a factor of $1/2$.

\subsection*{Different Spin-Resonator Coupling }

In Appendix \ref{sec:spin-qubit-QED}, we have shown how to realize
the prototypical Jaynes-Cummings Hamiltonian for SAW phonons interacting
with a DQD; see Eq.(\ref{eq:effective-Jaynes-Cummings-Hamiltonian}).
Alternatively, if one does not absorb the gradient $\Delta$ into
the definition of the qubit basis, one can identify the logical subspace
with the electronic states $\left|T_{0}\right\rangle $ and $\left|S\right\rangle $,
where $\left|S\right\rangle $ is one of the two hybridized singlets
(while the other one $\left|S'\right\rangle $ is far detuned and
neglected) \cite{bluhm11,foletti09,shulman14,dial13}. Here, the electronic
Hamiltonian reads $H_{0}=-J\left(\epsilon\right)\left|S\right\rangle \left\langle S\right|-\tilde{\Delta}\left(\left|T_{0}\right\rangle \left\langle S\right|+\mathrm{h.c.}\right)$,
where $\tilde{\Delta}=\left<S_{11}|S\right>\Delta$ and $J\left(\epsilon\right)$
describes the exchange interaction. In this regime, the spin-resonator
interaction takes on a form that is well known from other (localized)
implementations of mechanical resonators \cite{rabl10}, namely 
\begin{equation}
H_{\mathrm{int}}=g_{\mathrm{qd}}\left(a+a^{\dagger}\right)\otimes\left|S\right\rangle \left\langle S\right|,\label{eq:spin-resonator-QND-coupling}
\end{equation}
which can be viewed as a phonon-state dependent force, leading to
a shift of the qubit's transition frequency depending on the position
$\hat{x}=\left(a+a^{\dagger}\right)/\sqrt{2}$. Here, the single-phonon
Rabi frequency is $g_{\mathrm{qd}}=\kappa_{S}^{2}\eta_{\mathrm{geo}}e\phi_{0}\mathcal{F}\left(kd\right)$,
with $\kappa_{S}=\left<S_{02}|S\right>$. Based on the coupling of
Eq.(\ref{eq:spin-resonator-QND-coupling}), one can envisage a variety
of experiments known from quantum optics: For example, in the limit
of vanishing gradient $\Delta=0$, the $\hat{x}$ quadrature of the
phonon mode could serve as a quantum nondemolition variable, as it
is a integral of motion of the coupled system of phonon mode and electronic
meter.

\section{Generalized Definition of the Cooperativity Parameter\label{sec:cooperativity}}

In this Appendix we provide a generalized discussion of the cooperativity
parameter $C$ which in particular accounts for losses of the cavity
mode other than leakage through the non-perfect mirrors. \textcolor{black}{Furthermore,
we derive a simple, analytical estimate for the state transfer fidelity
$\mathcal{F}$ in terms of the parameter $C$ and undesired phonon
losses with a rate $\sim\kappa_{\mathrm{bd}}$. }

We consider a single qubit $\left\{ \left|0\right\rangle ,\left|1\right\rangle \right\} $
coupled to a cavity mode. The system is described by the Master equation
\begin{equation}
\dot{\varrho}=\underset{\mathcal{L}_{0}\varrho}{\underbrace{\left(\kappa_{\mathrm{gd}}+\kappa_{\mathrm{bd}}\right)\mathcal{D}\left[a\right]\varrho}}\underset{\mathcal{V}\varrho}{\underbrace{-i\left[H_{\mathrm{JC}},\varrho\right]+\Gamma_{\mathrm{deph}}\mathcal{D}\left[\left|1\right\rangle \left\langle 1\right|\right]\varrho}},
\end{equation}
where $\mathcal{D}\left[a\right]\varrho=a\varrho a^{\dagger}-\frac{1}{2}\left\{ a^{\dagger}a,\varrho\right\} $.
The first term describes decay of the cavity mode. The corresponding
decay rate can be decomposed into desired (leakage through the mirrors)
and undesired (bulk mode conversion etc.) contributions, labeled as
$\kappa_{\mathrm{gd}}$ and $\kappa_{\mathrm{bd}}$, respectively.
Thus, we write $\kappa=\kappa_{\mathrm{gd}}+\kappa_{\mathrm{bd}}$.
The second term with (on resonance) $H_{\mathrm{JC}}=g\left(S^{+}a+S^{-}a^{\dagger}\right)$
refers to the coherent interaction between qubit and cavity mode,
while the last term describes pure dephasing of the qubit with a rate
$\Gamma_{\mathrm{deph}}$. 

In the bad cavity limit (where $\kappa\gg g,\Gamma_{\mathrm{deph}}$),
one can adiabatically eliminate the cavity mode by projecting the
system onto the cavity vacuum, $\mathcal{P}\varrho=\mathrm{Tr_{cav}}\left[\varrho\right]\otimes\rho_{\mathrm{cav}}^{ss}=\rho\otimes\left|\mathrm{vac}\right\rangle \left\langle \mathrm{vac}\right|$.
Standard techniques (perturbation theory up to second order in $\mathcal{V}$,
compare Ref.\cite{kessler12}) then yield the effective Master equation
for the qubit's density matrix $\rho=\mathrm{Tr_{cav}}\left[\mathcal{P}\varrho\right]$
only, 
\begin{equation}
\dot{\rho}=\tilde{\kappa}\mathcal{D}\left[S^{-}\right]\rho+\Gamma_{\mathrm{deph}}\mathcal{D}\left[\left|1\right\rangle \left\langle 1\right|\right]\rho,\label{eq:effective-Master-equation-qubit-bad-cavity-limit}
\end{equation}
with the effective decay rate $\tilde{\kappa}=4g^{2}/\kappa$. 

For comparison, the same procedure in standard cavity QED, where $\kappa=\kappa_{\mathrm{gd}}$
and $\Gamma_{\mathrm{deph}}\mathcal{D}\left[\left|1\right\rangle \left\langle 1\right|\right]\rho\rightarrow\gamma\mathcal{D}\left[S^{-}\right]\rho$,
yields the effective Master equation for the atom only, $\dot{\rho}=\tilde{\kappa}\mathcal{D}\left[S^{-}\right]\rho+\gamma\mathcal{D}\left[S^{-}\right]\rho$.
Therefore, the atom decays with an effective spontaneous emission
rate $\gamma_{\mathrm{tot}}$ enhanced by the Purcell factor, $\gamma_{\mathrm{tot}}=\gamma+\tilde{\kappa}=\left(1+4g^{2}/\kappa\gamma\right)\gamma$.
Comparing good $\sim\tilde{\kappa}$ to bad $\sim\gamma$ decay channels,
here one defines the cooperativity parameter in a straightforward
way as $C_{\mathrm{atom}}=g^{2}/\kappa\gamma$. \textcolor{black}{This
is readily read as the cavity-to-free-space scattering ratio, since
the effective rate at which an excited atom emits an excitation into
the cavity is given by $\tilde{\kappa}\sim g^{2}/\kappa$. For $C_{\mathrm{atom}}>1$,
the atom is then more likely to decay into the cavity mode rather
than into another mode outside the cavity. In cavity QED, large cooperativity
$C_{\mathrm{atom}}\gg1$ has allowed for a number of key experimental
demonstrations such as an enhancement of spontaneous emission \cite{purcell46},
photon blockade \cite{birnbaum05} and vacuum-induced transparency
\cite{tanji-suzuki11}. }

The Master equation given in Eq.(\ref{eq:effective-Master-equation-qubit-bad-cavity-limit})
describes a two-level system subject to purely dissipative dynamics.
The dynamics can be fully described in terms of a set of three simple
rate equations for the populations $p_{k}=\left<k|\rho|k\right>$
$\left(k=0,1\right)$ and coherence $\rho_{10}=\left<1|\rho|0\right>$,
summarized as $\vec{p}=\left(p_{1},p_{0},\rho_{10}\right)$, 
\begin{equation}
\frac{d}{dt}\vec{p}=\left(\begin{array}{ccc}
-\tilde{\kappa} & 0 & 0\\
+\tilde{\kappa} & 0 & 0\\
0 & 0 & -\gamma_{\mathrm{eff}}/2
\end{array}\right)\vec{p},
\end{equation}
where $\gamma_{\mathrm{eff}}=\left(\tilde{\kappa}+\Gamma_{\mathrm{deph}}\right)$.
This allows for a simple analytical solution: For example, for a system
initially in the excited state $\rho\left(t=0\right)=\left|1\right\rangle \left\langle 1\right|$
it reads $p_{1}\left(t\right)=\exp\left[-\tilde{\kappa}t\right]$,
$p_{0}=1-\exp\left[-\tilde{\kappa}t\right]$ and $\rho_{10}\left(t\right)=0$. 

\begin{figure}[b]
\includegraphics[width=1\columnwidth]{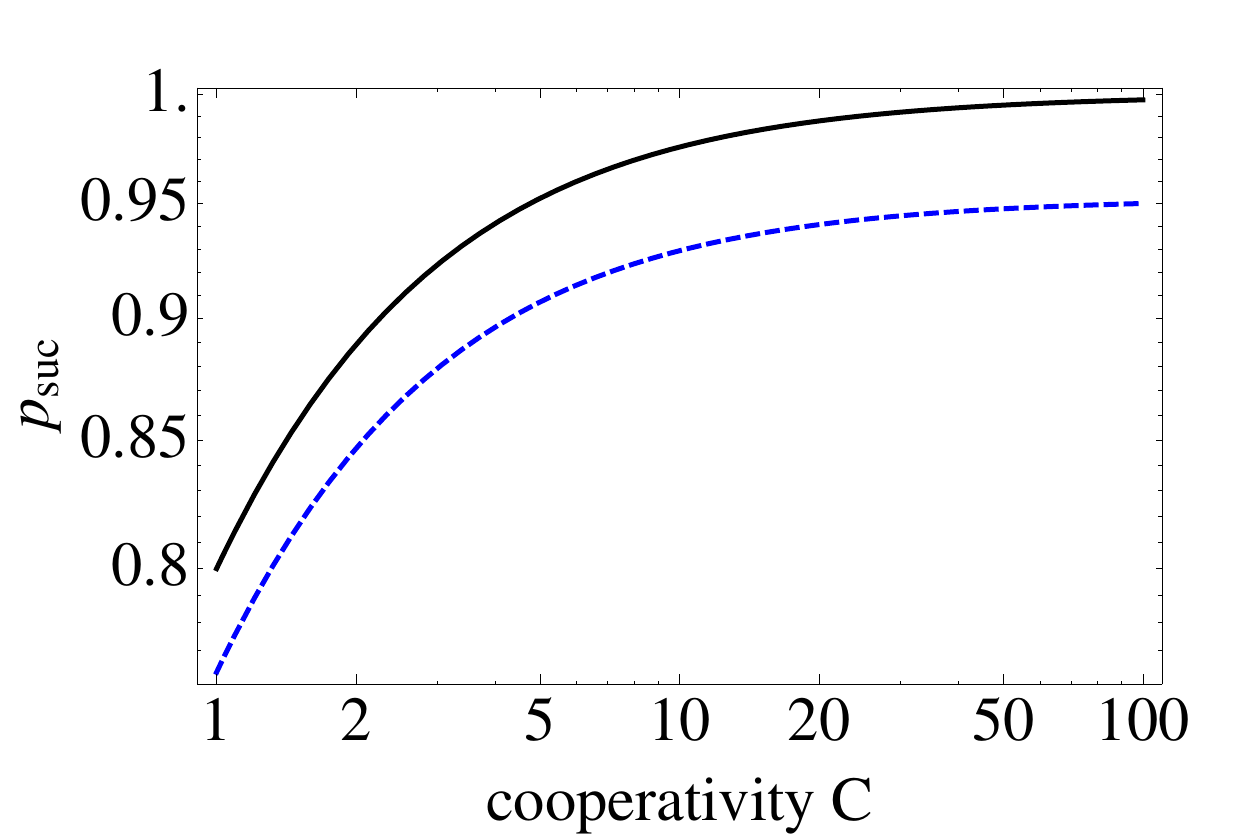}

\caption{\label{fig:coop-parameter_bad-decay_success-prob1}(color online).
Success probability $p_{\mathrm{suc}}$ for a qubit excitation to
leak through the mirror, as a function of the cooperativity $C$ for
$\varepsilon=\kappa_{\mathrm{bd}}/\kappa_{\mathrm{gd}}=0$ (black
solid) and $\varepsilon=5\%$ (blue dashed). }
\end{figure}

Here, we aim for a theoretical description that singles out the desired
trajectories, where phonon emission through the mirrors happens first,
from all others. To do so, we rewrite Eq.(\ref{eq:effective-Master-equation-qubit-bad-cavity-limit})
as 
\begin{equation}
\dot{\rho}=-iH\rho+i\rho H^{\dagger}+\mathcal{J}_{\mathrm{gd}}\rho+\mathcal{J}_{\mathrm{bd}}\rho,\label{eq:effective-Master-equation-Hamiltonian-jumps}
\end{equation}
where we have defined an effective (non-hermitian) Hamiltonian $H$
and jump operators according to 
\begin{eqnarray}
H & = & -\frac{i}{2}\gamma_{\mathrm{eff}}\left|1\right\rangle \left\langle 1\right|=-\frac{i}{2}\left(\tilde{\kappa}+\Gamma_{\mathrm{deph}}\right)\left|1\right\rangle \left\langle 1\right|,\\
\mathcal{J}_{\mathrm{gd}}\rho & = & \tilde{\kappa}_{\mathrm{gd}}S^{-}\rho S^{+},\\
\mathcal{J}_{\mathrm{bd}}\rho & = & \tilde{\kappa}_{\mathrm{bd}}S^{-}\rho S^{+}+\Gamma_{\mathrm{deph}}\left|1\right\rangle \left\langle 1\right|\rho\left|1\right\rangle \left\langle 1\right|.
\end{eqnarray}
Here, we have decomposed the effective decay rate $\tilde{\kappa}$
as 
\begin{equation}
\tilde{\kappa}=\tilde{\kappa}_{\mathrm{gd}}+\tilde{\kappa}_{\mathrm{bd}}=\frac{4g^{2}}{\kappa^{2}}\kappa_{\mathrm{gd}}+\frac{4g^{2}}{\kappa^{2}}\kappa_{\mathrm{bd}}.
\end{equation}
Formally solving Eq.(\ref{eq:effective-Master-equation-Hamiltonian-jumps})
gives 
\begin{equation}
\rho\left(t\right)=\mathcal{U}\left(t\right)\rho\left(0\right)+\int_{0}^{t}d\tau\mathcal{U}\left(t-\tau\right)\mathcal{J}\rho\left(\tau\right),\label{eq:effective-Master-equation-formal-solution}
\end{equation}
with the total jump operator $\mathcal{J}\rho=\mathcal{J}_{\mathrm{gd}}\rho+\mathcal{J}_{\mathrm{bd}}\rho$
and 
\begin{equation}
\mathcal{U}\left(t\right)\rho=e^{-iHt}\rho e^{iH^{\dagger}t}
\end{equation}
The exact solution given in Eq.(\ref{eq:effective-Master-equation-formal-solution})
can be iterated, giving an illustrative expansion in terms of the
jumps $\mathcal{J}$. It reads 
\begin{eqnarray*}
\rho\left(t\right) & = & \mathcal{U}\left(t\right)\rho\left(0\right)+\int_{0}^{t}d\tau_{1}\mathcal{U}\left(t-\tau_{1}\right)\mathcal{J}\mathcal{U}\left(\tau_{1}\right)\rho\left(0\right)\\
 &  & +\int_{0}^{t}d\tau_{2}\int_{0}^{\tau_{2}}d\tau_{1}\mathcal{U}\left(t-\tau_{2}\right)\mathcal{J}\mathcal{U}\left(\tau_{2}-\tau_{1}\right)\times\\
 &  & \mathcal{J}\mathcal{U}\left(\tau_{1}\right)\rho\left(0\right)+\dots
\end{eqnarray*}
Here, the $n$-th order term comprises $n$ jumps $\mathcal{J}$ with
free evolution $\mathcal{U}$ between the jumps. Now, we can single
out the desired events where the first quantum jump is governed by
$\mathcal{J}_{\mathrm{gd}}$. This leads to the definition 
\begin{equation}
\rho\left(t\right)=\mathcal{U}\left(t\right)\rho\left(0\right)+\rho_{\mathrm{gd}}\left(t\right)+\rho_{\mathrm{bd}}\left(t\right),
\end{equation}
where $\rho_{\mathrm{gd}}\left(t\right)$ subsumes all desired trajectories
\begin{eqnarray}
\rho_{\mathrm{gd}}\left(t\right) & = & \int_{0}^{t}d\tau_{1}\mathcal{U}\left(t-\tau_{1}\right)\mathcal{J}_{\mathrm{gd}}\mathcal{U}\left(\tau_{1}\right)\rho\left(0\right)\label{eq:definition-rho-good}\\
 &  & +\int_{0}^{t}d\tau_{2}\int_{0}^{\tau_{2}}d\tau_{1}\mathcal{U}\left(t-\tau_{2}\right)\mathcal{J}\mathcal{U}\left(\tau_{2}-\tau_{1}\right)\times\nonumber \\
 &  & \mathcal{J}_{\mathrm{gd}}\mathcal{U}\left(\tau_{1}\right)\rho\left(0\right)+\dots\nonumber 
\end{eqnarray}

We focus on a qubit, initially in the excited state, i.e., $\rho\left(0\right)=\left|1\right\rangle \left\langle 1\right|$.
Using the relations $\mathcal{U}\left(\tau_{1}\right)\rho\left(0\right)=e^{-\gamma_{\mathrm{eff}}\tau_{1}}\left|1\right\rangle \left\langle 1\right|,$
$\mathcal{J}_{\mathrm{gd}}\mathcal{U}\left(\tau_{1}\right)\rho\left(0\right)=\tilde{\kappa}_{\mathrm{gd}}e^{-\gamma_{\mathrm{eff}}\tau_{1}}\left|0\right\rangle \left\langle 0\right|$
and 
\begin{eqnarray}
\mathcal{J}_{\mathrm{gd}}\mathcal{U}\left(\tau_{2}-\tau_{1}\right)\mathcal{J}_{\mathrm{gd}}\mathcal{U}\left(\tau_{1}\right)\rho\left(0\right) & = & 0,\\
\mathcal{J}_{\mathrm{bd}}\mathcal{U}\left(\tau_{2}-\tau_{1}\right)\mathcal{J}_{\mathrm{gd}}\mathcal{U}\left(\tau_{1}\right)\rho\left(0\right) & = & 0,
\end{eqnarray}
the qubit's density matrix evaluates (to all orders in $\mathcal{J}$)
to 
\begin{eqnarray}
\rho\left(t\right) & = & e^{-\gamma_{\mathrm{eff}}t}\left|1\right\rangle \left\langle 1\right|+\rho_{\mathrm{gd}}\left(t\right)+\rho_{\mathrm{bd}}\left(t\right),\\
\rho_{\mathrm{gd}}\left(t\right) & = & \frac{\tilde{\kappa}_{\mathrm{gd}}}{\gamma_{\mathrm{eff}}}\left(1-e^{-\gamma_{\mathrm{eff}}t}\right)\left|0\right\rangle \left\langle 0\right|.
\end{eqnarray}
In the long-time limit $t\rightarrow\infty$, the system reaches the
steady state, $\rho\left(t\rightarrow\infty\right)=\rho_{\mathrm{gd}}+\rho_{\mathrm{bd}}$,
where $\rho_{\mathrm{gd}}=\frac{\tilde{\kappa}_{\mathrm{gd}}}{\gamma_{\mathrm{eff}}}\left|0\right\rangle \left\langle 0\right|$.
The associated success probability $p_{\mathrm{suc}}=\mathrm{Tr}\left[\rho_{\mathrm{gd}}\right]$
for faithful decay through the mirrors is then 
\begin{equation}
p_{\mathrm{suc}}=\frac{\tilde{\kappa}_{\mathrm{gd}}}{\tilde{\kappa}_{\mathrm{gd}}+\tilde{\kappa}_{\mathrm{bd}}+\Gamma_{\mathrm{deph}}}=\frac{1}{\frac{\kappa}{\kappa_{\mathrm{gd}}}+\frac{1}{4}\frac{\kappa^{2}\Gamma_{\mathrm{deph}}}{g^{2}\kappa_{\mathrm{gd}}}},\label{eq:p-succesful-general-expression}
\end{equation}
which is a simple branching ratio comparing the strength of the desired
decay channel $\sim\tilde{\kappa}_{\mathrm{gd}}$ to the undesired
ones $\sim\tilde{\kappa}_{\mathrm{bd}}+\Gamma_{\mathrm{deph}}$. In
the limit where $\kappa_{\mathrm{bd}}=0$, i.e., $\kappa=\kappa_{\mathrm{gd}}$,
the expression for $p_{\mathrm{suc}}$ simplifies to 
\begin{equation}
p_{\mathrm{suc}}=\frac{1}{1+\frac{1}{4}\frac{1}{C}}\overset{\underrightarrow{C\gg1}}{}1,
\end{equation}
with the usual definition found in the literature, $C=g^{2}/\left(\kappa\Gamma_{\mathrm{deph}}\right)=g^{2}T_{2}/\kappa$;
here, $\kappa=\omega_{\mathrm{c}}/Q_{\mathrm{eff}}=\left(\bar{n}_{\mathrm{th}}+1\right)\omega_{\mathrm{c}}/Q$
is understood to account for thermal occupation of the environment
$\bar{n}_{\mathrm{th}}$ in terms of a decreased mechanical quality
factor (compare for example Refs.\cite{kolkowitz12,ovartchaiyapong14}). 

It is instructive to rewrite the general expression for $p_{\mathrm{suc}}$
given in Eq.(\ref{eq:p-succesful-general-expression}) as 
\begin{equation}
p_{\mathrm{suc}}=\frac{1}{\left(1+\varepsilon\right)\left[1+\frac{1}{4C}\right]}.\label{eq:p-succesful-general-expression-2}
\end{equation}
with $\varepsilon=\kappa_{\mathrm{bd}}/\kappa_{\mathrm{gd}}$. Based
on this definition, it is evident that two conditions need to be satisfied
in order to reach $p_{\mathrm{suc}}\rightarrow1$ in the regime where
$\kappa_{\mathrm{bd}}>0$: (i) a low undesired loss rate, $\varepsilon=\kappa_{\mathrm{bd}}/\kappa_{\mathrm{gd}}\ll1$,
and (ii) high cooperativity, $C\equiv\frac{g^{2}}{\kappa\Gamma_{\mathrm{deph}}}=\frac{g^{2}T_{2}}{\kappa}\gg1$,
with $\kappa=\kappa_{\mathrm{gd}}+\kappa_{\mathrm{bd}}$. For an illustration,
compare Fig.\ref{fig:coop-parameter_bad-decay_success-prob1}. This
shows, that the usual definition and interpretation of the cooperativity
$C$ holds, provided that $\varepsilon\ll1$ is fulfilled.\textcolor{black}{{}
In order to quantify the cooperativity $C$ for SAW cavity modes both
in the $\sim\mathrm{MHz}$ $\left(\bar{n}_{\mathrm{th}}\gg1\right)$
and the $\sim\mathrm{GHz}$ $\left(\bar{n}_{\mathrm{th}}\ll1\right)$
regime, in the main text we take a (conservative) estimate as $C\equiv g^{2}T_{2}Q/\left[\omega_{c}\left(\bar{n}_{\mathrm{th}}+1\right)\right]$.
For artificial atoms (quantum dots, superconducting qubits, NV-centers,
$\dots$) with resonant frequencies $\sim\mathrm{GHz}$, at cryostatic
temperatures this definition reduces to $C\approx g^{2}T_{2}/\kappa$
as discussed above, whereas for a trapped ion with $\omega_{t}/2\pi\approx\omega_{c}/2\pi\sim\mathrm{MHz}$
it correctly gives $C\approx g^{2}T_{2}Q/\left(\omega_{c}\bar{n}_{\mathrm{th}}\right)$
with a decreased effective quality factor $Q_{\mathrm{eff}}=Q/\bar{n}_{\mathrm{th}}$
\cite{bennett13,kolkowitz12,bennett12}. }

\textit{\textcolor{black}{Fidelity estimate.}}\textcolor{black}{---For
small errors, the expression given in Eq.(\ref{eq:p-succesful-general-expression-2})
can be approximated as $p_{\mathrm{suc}}\approx1-\varepsilon-1/(4C)$.
Since the absorption process is just the time-reversed copy of the
emission process in the state transfer protocol for two nodes, we
can estimate the state transfer fidelity as $\mathcal{F}\geq p_{\mathrm{suc}}\times p_{\mathrm{suc}}$.
For small infidelities, we then find 
\begin{equation}
\mathcal{F}\gtrsim1-2\varepsilon-\frac{1}{2C},
\end{equation}
where the individual errors arise from intrinsic phonon losses $\sim\varepsilon$
and qubit dephasing $\sim C^{-1}\sim T_{2}^{-1}$, respectively. This
simple analytical estimate agrees well with numerical results presented
in Ref.\cite{stannigel10}, where (except for noise sources that are
irrelevant for our problem) $\mathcal{F}\approx1-(2/3)\varepsilon/\left(1+\varepsilon\right)-\mathcal{C}C^{-1}$
with a numerical coefficient $\mathcal{C}=\mathcal{O}\left(1\right)$
depending on the specific pulse sequence. Since $\varepsilon\ll1$,
this relation can be simplified to $\mathcal{F}\approx1-(2/3)\varepsilon-\mathcal{C}C^{-1}$.
For the state transfer of the coherent superposition $\left|\psi\right\rangle =\left(\left|0\right\rangle -\left|1\right\rangle \right)/\sqrt{2}$
as described in detail in Appendix \ref{sec:State-Transfer-Protocol},
we have explicitly verified the linear scaling $\sim\varepsilon$
and find numerically $\sim1/2\varepsilon$ for intrinsic phonon losses
{[}compare also Fig.\ref{fig:state-transfer-simulation-OHnoise} in
the main text, where $\mathcal{F}\approx95\%$ for $\varepsilon\approx10\%$
and $\sigma_{\mathrm{nuc}}=0${]} and take $\sim\varepsilon$ as a
simple estimate in Eq.(\ref{eq:fidelity-estimate}). Using experimentally
achievable parameters $\varepsilon\approx5\%$ and $C\approx30$,
we can then estimate $\mathcal{F}\approx90\%$.}

\section{Effects due to the Structure Defining the Quantum Dots\label{sec:QD-structure}}

In our analysis of charge and spin qubits defined in quantum dots,
we have neglected any potential effects arising due to the structure
defining the quantum dots, that is (i) the heterostructure for the
2DEG and (ii) the metallic top gates for confinement of single electrons.
In this Appendix, we give several arguments corroborating this approximate
treatment, showing that the QD structure does not negatively influence
the cavity nor the coupling between qubit and cavity. 

\textit{Heterostructure.}---Following the arguments given in Ref.\cite{simon96},
the 2DEG is taken to be a thin conducting layer a distance $d$ away
from the surface of a homogeneous $\mathrm{Al}_{x}\mathrm{Ga}_{1-x}\mathrm{As}$
crystal with typically $x\approx30\%$. This treatment is approximately
correct since the relevant material properties (elastic constants,
densities and dielectric constants) of $\mathrm{Al}_{x}\mathrm{Ga}_{1-x}\mathrm{As}$
and GaAs are very similar \cite{simon96}. The mode-functions and
speed of sound are largely defined by the elastic constants \cite{simon96}
which are roughly the same for both $\mathrm{Al}_{x}\mathrm{Ga}_{1-x}\mathrm{As}$
and pure GaAs; for example, the speed of the Rayleigh SAW for $\mathrm{Al}_{0.3}\mathrm{Ga}_{0.7}\mathrm{As}$
is $v_{s}\approx3010\mathrm{m/s}$ which differs from that of pure
GaAs by only 5\% \cite{simon96}. Moreover, the numerical values for
the material-dependent parameter $H$ entering the amplitude of the
mechanical zero-point motion $U_{0}$ according to Eq.(\ref{eq:estimate-zero-point-motion-amplitude-Simon})
differ by 2\% only \cite{simon96}; accordingly, the estimate for
our key figure of merit $U_{0}$ should be rather accurate. Also,
the piezo-electric coupling constants are rather similar, with $e_{14}\approx0.15\mathrm{C/m^{2}}$
for pure GaAs and $e_{14}\approx0.145\mathrm{C/m^{2}}$ for $\mathrm{Al}_{0.3}\mathrm{Ga}_{0.7}\mathrm{As}$
\cite{simon96,kornich14} yielding an accurate estimate for $\phi_{0}$
and $\xi_{0}$, respectively. Lastly, the heterostructure is not expected
to severely affect the $Q$-factor, since very high $Q$-values reaching
$Q>10^{4}$ have been observed in previous SAW experiments on AlN/diamond
heterostructures \cite{rodriguez-madrid12,benetti05}, where the differences
in material properties are considerably larger than for the heterostructure
making up the 2DEG.

\textit{Top gates.}---For the following reasons, we have disregarded
effects due to the presence of the metallic top gates: (i) In Ref.\cite{aizin98},
a closed form analytic solution for the piezoelectric potential $\phi\left({\bf x},t\right)$
accompanying a SAW on the surface of a GaAs/$\mathrm{Al}_{x}\mathrm{Ga}_{1-x}\mathrm{As}$
in the presence of a narrow metal gate has been obtained. In particular,
it is shown that $\phi\left({\bf x},t\right)$ is screened \textit{right}
below the gate, but remains practically unchanged with respect to
the ungated case outside of the edges of the gate. Since the QD electrons
are confined to regions outside of the metallic gates, they experience
the piezoelectric potential as calculated for the ungated case (see
Appendix \ref{sec:Piezoelectric-SAW} for details); therefore, our
estimates---where this screening effect has been neglected---remain
approximately valid. (ii) In Ref.\cite{naber06}, the coupling of
a traveling SAW to electrons confined in a DQD has been experimentally
studied. Here, in very good agreement to the experimental results,
the potential felt by the QD electrons has been taken as $V_{\mathrm{pe}}\sim\mathrm{sin}\left(kl\right)$
(where $l$ is the lithographic distance between the dots) confirming
the sine-like mode profile as used in our estimates. Moreover, with
$l=220\mathrm{nm}$ and $\lambda=1.4\mu\mathrm{m}$ the parameters
used in this experiment perfectly match the ones used in our estimates.
Intuitively (in the spirit of the standard electric dipole approximation
in quantum optics), since $\lambda\gg l$, the SAW mode cannot resolve
the dot structure and thus remains largely unaffected. (iii) In Ref.\cite{gustaffson12},
single phonon SAW pulses have been detected via a single electron
transistor (SET) directly deposited on the GaAs substrate with time-resolved
measurements clearly identifying the coupling as piezo-electric. Similarly
to a QD, the SET is defined by metallic gates. Here, a relation between
vertical surface displacement and surface charge induced on the SET
is theoretically derived. Using standard tabulated parameter values
for GaAs and neglecting any effects due to the presence of the metallic
gates, very good agreement with the experimental results is achieved.
In particular, based on the results given in Ref.\cite{gustaffson12},
a straightforward estimate gives $U_{0}\approx30\mathrm{am}$ for
the rather large cavity with $A\approx10^{6}\mu\mathrm{m}^{2}$, whereas
Eq.(\ref{eq:mechanical-zero-point-motion}) yields a smaller, conservative
estimate $U_{0}\approx2\mathrm{am}$, due to the averaging over the
quantization area $A$. (iv) The $Q$-factor of the cavity is not
expected to be severely affected by the presence of the metallic gates
since metallic Al cladding layers have been used on a GaAs substrate
to show basically dissipation-free SAW propagation over millimeter
distances \cite{deLima04,deLima05}. (v) Finally, there is a large
body of previous theoretical works on electron-phonon coupling in
gate-defined QDs (see for example Refs.\cite{kornich14,bruus93,brandes99,srinivasa14})
where any effects due to the structure defining the QDs have been
neglected as well. As a matter of fact, our description for the electron-phonon
coupling emerges directly from these previous treatments in the limit
where the continuum of phonon modes is replaced by a single relevant
SAW cavity mode (similar to cavity QED, other bulk modes are still
present and contribute to the decoherence of the qubit on a timescale
$T_{2}$). For example, the piezoelectric electron-phonon interaction
is given in Ref.\cite{srinivasa14} as 
\begin{equation}
H_{\mathrm{GaAs}}=\beta\sum_{k,\mu}\sqrt{\frac{\hbar}{2\rho Vv_{\mu}k}}\mathcal{M}_{k}\left[a_{k,\mu}+a_{-k,\mu}^{\dagger}\right],
\end{equation}
where $a_{k,\mu}^{\dagger}$ creates an acoustic phonon with wave
vector $k$ and polarization $\mu$ and $\mathcal{M}_{k}$ refers
to the matrix element for electron-phonon coupling; for free bulk
modes, $\mathcal{M}_{k}=\sum_{i,j}\sum_{\sigma}\left<i|\exp\left[i{\bf k}{\bf r}\right]|j\right>d_{i\sigma}^{\dagger}d_{j\sigma}$.
In agreement with our notation, the coupling constant is $\beta=ee_{14}/\epsilon$
and the square-root factor can be identified with $U_{0}$. Replacing
the sum by a single-relevant cavity mode $a$, we recover the Hamiltonian
describing the cavity-qubit coupling with $g\sim\beta U_{0}=e\phi_{0}$;
compare Eq.(\ref{eq:Hamiltonian-charge-qubit}).

\section{State Transfer Protocol\label{sec:State-Transfer-Protocol}}

\begin{figure}
\includegraphics[width=1\columnwidth]{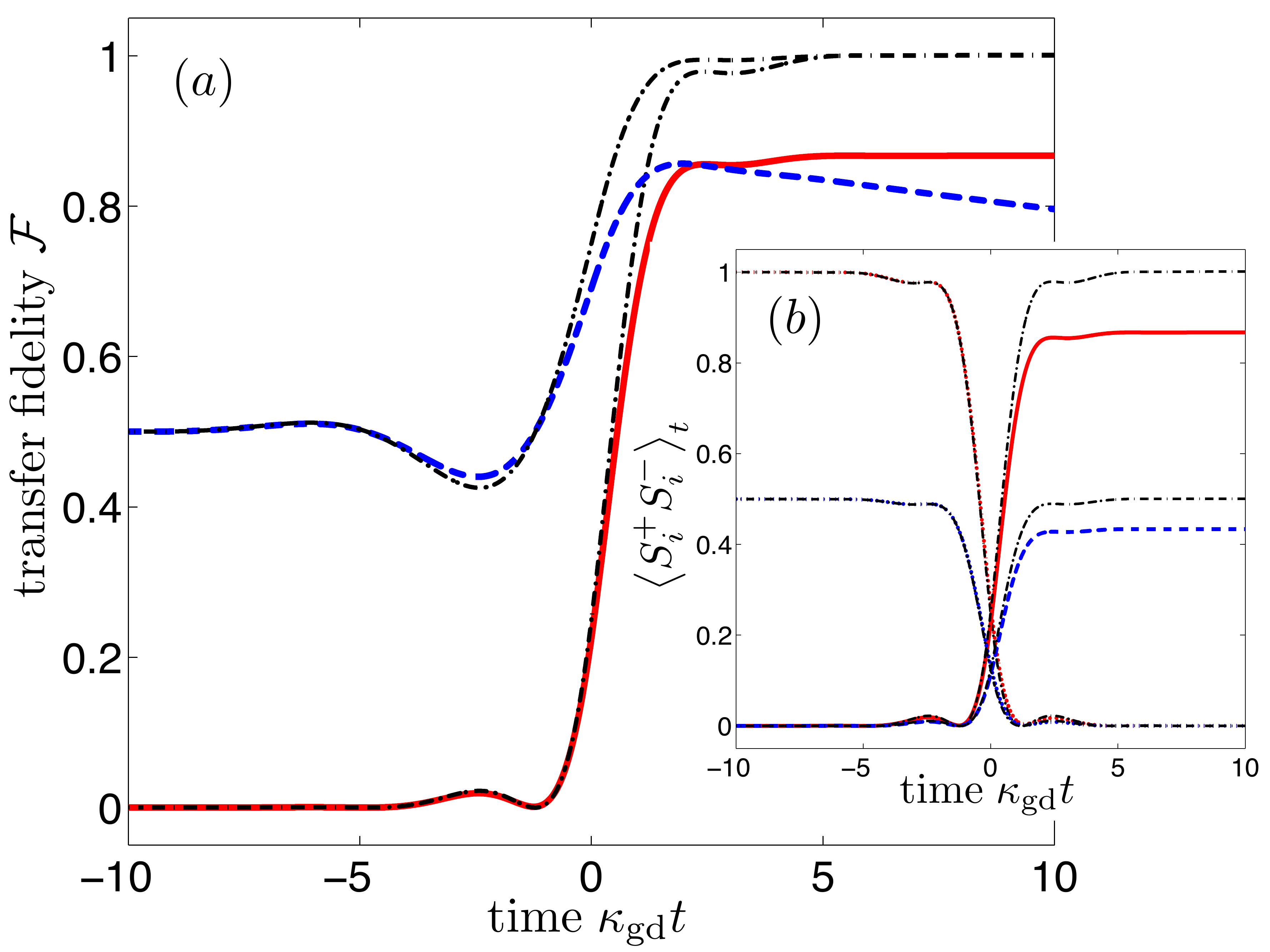}

\caption{\label{fig:state-transfer-simulation}(color online). Numerical simulation
of state transfer for two different initial states $\left|1\right\rangle _{1}$
(red solid line) and $\left(\left|0\right\rangle _{1}-\left|1\right\rangle _{1}\right)/\sqrt{2}$
(blue dashed line) in the presence of Markovian noise. Black (dash-dotted)
curves refer to the ideal, noise-free scenario $\left(\mathcal{L}_{\mathrm{noise}}=0\right)$
where perfect transfer is achieved, while colored curves take into
account decoherence processes. (a) Transfer fidelity $\mathcal{F}$;
(b) Excited-state occupation $\left\langle S_{i}^{+}S_{i}^{-}\right\rangle _{t}$
for first and second qubit. Numerical parameters: $g_{1}\left(t\geq0\right)=\kappa_{\mathrm{gd}}$,
$\kappa_{\mathrm{bd}}/\kappa_{\mathrm{gd}}=5\%$ and $\Gamma_{\mathrm{deph}}/\kappa_{\mathrm{gd}}=5\%$. }
\end{figure}

\begin{figure}
\includegraphics[width=1\columnwidth]{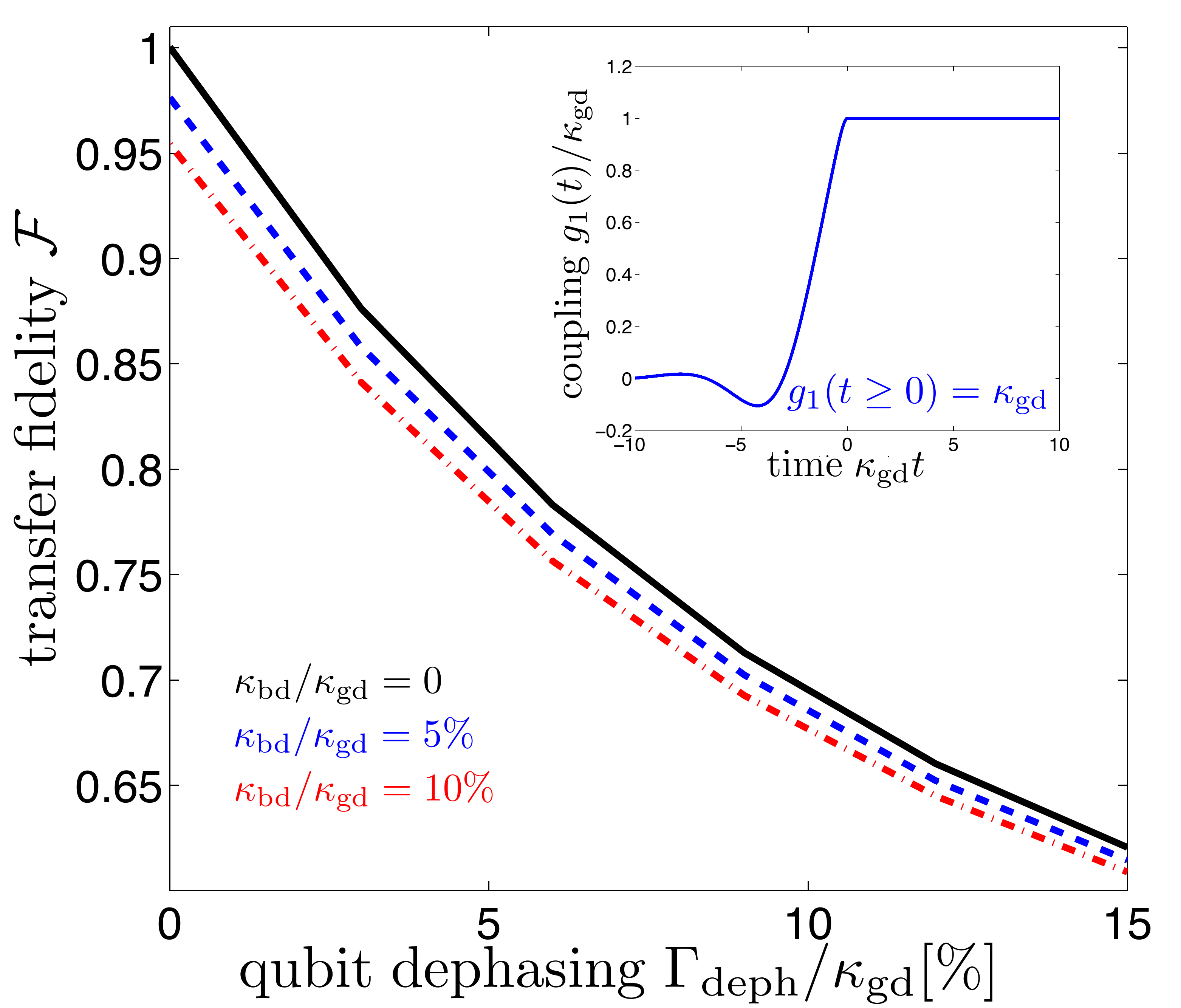}

\caption{\label{fig:state-transfer}(color online). Quantum state-transfer
protocol in the presence of Markovian noise. (a) State transfer fidelity
$\mathcal{F}$ for a coherent superposition $\left|\psi\right\rangle =\left(\left|0\right\rangle -\left|1\right\rangle \right)/\sqrt{2}$
as a function of the qubit's dephasing rate $\Gamma_{\mathrm{deph}}$
for different values of intrinsic phonon losses $\kappa_{\mathrm{bd}}/\kappa_{\mathrm{gd}}=0$
(black solid), $\kappa_{\mathrm{bd}}/\kappa_{\mathrm{gd}}=5\%$ (blue
dashed) and $\kappa_{\mathrm{bd}}/\kappa_{\mathrm{gd}}=10\%$ (red
dash-dotted). (b) Pulse shape $g_{1}\left(t\right)$ for first node. }
\end{figure}

In this Appendix, we provide further details on the numerical simulation
of the state transfer protocol as described by Eqs.(\ref{eq:Liouviilian-ideal})
and (\ref{eq:Liovillian-noise}), in the presence of Markovian noise.
In line with previous theoretical studies \cite{bennett13}, we show
that the simple approximate Markovian noise treatment results in a
pessimistic estimate for the noise transfer fidelity $\mathcal{F}$. 

We first provide results of the full time-dependent numerical simulation
of the cascaded Master equation given in Eqs.(\ref{eq:Liouviilian-ideal})
and (\ref{eq:Liovillian-noise}), including an exponential loss of
coherence for $\Gamma_{\mathrm{deph}}>0$; see Fig.\ref{fig:state-transfer-simulation}.
In contrast to the non-Markovian noise model discussed in the main
text, the qubits are assumed to be on resonance throughout the evolution,
that is $\delta_{i}=0\left(i=1,2\right)$, but experience undesired
noise as described by 
\begin{equation}
\mathcal{L}_{\mathrm{noise}}\rho=2\kappa_{\mathrm{bd}}\sum_{i=1,2}\mathcal{D}\left[a_{i}\right]\rho+\Gamma_{\mathrm{deph}}\sum_{i=1,2}\mathcal{D}\left[S_{i}^{z}\right]\rho.
\end{equation}
Here, the second term refers to a standard Markovian pure dephasing
term that leads to an exponential loss of coherence $\sim\exp\left(-\Gamma_{\mathrm{deph}}t/2\right)$.
As a reference we also show the results for the ideal, noise-free
scenario $(\mathcal{L}_{\mathrm{noise}}=0)$, where perfect state
transfer is achieved \cite{cirac97}. The results of this type of
time-dependent numerical simulations are then summarized in Fig.\ref{fig:state-transfer}.
For optimized, but experimentally achievable parameters $T_{2}^{\star}\approx3\mu\mathrm{s}$
\cite{shulman14}, and accordingly $\Gamma_{\mathrm{deph}}/\kappa_{\mathrm{gd}}\approx3\%$,
we then obtain $\mathcal{F}\approx0.85$, in the presence of realistic
undesired phonon losses $\kappa_{\mathrm{bd}}/\kappa_{\mathrm{gd}}=5\%$.

This shows that our non-Markovian noise model yields even higher state
transfer fidelities than the Markovian noise model. Intuitively, this
can be readily understood as follows: The simple Markovian noise model
gives a coherence decay $\sim\exp\left(-t/T_{2}\right)$, whereas
our non-Markovian noise model yields $\sim\exp\left(-t^{2}/T_{2}^{2}\right)$.
Therefore, for Markovian noise one can estimate the dephasing induced
error on the relevant timescale for state transfer $\sim\kappa^{-1}$
as $\sim\kappa^{-1}/T_{2}\approx\Gamma_{\mathrm{deph}}/\kappa\ll1$,
whereas non-Markovian noise leads to a considerably smaller error
$\sim\left(T_{2}^{-1}/\kappa\right)^{2}$.


\begin{thebibliography}{10}
\bibitem{JC-comment}Depending on the particular physical implementation
of the qubit, the Jaynes-Cummings type Hamiltonian given in Eq.(\ref{eq:generic-Jaynes-Cummings-Hamiltonian})
may emerge only as an effective description of a potentially driven,
two- or multilevel system. However, for e.g. charge or spin qubits
embedded in GaAs quantum dots it is a rather straightfoward way to
describe the coherent dynamics of a single node.
\bibitem{cooperativity}The cooperativity parameter $C$ compares
the coherent single phonon coupling strength $g$ with the geometric
mean of the relevant incoherent processes, i.e., the qubit\textquoteright{}s
decoherence rate and the cavity\textquoteright{}s effective linewidth;
in direct analogy to cavity QED, it is a key figure of merit in our
system as $C>1$ marks the onset of coherent quantum effects in a
coupled spin-oscillator system \cite{kolkowitz12}.
\bibitem{numerical-values-phi0}The numerical values we obtain for
$\phi_{0}$ agree very well with material-dependent constants $c_{\alpha}=\left|U_{\alpha}/\phi\right|\approx(0.1-1)\mathrm{fm/\mu V}$
for $\left(\alpha=x,z\right)$ tabulated in Refs.\cite{datta86,morgan07}.
\bibitem{spin-qubit-DQD}\textcolor{black}{This system is a particulary
promising physical realization as it provides a relatively high cooperativity
$C$ even in the absence of spin-echo pulses, the spin-resonator Hamiltonian
takes on the generic Jaynes-Cummings Hamiltonian Eq.(\ref{eq:generic-Jaynes-Cummings-Hamiltonian})
in a rather straight-forward way (Appendix \ref{sec:spin-qubit-QED})
{[}whereas Eq.(\ref{eq:Hamiltonian-trapped-ion}) and (\ref{eq:Hamiltonian-NV-center})
require further ingredients such as a MW drive (compare for example
Ref.\cite{habraken12}) to effectively obtain Eq.(\ref{eq:generic-Jaynes-Cummings-Hamiltonian}){]},
and, outside of the state-transfer time window, it provides better
control techniques such as spin-echo pulses than the simple charge-qubit
scenario.}
\bibitem{spin-coupling}Note, however, that more elaborate optimization
yields an optimal spot with an even larger coupling $g\approx0.3g_{\mathrm{ch}}$
\cite{taylor06}.
\bibitem{unidirectional-propagation}In a perfect realization of the
transfer scheme the system remains in a dark state throughout the
evolution and no field is scattered from the second mirror. Therefore,
the assumption of unidirectional propagation is not strictly needed;
compare Ref.\cite{cirac97}.
\end{thebibliography}
\end{document}